\documentclass[11pt]{article}
\usepackage{amsmath,amssymb}
\usepackage[margin=1in]{geometry}
\usepackage{bm}
\usepackage{expl3}
\usepackage{lmodern,amsmath,amssymb,amsthm,mathtools,microtype}
\usepackage{booktabs,tabularx,array}
\usepackage{xltabular}
\usepackage{tikz}
\usetikzlibrary{arrows.meta,decorations.pathreplacing}
\usepackage{tikz}
\usetikzlibrary{arrows.meta,decorations.pathreplacing}
\definecolor{hc}{RGB}{35,91,155}   
\definecolor{vc}{RGB}{172,62,44}   
\definecolor{safe}{RGB}{72,126,82}
\definecolor{capcol}{RGB}{112,167,199}
\definecolor{pending}{HTML}{F0ECFA}
\definecolor{passed}{HTML}{CBEBE2}
\definecolor{retained}{HTML}{F9DFB0}
\definecolor{ink}{HTML}{253848}
\definecolor{cubeedge}{HTML}{526377}
\definecolor{rectedge}{HTML}{7754A5}
\definecolor{passedge}{HTML}{16866F}

\usetikzlibrary{calc,decorations.pathreplacing,positioning}
\usepackage{titlesec}
\titlespacing*{\paragraph}{0pt}{1.5ex}{1em}

\usepackage{appendix}
\usepackage{subfigure}
\usepackage[english]{babel}
\usepackage{bbm}
\usepackage{graphicx}
\usepackage[center]{caption2}
\usepackage{amsfonts,amssymb,amsmath,latexsym,amsthm}
\usepackage{multirow}
\usepackage{geometry}
\usepackage[usenames,dvipsnames]{pstricks}
\usepackage{epsfig}
\usepackage{pst-grad} 
\usepackage{pst-plot} 
\usepackage[space]{grffile} 
\usepackage[shortlabels]{enumitem}
\usepackage{etoolbox} 
\makeatletter 
\patchcmd\Gread@eps{\@inputcheck#1 }{\@inputcheck"#1"\relax}{}{}
\makeatother

\newtheorem*{thm*}{Theorem}

\usepackage[colorlinks=true,linkcolor=blue,citecolor=blue]{hyperref}
\usepackage[capitalise,nameinlink]{cleveref}

\newtheorem{thm}{Theorem}[section]
\newtheorem{cor}[thm]{Corollary}
\newtheorem{lem}[thm]{Lemma}

\newtheorem{prop}[thm]{Proposition}

\newtheorem{quest}[thm]{Question}

\newtheorem{claim}[thm]{Claim}

\newtheorem{defn}[thm]{Definition}

\newtheoremstyle{plainupright}
  {\topsep}   
  {\topsep}   
  {\upshape}  
  {}          
  {\bfseries} 
  {.}         
  { }         
  {}          
\theoremstyle{plainupright}

\newtheorem{remark}{Remark}

\let\svthefootnote\thefootnote
\newcommand\blankfootnote[1]{%
	\let\thefootnote\relax\footnotetext{#1}%
	\let\thefootnote\svthefootnote%
}

\newcommand{\omt}[1]{}

\newcommand\supproof[1]{}

\newcommand{\X}{\mathcal X}
\newcommand{\LCA}{\operatorname{LCA}}
\newcommand{\Prot}{\operatorname{Prot}}

\newcommand{\collec}{{\mathcal{X}}}

\newcommand{\lbanach}{d_{banach}}

\newcommand{\Q}{\mathcal Q}

\newcommand{\cR}{\mathcal R}
\newcommand{\cP}{\mathcal P}
\newcommand{\Hist}{\mathcal H}

\newcommand{\EndLang}{\operatorname{end}}

\ExplSyntaxOn

\ExplSyntaxOff

\begin{document}
\title{Characterizing Local Breadth in Language Generation}

\author{
Jon Kleinberg\thanks{
Department of Computer Science and Information Science,
Cornell University, Ithaca, NY, USA.
}
\and
Fan Wei\thanks{
Department of Mathematics,
Duke University, Durham, NC, USA.
}
}
\date{}
\maketitle

\begin{abstract}
The rapid development of large language models makes a basic tension in generation especially important: a useful generator should avoid hallucination, but it should also avoid collapsing onto a narrow subset of valid outputs. We study this \emph{validity--breadth trade-off} in the Kleinberg--Mullainathan model of language generation in the limit, where the generator receives only positive examples from an unknown target language, revealed in an arbitrary and possibly adversarial order, with no distributional or computational assumptions.

We measure breadth by lower Banach density, which requires substantial coverage in every sufficiently large local region and therefore rules out arbitrarily large sparse holes. This local notion reveals structure that is invisible to lower asymptotic density, which measures coverage along prefixes of a fixed ordering.

\emph{(1) Exact optima in one dimension.}
For every target language $K$ in a countable family, we give an exact structural characterization of its optimal lower Banach density $b(K)$, with equivalent combinatorial
and topological formulations. Moreover, a single eventually valid generator attains $b(K)$ simultaneously for every target. For every $\alpha\in[0,1/2]$, we construct a countable language family containing a target
$K$ with $b(K)=\alpha$. Moreover, every target of finite Cantor--Bendixson point rank has $b(K)=1/2$, even when the ranks within the family are unbounded.

\emph{(2) Unavoidable trade-offs in higher dimensions.}
In dimensions \(d\ge2\), a qualitatively new phenomenon appears: there need not be a single best generator. A generator that achieves the best possible density when one language is the target may necessarily perform worse if the target turns out to be another language. Thus, unlike in one dimension, generator design becomes intrinsically multi-objective: one must decide which possible targets to prioritize. We make this conflict quantitative by constructing an explicit family with an unavoidable trade-off between the guarantees achievable on different targets.

\emph{(3) Complete characterization of density profiles.}
Since target-specific optima need not be compatible, the right object in higher dimensions is the full region of guarantees achievable simultaneously by one generator.
Once geometrically degenerate rectangles are excluded, we give an exact structural characterization of this region, with equivalent combinatorial and topological formulations. We also construct families whose exact achievable density regions are arbitrary prescribed products \(\prod_i[0,\alpha_i]\), with \(\alpha_i\in[0,1/2]\).

\emph{(4) Near-half guarantees and the sharp endpoint.}
For every $\varepsilon>0$, one generator simultaneously guarantees filtered density at least $1/2-\varepsilon$ at every target of finite Cantor--Bendixson point rank, without a uniform rank bound and under a near-logarithmic nondegeneracy condition, improving the previously known polylogarithmic threshold. The exact $1/2$ endpoint, however, can fail even for a rank-zero singleton family.
\end{abstract}

\clearpage
\pagestyle{empty}
\setcounter{tocdepth}{1}
\tableofcontents
\clearpage

\pagenumbering{arabic}  
\pagestyle{plain}  
\section{Introduction}\label{sec:intro}

The success of modern generative models raises a basic
information-theoretic question: what can be generated from positive
examples alone when the underlying language is unknown? A generator
should eventually avoid invalid outputs, but it should also cover the
target broadly rather than collapse onto a narrow subset.

We study this question in the framework of \emph{language generation
in the limit}
\cite{kleinberg2024limit}. Unlike work that analyzes the computational
power of particular architectures
\cite{chen2025transformer,peng-transformer,
sanford-transformer,strobl2024formal},
this framework allows arbitrary computation and imposes no probability
distribution on the observed examples. It therefore isolates the
limitations caused by incomplete information itself: impossibility
results are architecture-independent, while positive results describe
what is achievable in principle.

\paragraph{A minimal-assumption model: generation in the limit.}
Building on the classical theory
of language identification developed by Gold and Angluin
\cite{gold1967language,angluin1979finding,angluin1980inductive},
the newer framework of language generation in the limit introduced by Kleinberg and Mullainathan \cite{kleinberg2024limit} takes a \textbf{minimal-assumption, non-probabilistic}
approach to language generation.

Consider a countable collection of infinite languages
$\collec=\{L_1,L_2,\ldots\}$,
where each language is an infinite set of finite strings.
An adversary chooses an unknown target language
$K\in\collec$ and enumerates its strings one at a time,
in an arbitrary order.

At time $t$, the algorithm has observed the finite set
$S_t=\{x_1,\ldots,x_t\}\subseteq K$ of strings revealed
by the adversary. The generator should in response produce a string
$o_t$ that has neither been revealed by the adversary 
nor previously generated by the algorithm. 
The algorithm achieves \emph{generation in the limit}
if there exists a finite time $t^*$, possibly depending
on the adversary's enumeration, such that $o_t \in K$
for every $t\ge t^*$. Thus, the algorithm may
initially produce incorrect outputs, but must eventually
generate only new strings from the true language.

The model is non-probabilistic and uses minimal assumptions in the following sense: the
generator receives only positive examples, with no probability
distribution imposed on the strings or their order of appearance, and
we place no computational restriction on the generator. It therefore
isolates the informational foundations of language generation
from statistical assumptions and implementation details. The objective
is weaker than classical identification in the limit, where the learner
must eventually identify \(K\) exactly.
Although identification is known to be impossible for some countable language
families \cite{gold1967language,angluin1979finding,angluin1980inductive}, Kleinberg and Mullainathan proved that eventual generation is
possible for every countable family of infinite languages
\cite{kleinberg2024limit}. This foundational result has motivated a
growing line of work asking which stronger guarantees remain possible
under  this framework. For an incomplete list, see
\cite{kalavasis-stoc25,kalavasis2024breadth,
charikar-pabbarju,bai2025noise,raman2025noisy,
li2025learning,hanneke2025union, cai2026dense,mehrotra2026contamination,ganju2026timely,
kleinberg2026memory,panigrahi2026hallucination}. Our question is how much
\emph{breadth} can be guaranteed in addition to eventual validity.

\subsection{Tension between validity and breadth in language generation}

Generation in the limit guarantees eventual validity, but says nothing about how much of the target language $K$ is generated.
Indeed, in order to guarantee generation in the limit, the algorithm of Kleinberg and Mullainathan \cite{kleinberg2024limit} repeatedly refines provisional languages to increasingly restrictive sets.
Since the true language cannot generally be identified, this conservative strategy is natural: it eventually ensures validity, but may confine generation to a very narrow subset of $K$.

This illustrates a fundamental tension, also central to practical LLM design
\cite{kalai2023calibrated,holtzman2020degeneration,
pillutla2021mauve}:
\emph{validity} requires avoiding hallucinations, while
\emph{breadth} requires generating diverse valid outputs rather than suffering mode collapse.
But is sacrificing breadth necessary to ensure validity?

This leads to the central question on the \textbf{tension between validity and breadth}: 
\begin{quest}[Kleinberg and Mullainathan \cite{kleinberg2024limit}] \label{mainq}
\emph{For a given collection $\collec$, how broadly can an algorithm generate from an unknown target language while guaranteeing eventual validity?}
\end{quest}

To address this question quantitatively, we follow recent work that formulates density as a measure of breadth.
As we explain next, two distinct notions of density capture different interpretations of broad generation.

\subsection{Two notions of breadth: asymptotic density and
Banach density}

In our previous work
\cite{kleinberg2025density,kleinberg2026density,KWBanach},
we formulated Question~\ref{mainq} mathematically
by measuring the breadth of language generation
through two notions of density:
asymptotic density and Banach density.
These notions capture two different ways
in which a generator can provide broad coverage
of the target language.

\paragraph{Breadth measured by ranking: prefix average.}

Suppose the strings of a language  are ranked
by frequency or importance \cite{newman2005power}. A natural measure of breadth is the fraction of the first $n$ strings that a generator produces as $n$ grows.

\begin{defn}[Lower asymptotic density]
The \emph{lower asymptotic density}
of a set $A$ in $B = \{b_1,b_2,\ldots\}$ is
\begin{equation}
    \underline d(A,B)
    =
    \liminf_{n\to\infty}
    \frac{|A\cap\{b_1,\ldots,b_n\}|}{n}.
    \label{def:intro-asymptotic}
\end{equation}
\end{defn}
For a generator with output set $O$ and target language $K$, the quantity $\underline d(O,K)$ measures coverage of increasingly large ranked prefixes of $K$.

\paragraph{Breadth measured across spatial regions: local density.}

A different notion of breadth is needed to detect local failures of
coverage. Strings often have meaningful relationships that cannot be
represented by a single ranking, and modern language models commonly
represent text by vectors in \(\mathbb R^d\). (Note that current approaches in practice typically map arbitrarily long strings to representations in a fixed dimension $d$, rather than having $d$ grow with the length of the string \cite{cer2018universal,pillutla2021mauve,reimders2019sentence}; we therefore take $d$ as a fixed quantity rather than something that depends on the string being output.)
A generator may cover
many strings overall while systematically neglecting certain topics,
styles, or combinations of attributes. Such an underrepresented region
is a \emph{sparse hole}, representing a local form of mode collapse.

We measure coverage inside axis-parallel boxes. Although embeddings
naturally lie in \(\mathbb R^d\), which embedded points lie in a box
depends only on coordinatewise order. For finite-precision embeddings,
replacing each realized coordinate value by its integer rank preserves
all box traces. We therefore study local coverage on \(\mathbb Z^d\),
where the filtered theory additionally depends on the lattice volume
\(V(R)\).

\begin{defn}[Lower Banach density]
\label{def:banach-intro}
For $A, B\subseteq\mathbb Z^d$,
where $B$ is infinite, define
\[
    \lbanach(A,B)
    =
    \liminf_{|R\cap B|\to\infty}
    \frac{|A\cap B \cap R|}{|B\cap R|},
\]
where $R$ ranges over all finite
axis-parallel boxes in $\mathbb Z^d$.
\end{defn}

The boxes may have arbitrary side lengths, including long, thin ones.
A guarantee $\lbanach(O,K)\ge\alpha$ means that boxes containing increasingly many points of \(K\) must contain at least an $(\alpha-o(1))$ fraction of valid generated strings, uniformly over their locations and shapes.

In this sense, lower Banach density is more demanding than lower asymptotic density.
High coverage elsewhere cannot compensate for sparse holes: arbitrarily large regions with a vanishing fraction of generated strings force the lower Banach density to zero.

\subsection{Previous landscape: topology, geometry, and the open problem}
\label{subsec:prior-results}

For lower asymptotic density, we proved that one eventually valid
generator always guarantees density \(1/2\), and that this is optimal
in general
\cite{kleinberg2025density,kleinberg2026density};
see also \cite{cai2026dense}. The value \(1/2\) is the natural
one-output-per-input benchmark: even if the target were known, the
adversary can race the generator for the available strings and claim
roughly half of them. More generally, with \(C\) outputs per input, the
benchmark becomes \(C/(C+1)\).

The striking feature of the asymptotic-density theorem is its
universality: the optimal guarantee is independent of the structure of
the candidate family \cite{kleinberg2025density,kleinberg2026density}. \textbf{Lower Banach density behaves fundamentally
differently. Already in one dimension, some countable families force
density zero. However finite Cantor--Bendixson rank restores the optimal
value \(1/2\)} \cite{KWBanach}.

To capture why lower Banach density is more sensitive to the structures of $\collec$, in previous works we defined the \emph{finite-evidence topology}
on \(\collec\), whose basic neighborhoods are
\[
U_{L,F}
=
\{M\in\collec:F\subseteq M\subseteq L\},
\]
where \(F\subseteq L\) is finite. These are precisely the refinements of
\(L\) that remain possible after the positive examples in \(F\) have
been observed. \textbf{Thus topology is irrelevant to the universal asymptotic-density
guarantee, but can determine the optimal lower Banach-density
guarantee.}

In dimensions \(d\ge2\), our previous work identified an independent
\emph{geometric obstruction}: ordinary lower Banach density can be zero
even when the target is completely known \cite{KWBanach}. The
obstruction comes from rectangles with enormous geometric volume but
very few target points, where Ramsey and discrepancy phenomena can
defeat uniform coverage

To separate this geometric degeneracy from uncertainty about the target
in higher dimensions, we use a threshold function that excludes
rectangles containing too few target points relative to their geometric
volume.

\begin{defn}[Filtered lower Banach density]
\label{def:filtered-banach-h}
Let 
$h:\mathbb N_{\ge1}\longrightarrow(0,\infty)$
be nondecreasing.  For \(A,B\subseteq\mathbb Z^d\), with \(B\)
infinite, write
$V(R)=|R|$ and $
m_B(R)=|B\cap R|$.
The \emph{\(h\)-filtered lower Banach density} of \(A\) relative to
\(B\) is
\[
\delta_h(A,B)
=
\lim_{s\to\infty}
\inf_{\substack{
R\text{ a finite axis-parallel rectangle}\\
m_B(R)\ge s\,h(V(R))
}}
\frac{|A\cap B\cap R|}{m_B(R)}.
\]
An empty infimum is interpreted as \(1\).
\end{defn}

Equivalently, \(\delta_h(A,B)\) is the lower limit of the relative
density over rectangles satisfying
$\frac{m_B(R)}{h(V(R))}\to\infty$.
We call \(B\) \emph{\(h\)-vacuous} when
$\sup_R\frac{m_B(R)}{h(V(R))}<\infty.$
For a vacuous target, the family of rectangles in the infimum is
empty for all sufficiently large \(s\).  Thus, under the convention
\(\inf\emptyset=1\),
$\delta_h(A,B)=1$
for every \(A\subseteq\mathbb Z^d\).  In particular, a vacuous target
imposes no constraint on a density guarantee.

When \(h\equiv1\), this definition reduces to ordinary lower Banach
density:
$\delta_1(A,B)=\lbanach(A,B)$.

Our previous finite-rank theorem~\cite{KWBanach} required a
polylogarithmic nondegeneracy threshold, with no sharp exponent.
The present paper reaches
\[
h_\tau(V)=\log^{1+\tau}(2+V)\]
for every $\tau>0$,
bringing the threshold arbitrarily close to logarithmic.

The earlier results thus revealed two independent obstructions:
a \emph{topological obstruction} caused by uncertainty about the
target, and a \emph{geometric obstruction} caused by rectangular
Ramsey/discrepancy phenomena. But they did not determine the optimal
density for an arbitrary family.

Three basic questions remained:

\begin{enumerate}[label=(\arabic*),leftmargin=*]
\item What is the exact optimal lower Banach density for an arbitrary
one-dimensional target, without a finite-rank assumption?
\item Can all target-specific optima be attained simultaneously by one
generator?
\item When simultaneous optimality fails in higher dimensions, what is
the complete region of filtered lower Banach density guarantees that can be achieved together?
\end{enumerate}

\noindent\textbf{This paper answers all three questions.}

\subsection{Main results: Exact local breadth and its limits}

{\noindent\textbf{In one dimension, all target-specific optima can be achieved simultaneously by one generator. In higher dimensions, geometry can force unavoidable trade-offs between targets.}}

In one dimension, we give an exact structural characterization of the
optimal lower Banach density for every target language \(K\).
We define a quantity \(b(K)\) purely from the structure of the candidate
family, without reference to any generator or adversarial enumeration, 
and prove that \(b(K)\) is exactly the largest density that can be
guaranteed at \(K\) by an eventually valid generator.  Moreover, one
generator attains these target-specific optima simultaneously for every
target in the family.

In dimensions \(d\ge2\), this scalar picture breaks down. A generator
that is optimal when one language is the target may necessarily perform
worse when the target is another language. Thus the target-specific
optima need not be jointly attainable. The correct object is therefore the entire region of achievable density
profiles \(\alpha:\collec\to[0,1]\). For filtered lower Banach density
under the threshold condition stated below, we give an exact
topological--combinatorial characterization of this region and derive
sharp finite-rank consequences. Moreover, if \(\sup_N N/h(N)=\infty\), every
\(\prod_i[0,\alpha_i]\), \(\alpha_i\in[0,1/2]\),
is the \emph{exact achievable} \(h\)-filtered profile region
of some family (Lemma~\ref{prop:profile-realization}).

Table~\ref{tab:intro-results} summarizes the main conclusions, where every generator must be
eventually valid on the entire candidate family.
Precise threshold and rank hypotheses appear below.

{\small
\setlength{\tabcolsep}{5pt}
\renewcommand{\arraystretch}{1.15}

\begin{xltabular}{\textwidth}{@{}
>{\raggedright\arraybackslash}p{0.19\textwidth}
>{\raggedright\arraybackslash}X
>{\raggedright\arraybackslash}p{0.1\textwidth}@{}}

\caption{Main results, informally. }
\label{tab:intro-results}\\

\toprule
Setting & Main conclusion & Reference \\
\midrule
\endfirsthead

\toprule
Setting & Main conclusion & Reference \\
\midrule
\endhead

One dimension &
\textbf{Exact optima, attained simultaneously.}
Two equivalent structural formulas determine every
target's optimal lower Banach density, and a single
generator attains them all. &
Thm.~\ref{thm:intro-one-dimensional} \\

\addlinespace
Higher dimensions &
\textbf{Unavoidable trade-offs and a new design principle.}
A generator attaining the best guarantee at one possible target may
necessarily sacrifice its guarantee at another. Thus no single generator need be best for all
targets: one must decide which targets to prioritize. &
Thm.~\ref{thm:intro-higher-nonsimultaneous} \\

\addlinespace
Higher-dimensional profiles &
\textbf{Complete characterization of achievable profiles.}
Equivalent topological and combinatorial criteria determine exactly
which filtered-density guarantees one generator can achieve simultaneously.
Moreover, when \(\sup_N N/h(N)=\infty\), every
\(\prod_i[0,\alpha_i]\), \(\alpha_i\in[0,1/2]\),
is the \emph{exact achievable region} of some countable family. &
Thm.~\ref{thm:main};
Lemma~\ref{prop:profile-realization} \\

\addlinespace
Finite Cantor--Bendixson point ranks &
\textbf{Near-half guarantees with a sharp endpoint.}
In dimension one, one generator attains the optimal value \(1/2\).
In higher dimensions, for every \(\varepsilon>0\), one generator
guarantees \(1/2-\varepsilon\) simultaneously for all such targets.
In higher dimensions, exact \(1/2\) can fail even at rank zero, so the
loss \(\varepsilon\) cannot in general be removed. &
Cor.~\ref{cor:intro-one-dimensional-rank};
Thms.~\ref{thm:finite-rank-near-half}, \ref{end:main} \\
\bottomrule
\end{xltabular}
}

\subsubsection{One-dimensional}

We first consider the one-dimensional case, where the
optimal lower Banach density admits an explicit 
characterization.
The key objects are finite descending chains of candidate
languages
\[
H=(K=L_0\supsetneq L_1\supsetneq\cdots\supsetneq L_r).
\]
Such a chain partitions $K$ into layers, where each layer
$D_i$ consists of the new strings in $L_i$ that are not
present in $L_{i+1}$ (See Figure \ref{fig:language-layers} in Section \ref{subsec:setup-chains}):
$D_i=L_i\setminus L_{i+1}$ for $0\le i<r$, and  $D_r=L_r.$

A \emph{relative interval} $J$
is a finite set of consecutive points of $K$. Given  a chain $H$ and a relative interval $J$, the \emph{pairing score}
counts one unit for every pair of points in each layer in $J$. 
\begin{equation}
  p(H,J)=
\frac{1}{|J|}
\sum_{i=0}^{r}
\left\lfloor\frac{|D_i\cap J|}{2}\right\rfloor.  \label{pairing-intro} 
\end{equation}
Each layer represents points that become safer only after the adversary
rules out a smaller candidate language. The adversary can claim the first point in each layer in the
associated alternating game, so the generator can secure at most
\(\lfloor |D_i\cap J|/2\rfloor\) points of that layer. Equivalently, the deficit from
\(1/2\) is one half-point for every layer meeting \(J\) in an odd
number of points.

We test the worst score on  relative intervals with at least $n$ points arbitrarily far
from the origin:
\begin{equation}\label{eq:intro-cn}
c_n(H)=
\lim_{R\to\infty}
\inf_{\substack{
J\text{ a relative interval of }K\\
|J|\ge n,\ J\cap[-R,R]=\emptyset
}}
p(H,J).
\end{equation}
Here, letting $R\to\infty$ moves the intervals
arbitrarily far from the origin, while later increasing
$n$ restricts attention to  larger intervals.

Our central one-dimensional result identifies the same optimal density
value in two strikingly different ways: one combinatorial and one topological.

\paragraph{Combinatorial formula.}
An \emph{admissible obstruction tree} is a well-founded tree
of strictly descending candidate languages, in which every
finite set of positive examples remains consistent with a
proper refinement at each non-leaf vertex
(Definition~\ref{def:aot}).

Let $\mathfrak T(K)$ denote the admissible obstruction trees
rooted at $K$. Define
\begin{equation}\label{eq:intro-bn-combinatorial}
b_n^{\mathrm{comb}}(K)
=\inf_{T\in\mathfrak T(K)}
\ \sup_{{H\text{ a root-to-leaf path of }T}}c_n(H).
\end{equation}
This quantity depends only on the refinement poset structure of the candidate family and the explicit pairing scores associated with descending chains. In particular, its definition involves neither generation algorithms nor adversarial enumerations. The theorem below shows that this structural invariant determines the exact optimal density in the original online problem. 

Although every fixed finite history has
$c_n(H)\to 1/2$, the obstruction tree may depend
on $n$. This lack of uniformity allows the
optimal value to be strictly smaller than $1/2$.

\paragraph{Topological characterization.}
Our second characterization reveals a more intrinsic
topological structure underlying the optimal density.

The pairing score depends on the entire refinement history, not merely
on its final language: two histories ending at the same language may
have different layer decompositions and hence different scores. We
therefore lift the finite-evidence topology from the language family
\(\collec\) to the space \(\mathcal H\) of finite descending histories.

The resulting \emph{finite-evidence history topology} (Section~\ref{sec:history-topology}) is the canonical and finest 
topology preserving finite-evidence convergence under continuation:
when proper refinements \(M_j\subsetneq L\) converge to \(L\), appending
them to any history ending at \(L\) produces histories converging back
to that history. This characterization is formalized in
Lemma~\ref{lem:finest-history-topology}.

Let \(\mathcal H_K\) denote the histories beginning at \(K\). Define a topological invariant of the history
space equipped with its pairing-score function.
\begin{equation}\label{eq:intro-bn-topological}
b_n^{\mathrm{top}}(K)
=
\sup_{
\mathcal U\subseteq\mathcal H_K\text{ open}, 
(K)\in\mathcal U
}
\ \inf_{H\in\mathcal U}c_n(H).
\end{equation}
This $b_n^{\mathrm{top}}(K)$ has a natural topological
interpretation: the topological lower limit of
\(c_n\) at the initial history \((K)\): the supremum of the lower bounds that hold throughout a finite-evidence neighborhood of $(K)$.

The combinatorial and topological formulations
are quite different in nature. Nevertheless,
our main theorem establishes that they coincide
and, more remarkably, that their common value
determines the exact optimal lower Banach density
in the original language generation problem.

\begin{thm}[Exact one-dimensional value and simultaneous optimality]
\label{thm:intro-one-dimensional}
Let $\collec$ be a nonempty countable family
of infinite subsets of $\mathbb Z$.
Then for every $K\in\collec$ and every $n\ge1$,
$b_n^{\mathrm{comb}}(K)
=
b_n^{\mathrm{top}}(K)$.
Denote their common value by $b_n(K)$, and let 
$b(K)=\lim_{n\to\infty}b_n(K)$.

\begin{enumerate}
    \item (Upper bound) For every generator that is eventually
valid on $\collec$ and every $K\in\collec$,
there exists an adversarial enumeration of $K$
for which
$\lbanach(O,K)\le b(K)$.
\item {(Lower bound.)} There exists a single eventually valid generator
such that, for every $K\in\collec$ and every
adversarial enumeration of $K$, its output set
$O$ satisfies 
$\lbanach(O,K)\ge b(K)$.
\end{enumerate}

Consequently, $b(K)$ is exactly the optimal
worst-case lower Banach density for the target
$K$, and these optima are attained \textbf{simultaneously}
by a single generator.
\end{thm}

The theorem converts an infinite online problem into a static
combinatorial--topological duality. The obstruction-tree formula gives
global certificates of impossibility, while the topological formula
describes the strongest score that finite positive evidence can protect.
Neither formula quantifies over generators or adversarial enumerations,
yet their common value is exactly the online optimum.

\paragraph{This characterization recovers both extremal regimes.}

The same invariant unifies the two opposite phenomena from our previous
work.

At the zero-density extreme, the family constructed
in~\cite[Theorem~2.4]{KWBanach} contains a target
$K_0=\mathbb N$ with $b_n(K_0)=0$ for every $n$
(Proposition~\ref{prop:prior-zero-value}).
At the opposite extreme, every target of finite
Cantor--Bendixson point rank satisfies $b(K)=\lim b_n^{\mathrm{top}}(K) =1/2$,
even when the point ranks are not uniformly bounded
(Corollary~\ref{cor:intro-one-dimensional-rank}).
In fact, every value between these two extremes $0$ and $1/2$ occurs as an exact
optimum (Lemma \ref{cor:full-range}).

A particularly useful consequence is the following strict strengthening
of our previous finite-rank theorem~\cite{KWBanach}.

\begin{cor}
\label{cor:intro-one-dimensional-rank}
Let $\collec$ be a countable family of infinite
subsets of $\mathbb Z$, and suppose that every
language in $\collec$ has finite
Cantor--Bendixson point rank.
Then a single eventually valid generator achieves
the optimal lower Banach density $1/2$
for every target language.
\end{cor}
Unlike the previous theorem, the point ranks need not be uniformly
bounded, so the family itself may have infinite Cantor--Bendixson rank.
Moreover, whereas the earlier \(1/2\) theorem required a separate and
involved argument, this stronger statement is now a short corollary, by the definitions of isolated points and 
 Cantor--Bendixson derivatives.
Thus the characterization is the core result determining the optimal
density; once it is in place, the finite-rank theorem follows
immediately.

\vspace{-0.1in}
\subsubsection{Higher dimensions}

\paragraph{1. Failure of simultaneous optimality.} 

\textbf{In higher dimensions, there need not be a single best
generator.}
In one dimension, a single generator attains every
target's optimum.  In dimensions \(d\ge2\), this can fail fundamentally:
optimizing
for one possible target may force a loss at another. 
\textbf{Thus the generator design problem in higher dimensions becomes inherently multi-objective: one must decide which target languages to
prioritize. } 

This mirrors a basic issue in practical generative-system
design, where improving coverage for one domain or class of outputs may
come at the expense of another. This phenomenon has a close analogue in current LLM alignment, where a
single model may be asked to balance several potentially conflicting
rewards or user preferences, and the design goal is often expressed as
a Pareto frontier rather than a single scalar optimum
\cite{rame2023rewarded,agnihotri2026multiobjective}.
Our theorem leads to a parallel design lesson: the priorities among competing
targets must be specified before the generator is optimized.

The following theorem makes this conflict quantitative.  It already
occurs for ordinary lower Banach density, before any filtering is
introduced.

\begin{thm}[Failure of simultaneous optimality, informal versions of Theorems~\ref{thm:general} and
\ref{thm:main-not-simul}]
\label{thm:intro-higher-nonsimultaneous}
For every $d\ge2$, there exists a countable
language family $\collec$ with a distinguished
language $K_0$ such that the following hold:
\begin{itemize}
\item Some eventually valid generator guarantees
$\lbanach(O,K_0)\ge1/4$.

\item Another eventually valid generator guarantees
$\lbanach(O,K)\ge1/2$ simultaneously for every
$K\in\collec\setminus\{K_0\}$.
\end{itemize}
Both generators are eventually valid on the entire
family $\collec$. Nevertheless, any single
eventually valid generator guaranteeing density
$x$ at $K_0$ and density $y$ at every other
language must satisfy
$x+y\le\frac12$.
In particular, simultaneous optimality fails.
\end{thm}

The same family witnesses this incompatibility for every positive
nondecreasing threshold \(h\) satisfying
\(\sup_N N/h(N)=\infty\): the individual optima and the trade-off
\(x+y\le1/2\) remain unchanged. The obstruction is geometric:
horizontal and vertical rectangle requirements compete for the same
outputs before finite evidence reveals which target is present. One very rough heuristic is:  each
four-point square cell supplies at most one useful output: one choice helps a
horizontal rectangle test, while the other helps a vertical test, and
no output is counted in both directions.  Summing this competition over
a large block gives \(x+y\le1/2\).

\paragraph{2. Exact characterization of achievable density profiles.}

The failure of simultaneous optimality raises a natural
question: 
\begin{quote}
\emph{Which target-wise density guarantees can one generator achieve
simultaneously?}
\end{quote}
Therefore we consider \emph{density profiles}
$\alpha:\collec\to[0,1]$.
It is \emph{achievable at threshold \(h\)} if one eventually valid
generator guarantees
$\delta_h(O,K)\ge\alpha(K)$
for every \(K\in\collec\) and every full enumeration of \(K\).

\begin{thm*}[Informal versions of Theorems~\ref{thm:main} and
\ref{thm:topological-protected-tree}, and
Corollary~\ref{cor:hd-obstruction}]
Fix a nondecreasing threshold function \(h\) such that
$\frac{h(V)}{\log(2+V)}$
is nondecreasing.  For every countable collection \(\collec\), we give
an exact structural characterization of all density profiles achievable
at threshold \(h\).
\end{thm*}

The theorem replaces the original infinite online interaction by static
certificates built from finite refinement histories, finite positive
tests, and finite geometric games.

As in one dimension, it has equivalent combinatorial and topological
formulations, extending the duality between obstruction structures and
finite-evidence neighborhoods. Thus, topology remains intrinsic to the
optimal breadth problem.

Our structural theory also shows that, whenever
\(\sup_N N/h(N)=\infty\), every product
\(\prod_i[0,\alpha_i]\), \(\alpha_i\in[0,1/2]\),
is the \emph{exact achievable density region} of some
countable family (Lemma~\ref{prop:profile-realization}).

The higher-dimensional theorem uses the same broad architecture but
requires a new local quantitative object.  The one-dimensional pairing
score no longer suffices: overlapping rectangles can place competing
demands on the same outputs.  We therefore use a {\it finite geometric 
game} to test whether one strategy can meet all
rectangle requirements along a refinement history.
For the case of finite point rank, a new multiscale discrepancy
argument certifies these geometric games, yielding
simultaneous near-half guarantees at the
near-logarithmic threshold.

\paragraph{3. Finite Cantor--Bendixson point rank:
simultaneous near-half density and sharpness.}
Similar to the one-dimensional case, our characterization also yields a sharp
strengthening of the finite-rank result
of~\cite{KWBanach}.

Although simultaneous optimality fails in general
in higher dimensions, finite Cantor--Bendixson
point rank restores a strong simultaneous guarantee:
a single generator achieves density arbitrarily
close to $1/2$ for every target.
Moreover, this guarantee is essentially sharp.

\begin{thm*}[Finite point ranks and sharpness, informal versions of Theorem~\ref{thm:finite-rank-near-half} and Theorem \ref{end:main}]
Fix \(d\ge2\), \(\tau>0\), and \(\varepsilon>0\), and let
$h_\tau(V)=\log^{1+\tau}(2+V)$.
\begin{enumerate}[label=(\roman*),leftmargin=*]
\item One eventually valid generator guarantees
$\delta_{h_\tau}(O,K)\ge\frac12-\varepsilon$
simultaneously for every target \(K\) of finite Cantor--Bendixson point rank.
 The point ranks need not have a common
bound.

\item The endpoint \(1/2\) is not always attainable.  Even for the
rank-zero singleton family \(\{\mathbb Z^d\}\), no generator guarantees
$\delta_{h_\tau}(O,\mathbb Z^d)\ge\frac12$
against every full enumeration.
\end{enumerate}
\end{thm*}

Part~(i) strengthens both components of our previous theorem
\cite{KWBanach}: it removes the uniform rank bound and improves the
nondegeneracy requirement from an unspecified polylogarithmic threshold
to
$h_\tau(V)=\log^{1+\tau}(2+V)$
for every \(\tau>0\). Part~(ii) shows that \(1/2\) is the
sharp endpoint but need not be attained.

\subsection{Proof challenges and strategy, comparison with previous approaches}
\label{subsec:proof-strategy}

\paragraph{One dimension: from bounded rank to arbitrary refinement histories.}

Our approach builds on our three previous papers
\cite{kleinberg2025density,kleinberg2026density,KWBanach}.
However, extending the earlier methods to obtain
 lower Banach density guarantees has
fundamental challenges and requires several new ideas.
Our first two papers studied lower asymptotic
density, for which a universal $1/2$ guarantee
is possible. A key advantage of asymptotic density is that losses
can be compensated elsewhere in a sufficiently
long prefix, so the proofs are much simpler. Lower Banach density is much less
forgiving: losses concentrated in a distant
interval cannot be compensated by outputs elsewhere.

In our previous work~\cite{KWBanach}, finite
Cantor--Bendixson rank allowed us to build a static inclusion tree that
guided the generator's successive language guesses. The density
analysis then reduced to a local finite game along descending paths in
that tree. The rank bounded the tree depth and hence the number of
potentially costly changes of guess.

Without finite rank, descending paths can have arbitrarily large finite
length. The present proof replaces bounded depth by a quantitative
analysis of each refinement history and replaces one global spatial
state by localized states that cannot be repeatedly disrupted by
inputs arriving far away.

\paragraph{The guiding observation: the cost of a chain
is not its length.}

To understand the difficulty, consider how the
adversary can exploit a descending chain
$K=L_0\supsetneq L_1\supsetneq\cdots
\supsetneq L_r$.
The adversary may initially reveal only strings from a smaller language
\(L_r\). Because these inputs are compatible with \(L_r\) being the
true target, eventual validity forces the generator, after sufficiently
long continuation inside \(L_r\), to avoid producing strings outside
\(L_r\), even when the actual target is the larger language \(K\).

The crucial loss occurs when the adversary begins revealing that the
narrower candidate languages were incorrect.  Suppose, for the moment,
that the adversary can force the generator to move only along the
chosen chain.  When the adversary first reveals a point in the layer 
$D_i=L_i\setminus L_{i+1}$,
the generator can safely begin producing from this newly exposed layer,
but the adversary has already claimed its first point.  In the resulting
local pairing game, if an interval \(J\) contains \(m\) points of
\(D_i\), then the generator can in general secure only
$\left\lfloor m/2\right\rfloor$
of them.

Establishing this confinement is itself a non-trivial part of the proof: the
obstruction-tree construction and eventual-validity argument ensure
that the generator cannot escape to unrelated candidate languages while
the adversary exposes the layers of the selected chain.

This leads to a simple but important observation:
\emph{a long refinement chain need not cause
a large density loss.}
For example, suppose $J$ contains exactly
one point from each of $|J|$ different layers.
The adversary can claim the first point of
every layer, leaving the generator with
pairing score zero.
This is the mechanism behind our earlier
zero-density construction in \cite{KWBanach}. In contrast, suppose every layer meets $J$
in an even number of points.
The generator can then pair the points within
each layer and achieve a score of $1/2$,
regardless of how many layers occur.

Without finite Cantor--Bendixson rank, this
approach no longer provides a depth bound.
However, the preceding examples reveal that
\emph{the length of a refinement chain is not
the essential issue}. What matters is how
its layers intersect the interval being measured.
This motivates our pairing score $p(H,J)$ in 
\eqref{pairing-intro} and its asymptotic
version $c_n(H)$, which turns out to measure the actual
local cost of a descending chain.

The first new ingredient is therefore to
\emph{replace bounded depth by quantitative
control of descending chains}.
This allows us to distinguish long chains
that are intrinsically harmful from those that
preserve good density.

\paragraph{Finite protection: why topology enters naturally.}

Understanding the cost of a fixed chain is
only the beginning.
A good pairing score for a descending chain
is not enough: future refinements may have
much worse scores.
Our second new idea is \emph{finite protection}.
Given a descending chain
$H=(L_0\supsetneq L_1\supsetneq\cdots
\supsetneq L_r),$
we seek a carefully chosen finite set
$F\subseteq L_r$ such that every further
refinement containing $F$ preserves the
desired pairing-score guarantee.

The key observation is that once the adversary
has revealed every element of $F$, all
subsequent refinements consistent with the
observed examples must contain $F$.
Thus, a finite amount of positive evidence
can protect the generator against harmful
refinements.

The difficulty is that such a protected set
need not exist.
The key insight is a duality we proved: \emph{either a finite set
protects the desired guarantee, or an
obstruction tree witnesses why every finite
attempt at protection fails}
(Corollary~\ref{cor:protected-set}).

An obstruction tree describes global possibilities
for successive refinements, whereas a protected
set controls refinements through finite positive
evidence. It is not immediate that these two forms
of control have the same strength. The
obstruction--closure duality
(Lemma~\ref{lem:obstruction-closure})
establishes their equivalence.

This duality reveals the intrinsic role of
topology.
Our earlier finite-evidence topology describes
which languages can be distinguished using
finite positive evidence.
Since pairing scores depend on entire descending
chains rather than just their final languages,
we naturally extend this topology to the space
of descending chains
(Lemma~\ref{lem:finest-history-topology}) by a canonical ``lifting".

\emph{The resulting open neighborhoods capture
which collections of descending chains can be
protected by finite positive evidence.}
Our characterization then establishes the
nontrivial connection: the topological lower
limit of pairing scores equals the exact
optimal density in the infinite online
generation problem.

\paragraph{Turning protected descending chains into one generator.}

The next challenge is to turn the topological
characterization into a single online generator
that is eventually valid and attains the
optimal density at every target.

Our third new ingredient is a static
\emph{protected forest}.
Like the static tree built through Cantor--Bendixson rank in
\cite{KWBanach}, its goal is to provide a fixed structure
that organizes the generator's possible
changes of language guesses.
The earlier static tree in~\cite{KWBanach} crucially
relied on finite Cantor--Bendixson rank for both the construction and the analysis because of the bound on its depth; here no such bound exists.

Moreover, the forest must balance two
competing requirements.
If the true target is $K$, the generator's
guesses must eventually remain in the
subtree rooted at $K$ to ensure validity.
At the same time, the lowest common ancestor (or pullbacks) 
of recent guesses must return to $K$
infinitely often to ensure breadth.
Thus, restricting the guesses too much
can jeopardize density, while allowing
them too much freedom can jeopardize validity.

Our solution is to protect not merely the
current language, but every relevant
refinement history (descending  chains).
The forest preserves the pairing-score
requirements of all relevant ancestors.
Since each finite history has only finitely
many ancestors, their finite protected
sets can be combined.
This allows a single forest, potentially
of infinite depth, to support the
guarantees for all targets.

A second difficulty is spatial.
The earlier algorithm \cite{KWBanach} used a single
evolving pullback and pool of candidate
outputs, whose changes were controlled
by finite rank.
Without a depth bound, examples arriving
far from an interval could repeatedly
alter this global state and disrupt
the density guarantee inside that interval.

Our fourth new idea is to
\emph{localize the pullback mechanism}.
Each finite spatial \emph{board} maintains its own
pullback record and pool of candidate outputs,
updated only by inputs from that board.
Crucially, upward pullbacks merge language
layers and therefore cannot decrease
the pairing score.
The resulting boardwise guarantees,
together with control of boundary losses,
yield the desired lower Banach density.

Finally, once the true target is captured
in the forest, newly initialized boards
generate only from its subsets.
The finitely many earlier exceptional boards
can produce only finitely many invalid outputs,
by freshness.
Thus, the construction achieves eventual
validity and optimal density simultaneously
for every target, without any finite-rank
assumption.

\paragraph{Proof challenges and strategy in higher dimensions.}

The higher-dimensional proof retains the
finite-evidence and topological framework
of the one-dimensional case, but geometry
introduces two fundamental difficulties.

First, in sharp contrast to
the one-dimensional case, the individually optimal guarantees
for different target languages need not be simultaneously
attainable by one generator
(Theorem~\ref{thm:intro-higher-nonsimultaneous}). Competing
rectangle directions can place incompatible demands on the
same outputs. Thus, we must characterize the
entire region of achievable density profiles,
rather than a separate optimum for each target.

Second, the one-dimensional pairing score
no longer suffices. In one dimension, points
can be paired independently within each
language layer. In higher dimensions,
rectangles overlap, and the adversary can
stop at different stages of a refinement
history. A strategy that works for each
rectangle separately need not work for
all rectangles simultaneously.

Our earlier work~\cite{KWBanach} already
identified geometric obstructions even
for a known target, motivating filtered
lower Banach density, and developed local
discrepancy games under finite-rank assumptions.
Our new idea is to replace the pairing
score by a \emph{finite geometric stopping
game}, in which one causal strategy must
satisfy all relevant rectangle requirements
at every possible stopping stage.
Crucially, these local games must be
coordinated across arbitrary refinement
histories, without a finite-rank bound.
Here topology again provides the organizing
principle: finite protected sets control
compatible refinements, while protected trees
and finite-evidence history neighborhoods
coordinate the geometric games.
Obstruction trees give the dual impossibility
certificates, yielding the characterization
of achievable density profiles
(Theorems~\ref{thm:main} and
\ref{thm:topological-protected-tree},
and Corollary~\ref{cor:hd-obstruction}).

Finally, finite Cantor--Bendixson rank controls the number of language layers, but
it does not resolve the geometric interaction among rectangles at many
different scales.  We handle this with a multiscale discrepancy
argument on a sparse dyadic hierarchy.

When a smaller region becomes endangered, it takes responsibility for
its own points, and its ancestors stop relying on the unused points
inside it.  A coarse-scale potential must therefore reserve an
allowance for possible future transfers of responsibility to its
descendants.  This recursive descendant correction is the main new
ingredient beyond a standard discrepancy-game potential.  Combined
with a ``most dangerous point'' response and the logarithmic sewing
lemma, it gives the \(1/2-\varepsilon\) guarantee at the threshold
\(h_\tau(V)=\log^{1+\tau}(2+V)\).

Thus, \emph{topology determines which refinement histories finite
evidence can exclude, while geometry determines whether the remaining
density requirements can be satisfied simultaneously.}

\part{Dimension one}

We begin with the one-dimensional problem. We first state the main characterization
theorem for orientation. We then define the structural invariant
\(b(K)\), prove matching adversarial and constructive bounds, and
identify \(b(K)\) topologically through the finite-evidence history
space. Throughout this part we use the following conventions.
\paragraph{Model conventions.}
A \emph{full enumeration} by an adversary of a target \(K\) is an infinite sequence
\((x_t)_{t\ge1}\) of elements of \(K\) whose range is exactly \(K\);
repetitions are permitted.  Write
\[
S_t=\{x_1,\ldots,x_t\},
\qquad
O_{t-1}=\{o_1,\ldots,o_{t-1}\}.
\]
After receiving \(x_t\), the generator produces one output \(o_t\).
Throughout the paper we require the global freshness condition
$o_t\notin S_t\cup O_{t-1}$
for every  $t$.
We call a point satisfying this condition \emph{available}.
A later input may equal an earlier generator output.

A generator is \emph{universally eventually valid} on \(\X\) if,
for every \(K\in\X\) and every full enumeration of \(K\), all but
finitely many outputs belong to \(K\).

\begin{thm}[Exact structural characterization]
\label{thm:main-oneD}
Let \(\X\) be a nonempty countable collection of infinite subsets of
\(\mathbb Z\). For every \(K\in\X\), let \(b(K)\) be the structural
quantity defined combinatorially in \eqref{eq:bn}, or equivalently topologically as in Theorem \ref{thm:topological-bn}. Then the following hold.

\begin{enumerate}
\item \emph{Achievability.}
There is a single generator such that, for every \(K\in\X\) and every
full enumeration of \(K\), every output is available, all but finitely
many outputs belong to \(K\), and its output set \(O\) satisfies
\(\lbanach(O,K)\ge b(K)\).

\item \emph{Optimality.}
Let \(\mathcal A\) be any generator that is eventually valid on every
full enumeration of every language in \(\X\). Then, for every
\(K\in\X\), there is a full enumeration of \(K\) such that the output set
\(O\) produced by \(\mathcal A\) satisfies
\(\lbanach(O,K)\le b(K)\).
\end{enumerate}

Consequently, \(b(K)\) is the optimal worst-case lower Banach-density
guarantee for the target language \(K\).
\end{thm}

\section{The combinatorial and topological formulations of the optimal value}
\label{sec:statement}

This section develops the structural quantity \(b(K)\) appearing in
Theorem~\ref{thm:main-oneD}, which is exactly the optimal lower
Banach-density guarantee achievable at the target language \(K\).
We first define the local pairing score of a refinement history and use
it to give a combinatorial min--max formula over obstruction trees.
We then give an equivalent topological formulation by formally
developing the finite-evidence history topology introduced in the
Introduction.

\subsection{Setup: Chains and their pairing scores}
\label{subsec:setup-chains}

We consider finite chains of languages, strictly decreasing and
starting at some language $A$, written as
$H=(A=L_0\supsetneq L_1\supsetneq\cdots\supsetneq L_d)$. The letter $H$
denotes the whole list. (The word \emph{pod} is reserved for a set of
candidate output points, introduced later, and should not be confused
with a chain.)

\begin{figure}[ht]
\centering
\begin{tikzpicture}[
    x=0.52cm,
    y=0.82cm,
    font=\small
]

\foreach \y in {0,1,2,3} {
    \draw[gray!45, thin] (-9.5,\y) -- (9.5,\y);
}

\node[anchor=east] at (-10,3) {$L_0=K$};
\node[anchor=east] at (-10,2) {$L_1$};
\node[anchor=east] at (-10,1) {$L_2$};
\node[anchor=east] at (-10,0) {$L_3$};

\foreach \x in {-9,-7,-6,-4,-3,-1,0,2,3,5,6,8,9} {
    \draw[gray!20, densely dotted]
        (\x,-0.12) -- (\x,3.12);
}

\foreach \x in {-9,-3,2,8} {
    \fill[blue!75!black] (\x,3) circle (3pt);
}

\foreach \x in {-7,-1,5,9} {
    \foreach \y in {2,3} {
        \fill[green!55!black] (\x,\y) circle (3pt);
    }
}

\foreach \x in {-6,0,6} {
    \foreach \y in {1,2,3} {
        \fill[orange!85!black] (\x,\y) circle (3pt);
    }
}

\foreach \x in {-4,3} {
    \foreach \y in {0,1,2,3} {
        \fill[red!75!black] (\x,\y) circle (3pt);
    }
}

\draw[->, gray!75!black]
    (-9.6,-0.65) -- (10,-0.65)
    node[right] {$\mathbb Z$};

\foreach \x in {-9,-8,...,9} {
    \draw[gray!65]
        (\x,-0.59) -- (\x,-0.71);
}

\node[below] at (-9,-0.73) {$-9$};
\node[below] at (0,-0.73) {$0$};
\node[below] at (9,-0.73) {$9$};

\fill[blue!75!black] (-6.5,-1.65) circle (3pt);
\node[anchor=west] at (-6.1,-1.65) {$D_0$};

\fill[green!55!black] (-2.5,-1.65) circle (3pt);
\node[anchor=west] at (-2.1,-1.65) {$D_1$};

\fill[orange!85!black] (1.5,-1.65) circle (3pt);
\node[anchor=west] at (1.9,-1.65) {$D_2$};

\fill[red!75!black] (5.5,-1.65) circle (3pt);
\node[anchor=west] at (5.9,-1.65) {$D_3$};

\end{tikzpicture}

\caption{A descending chain
$K=L_0\supsetneq L_1\supsetneq L_2\supsetneq L_3$
of subsets of $\mathbb Z$.
Each row displays the points belonging to that language
in a finite window, with their positions preserved.
The colors distinguish the disjoint layers
$D_i=L_i\setminus L_{i+1}$ for $i<3$ and
$D_3=L_3$.}
\label{fig:language-layers}
\end{figure}
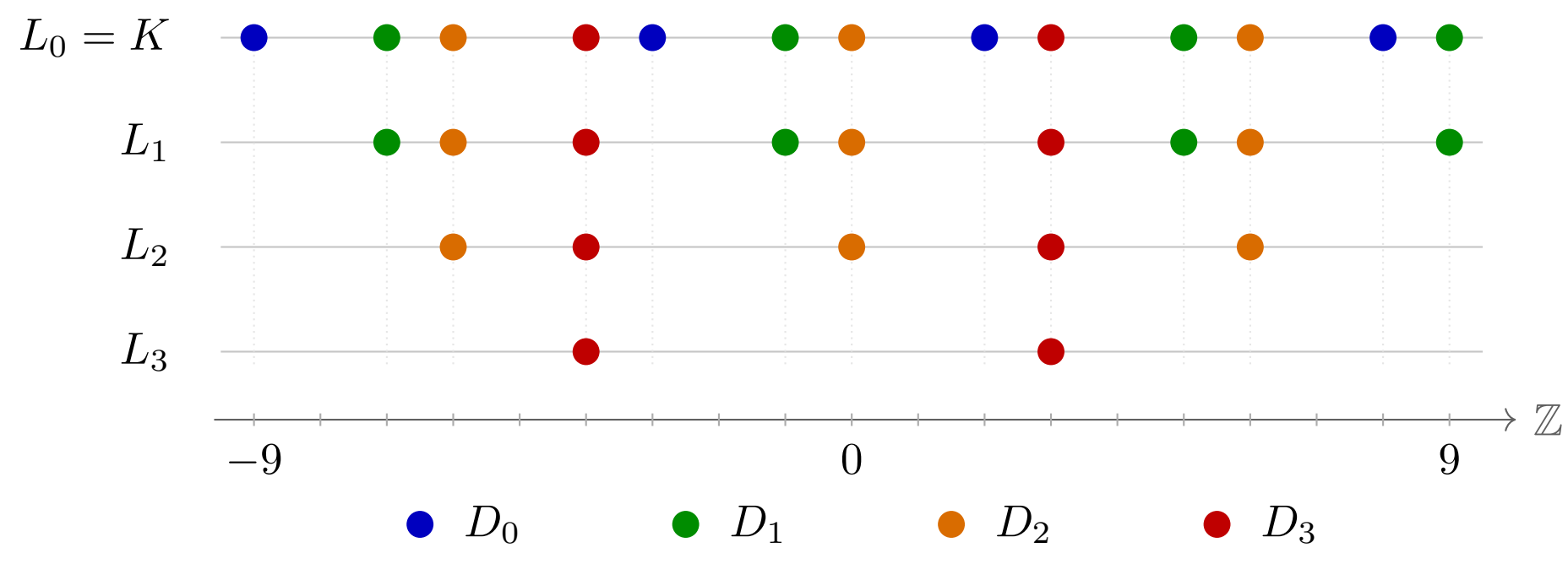

The \emph{layers} of $H$ are the sets $L_d$,
$L_{d-1}\setminus L_d$, \ldots, $L_0\setminus L_1$. They are nonempty and partition $A$. The order in which we list them reflects how an
adversary can reveal $K$. Suppose it first gives points of the smallest
language $L_d$, then points of $L_{d-1}$, and so on back to $K$. The
layers then appear one at a time: passing from $L_i$ to $L_{i-1}$
after the adversary reveals points of the new layer $L_{i-1}\setminus L_i$ the first time. As long as only
points of $L_i$ have been given, the inputs are consistent with the
target being $L_i$. (See Figure \ref{fig:language-layers}) Long chains are therefore typically the hardest
case for the algorithm to guarantee good density. Each additional layer is one more stage at
which the algorithm can be misled.

\begin{defn}
For a relative interval $J$ of a language $A$, and a chain $H$ starting at $A$,  we define the {\it raw pairing score }
\begin{equation}
 p(H,J)=\frac{1}{|J|}
       \sum_{\text{layers }E\text{ of }H}
                 \left\lfloor\frac{|E\cap J|}{2}\right\rfloor.
 \label{eq:pairing}
\end{equation}
\end{defn}

In words, each layer contributes half of its points in $J$, rounded
down, and $p(H,J)$ is the total contribution as a fraction of $|J|$.
The name comes from the following picture. Split the points of $J$
into pairs, never pairing two points from different layers, and count
one point for each pair. Since $\lfloor m/2\rfloor=(m-1)/2$ when $m$ is
odd and $m/2$ when $m$ is even, we can also write
\begin{equation}
 p(H,J)=\frac12-
 \frac{\#\{E:\ |E\cap J|\text{ is odd}\}}{2|J|}.
 \label{eq:odd}
\end{equation}
So $p(H,J)$ falls short of $1/2$ only because of the layers that meet
$J$ in an odd number of points, and each of them costs half a point.

The zero-density construction in our earlier work \cite{KWBanach} suggests that sparse output can arise when the learner moves up a chain of increasingly large guesses for the target language. In the online game, such a move occurs when Black reveals a point outside the previously encountered layers. The first point in the new layer \(E\) thus belongs to Black, after which Black and White can alternate in occupying points of \(E\). Consequently, White gets at most \(\lfloor |E\cap J|/2\rfloor\) points from each layer \(E\), which motivates the pairing score \(p(H,J)\).

For a positive integer $n$ we define the \emph{pairing score}
\begin{equation}
 c_n(H)=\lim_{R\to\infty}
 \inf_{\substack{J\text{ a relative interval of }A\\
                 |J|\ge n,\ J\cap[-R,R]=\emptyset}}p(H,J).
 \label{eq:cn}
\end{equation}
The two parameters play different roles. The number $n$ is a lower
bound on the number of points of $K$ in $J$, and it stays fixed. The
number $R$ only removes the intervals that meet the bounded region
$[-R,R]$. So letting $R$ grow does not force $|J|$ to grow, and short
intervals far from the origin still count. As $R$ grows the infimum is
taken over fewer intervals, so it can only increase, and the limit
exists. The infimum is never over an empty set. When the language $A$ is infinite,
it contains $n$ consecutive points lying outside $[-R,R]$, for every
$R$ and $n$.

We will use the following direct consequence of the definition.

\begin{claim}\label{claim:cn-eventual-interval-bound} For a chain $H$ starting from language $A$, if
$c_n(H)>q$, then there is a bounded region outside which every
relative interval of $A$ with at least $n$ points has pairing score
greater than $q$. This region may depend on the chain $H$.
\end{claim}

\subsection{Combinatorial Statement for the Main Theorem}

We now define the combinatorial invariant used in the
characterization. It is a min--max over refinement trees, motivated by
the zero-density construction in our previous work \cite{KWBanach}. The equivalent
topological formulation will be given afterward.

\begin{defn}\label{def:aot}
    Fix \(K\in\X\). An \emph{admissible obstruction tree} is a countable
rooted tree with language labels satisfying:
\begin{enumerate}
\item The root is labelled \(K\).
\item A child's label is a language in $\X$ which is a proper subset of its parent's label language.
\item If a vertex labelled \(L\) has children, then for every finite
\(F\subseteq L\), there is some child with label \(M\) such that
\[
                         F\subseteq M\subsetneq L.
\]
\item There is no infinite path starting at the root.
\end{enumerate}
The single-vertex tree is permitted.
\end{defn}

The third condition expresses persistence of ambiguity: whatever
finite evidence from \(L\) has appeared, some smaller child remains
consistent. The child is allowed to depend on the finite evidence.
For example, the sets \(\mathbb N\setminus\{a\}\), one for each
\(a\), accommodate every finite subset of \(\mathbb N\), although
each is a proper subset.

The last condition allows arbitrarily long finite paths, but forbids
an infinite one. Every vertex has a leaf below it. These are trees
used to define the structural number, not the forest maintained
by the generator. The same language may label vertices on different
branches.

Define
\begin{equation}
 b_n(K)=\inf_{\substack{T\text{ admissible}\\
                        \text{root label }K}}
                  \ \sup_{H\text{ a root-to-leaf path of }T}c_n(H),
 \qquad
 b(K)=\lim_{n\to\infty}b_n(K).
 \label{eq:bn}
\end{equation}
The scores lie in \([0,1/2]\). Increasing \(n\) increases every
\(c_n\), hence increases \(b_n\), so the last limit exists.

The following immediate consequence will be used in constructing the
generator.

\begin{claim}\label{claim:bn-tree-path-bound}
    For any threshold $q < b_n(K)$, every admissible tree has at least one root-to-leaf path with pairing score above $q$. 
\end{claim}

Now we are ready to state the main theorem lower bound. 
\begin{thm}\label{thm:lower}
For every nonempty countable class \(\X\) of infinite subsets of
\(\mathbb Z\), there is one generator such that, on every full enumeration of every \(K\in\X\):
\begin{enumerate}
\item every output is available, and all but finitely many outputs
belong to \(K\);
\item its output set \(O\) satisfies
$\lbanach(O,K)\ge b(K)$.
\end{enumerate}
\end{thm}

This is an information-theoretic existence theorem. The generator
may use the whole language class, language inclusions, and the
finite witnesses proved to exist below. We do not claim that those
tests or witnesses can be computed from arbitrary descriptions of
the languages.

\subsection{Topological formulation of the characterization theorem }
The combinatorial obstruction-tree definition of \(b_n(K)\) is global: it organizes
all refinements that can remain possible under finite positive evidence.
We now give an equivalent local formulation in terms of neighborhoods of
refinement histories.  This requires lifting the finite-evidence topology \cite{KWBanach}
from individual languages to finite descending histories.  The resulting
finite-evidence history topology is defined formally in
Definition~\ref{opensets} and developed in
Subsection~\ref{sec:history-topology}; the corresponding topological
form of the main characterization is stated in
Theorem~\ref{thm:topological-main}. 

Our earlier works \cite{kleinberg2025density, KWBanach} equipped \(\collec\) with the {\it finite-evidence topology},
whose basic neighborhoods are
\[
U_{L,F}=\{M\in\X:F\subseteq M\subseteq L\},
\]
where \(F\subseteq L\) is finite.
$L$ runs through all possible languages $L \in \X$ and $F$ all finite subsets of $L$. 

The topology on individual languages is no longer sufficient, because
the score \(c_n(H)\) depends on the entire descending chain (called history) \(H\),
not only on its final language. We therefore lift the finite-evidence
topology to the space of finite descending histories. The resulting
\emph{finite-evidence history topology}, defined in
Definition~\ref{opensets}, is by
Lemma~\ref{lem:finest-history-topology} the finest topology preserving
finite-evidence convergence under continuation. 

Let \(\mathcal H_K\) denote the finite descending chains (histories) beginning at \(K\). The
obstruction-tree definition of \(b_n(K)\) has the following equivalent
description in terms of open neighborhoods of the initial history
\((K)\). Thus we also obtain an equivalent topological formulation for Theorem \ref{thm:lower}.

\begin{thm}
\label{thm:topological-bn}
For every \(K\in\X\) and every positive integer \(n\),
\begin{equation}
   b_n(K)
 =
 \sup_{\substack{\mathcal U\subseteq\mathcal H_K\text{ open}\\
                 (K)\in\mathcal U}}
 \inf_{H\in\mathcal U}c_n(H). \label{eq:bn-topo}  
\end{equation}
\end{thm}
The proof of this theorem is in Subsection \ref{subsec:topo1D}. 

\subsection{Intuition behind the combinatorial and topological formulations}

The combinatorial and topological formulas organize the same refinement
phenomenon at two different scales.

An admissible obstruction tree is a global certificate. At every
nonterminal history and for every finite set of positive examples, it
provides some proper refinement that remains consistent with those
examples. These choices must be organized into one well-founded tree,
and every terminal branch is evaluated by its pairing score.

The topological formulation organizes the same ambiguity locally.
At each history, one chooses a finite positive test and retains every
proper refinement that contains it. Repeating this rule along all
surviving histories produces an open neighborhood. Its score is the
smallest pairing score among all histories that survive this coherent
system of finite tests.

Lemma~\ref{lem:obstruction-closure} identifies the relation between
these two structures: a history lies in the closure of a set
\(\mathcal D\) precisely when the low-score continuations in
\(\mathcal D\) can be organized into one admissible obstruction tree
below that history. Applying this lemma to
$\mathcal D_q=\{\text{descending chain} H:c_n(H)\le q\}$
gives Theorem~\ref{thm:topological-bn} and identifies the
obstruction-tree min--max with the topological lower limit of \(c_n\)
at the initial history.

\section{Finite Cantor--Bendixson rank as a corollary}
\label{sec:finite-rank-corollary}

We conclude by showing that the finite Cantor--Bendixson rank theorem
from \cite{KWBanach} is an immediate corollary of the structural
characterization.

\begin{defn}\label{def:CBrank}
    Set $\collec^{(0)}=\collec$ and, for each integer $r\ge 0$,
let $\collec^{(r+1)}$ be obtained from $\collec^{(r)}$
by removing its isolated points.
A language $L$ has \emph{Cantor--Bendixson point rank} $r$
if $L\in\collec^{(r)}\setminus\collec^{(r+1)}$.
We say that $\collec$ has \emph{finite Cantor--Bendixson rank}
if $\collec^{(r)}=\emptyset$ for some finite $r$.
\end{defn}
A language of rank
zero is isolated in the original space.

The following elementary consequence of the topology is the only
rank property that we need.

\begin{lem}
\label{lem:finite-rank-witness}
Suppose that \(L\) has finite rank \(r\). Then there is a finite
set \(F\subseteq L\) such that every proper language \(M\in\X\)
satisfying \(F\subseteq M\subsetneq L\) has rank strictly less
than \(r\).

In particular, if \(r=0\), no proper language \(M\in\X\) can satisfy
\(F\subseteq M\subsetneq L\).
\end{lem}

\begin{proof}
Since \(L\in\X^{(r)}\setminus\X^{(r+1)}\), the language \(L\) is an 
isolated point in the subspace \(\X^{(r)}\). Since the sets
\(U_{L,F}\) form a basis, there is a finite
\(F\subseteq L\) such that
\(U_{L,F}\cap\X^{(r)}=\{L\}\).

Now let \(M\in\X\) satisfy \(F\subseteq M\subsetneq L\). Since \(M\in U_{L,F}\setminus\{L\}\) and
\(U_{L,F}\cap\mathcal X^{(r)}=\{L\}\), we have
\(M\notin\mathcal X^{(r)}\). Therefore \(M\) was removed at some
stage \(s<r\), so it has finite point rank \(s\).
\end{proof}

The connection with obstruction trees is direct. At a vertex labelled
by a finite-rank language \(L\), the preceding lemma gives finite
evidence \(F\) that forces every proper refinement containing \(F\) to
have smaller rank. If the obstruction-tree vertex is not a leaf, its
child condition supplies such a refinement. Thus one can move down the
tree while strictly decreasing rank.

\begin{cor}
\label{cor:finite-rank-b}
Suppose that \(K\) has finite rank \(r\). Then, for every
positive integer \(n\),
\(b_n(K)\ge \frac12-\frac{r+1}{2n}\).
Consequently, \(b(K)=1/2\).

In particular, if \(\X\) has finite Cantor--Bendixson rank, then the
generator in the structural theorem achieves lower Banach density
\(1/2\) on every target language.
\end{cor}

\begin{proof}
Fix \(n\), and let \(T\) be an arbitrary admissible obstruction tree
rooted at \(K\).

We construct a root-to-leaf path in \(T\). Suppose that the current
vertex is labelled by a language \(L\) of rank \(s\). If the
vertex is a leaf, stop. Otherwise, apply
Lemma~\ref{lem:finite-rank-witness} to obtain a finite set
\(F\subseteq L\) such that every proper refinement of \(L\) containing
\(F\) has rank less than \(s\).

By the child condition for an admissible obstruction tree, the current
vertex has a child labelled by some \(M\) with
\(F\subseteq M\subsetneq L\). The lemma gives that \(M\) has point
rank less than \(s\). Move to this child and continue.

The rank strictly decreases at every step. Starting from rank
\(r\), the process therefore reaches a leaf after at most \(r\)
edges. Let \(H\) be the resulting root-to-leaf chain. It has at most
\(r+1\) layers.

Let \(J\) be any relative interval of \(K\). If the layers of \(H\)
are \(D_1,\ldots,D_s\), where \(s\le r+1\), then
\(\lfloor |D_i\cap J|/2\rfloor\ge
|D_i\cap J|/2-1/2\) for every \(i\). Since the layers partition \(K\),
their intersections with \(J\) partition \(J\). It follows that
\(p(H,J)\ge \frac12-\frac{s}{2|J|}
\ge \frac12-\frac{r+1}{2|J|}\).

In particular, whenever \(|J|\ge n\), we have
\(p(H,J)\ge \frac12-\frac{r+1}{2n}\). Therefore
\(c_n(H)\ge \frac12-\frac{r+1}{2n}\).
We have shown that the arbitrary obstruction tree \(T\) contains a
root-to-leaf path \(H\) with this lower bound. Hence
\[
    \sup_{H\text{ a root-to-leaf path of }T} c_n(H)
    \ge
    \frac12-\frac{r+1}{2n}.
\]
Taking the infimum over all admissible obstruction trees gives
\(b_n(K)\ge \frac12-\frac{r+1}{2n}\).

On the other hand, every pairing score is at most \(1/2\), so
\(b_n(K)\le1/2\). Letting \(n\to\infty\) yields \(b(K)=1/2\).

The structural theorem now gives one eventually valid generator whose
output set satisfies
\(\lbanach(O,K)\ge b(K)=1/2\). If \(\X\) has finite
Cantor--Bendixson rank, every language in \(\X\) has finite rank,
so the conclusion holds simultaneously for every target.
\end{proof}

\section{The matching adversarial upper bound}
\label{sec:upper}

We now prove that the structural number \(b(K)\) is also an upper
bound. Thus no generator that is eventually valid on every language
can guarantee lower Banach density strictly larger than \(b(K)\) on
the target \(K\).

The proof is organized into mega steps, following the same general adversarial strategy as Theorem~4.1 of our previous work~\cite{KWBanach}. One mega step
produces one relative interval of \(K\) on which the generator's
output density is controlled by the pairing score of a root-to-leaf
path in an obstruction tree. Repeating the mega step at larger and
larger scales gives the desired Banach-density upper bound.

\begin{thm}
\label{thm:upper}
Let \(\mathcal A\) be a generator that is eventually valid on every
full enumeration of every language in \(\X\). Then, for every \(K\in\X\),
there is a full enumeration of \(K\) such that the output set \(O\) of
\(\mathcal A\) satisfies
\[
                         \lbanach(O,K)\le b(K).
\]
\end{thm}

We begin with the elementary race that will be used on each layer of
a path.

\begin{lem}
\label{lem:upper-race}
Let \(A\) be a finite set none of whose points has yet been used.
Suppose that, as long as \(A\) has an available point, the adversary
inputs one available point of \(A\). When the race ends, the algorithm
has generated at most
$\left\lfloor\frac{|A|}{2}\right\rfloor$
points of \(A\).
\end{lem}

\begin{proof}
Let \(a\) be the number of points of \(A\) input by the adversary and
let \(o\) be the number generated by the algorithm. At the end of the
race every point of \(A\) has been used, so $a+o=|A|$
There were \(a\) rounds in the race, and the algorithm produces only
one output per round. Hence \(o\le a\), and therefore $ o\le \left\lfloor\frac{|A|}{2}\right\rfloor$.
\end{proof}

\subsection{One mega step}

Fix a positive integer \(n\), a number \(\varepsilon>0\), and an
admissible obstruction tree \(T\) rooted at \(K\). Suppose that only
finitely many rounds have occurred so far, and that all inputs revealed
so far belong to \(K\).

We describe a \(T\)-mega step. It will produce a root-to-leaf path $ H=(K=L_0\supsetneq L_1\supsetneq\cdots \supsetneq L_d)$
and a relative interval \(J\) of \(K\) such that $ |J|\ge n$ and
\begin{equation}
\label{eq:upper-mega-score}
                    \frac{|O\cap J|}{|J|}
                    \le p(H,J)
                    \le c_n(H)+\varepsilon.
\end{equation}
The interval can moreover be required to lie beyond any prescribed
finite central region.

The mega step is defined recursively on the obstruction tree. For a
vertex \(v\), write \(L(v)\) for its language label. Whenever the
procedure is called at \(v\), every input revealed so far belongs to
\(L(v)\).

\paragraph{Refill step.}
At the beginning of each new attempt at \(v\), the adversary reveals
the first point of \(L(v)\), in some fixed ordering of \(L(v)\), that
has not yet appeared as an input.

This refill step will only be important if infinitely many attempts
are made at \(v\). In that case it ensures that the resulting input
sequence is a full enumeration of \(L(v)\).

\paragraph{Round at a leaf.}
Suppose first that \(v\) is a leaf. Let $H=(K=L_0\supsetneq L_1\supsetneq\cdots
\supsetneq L_d=L(v))$
be the path from the root to \(v\).

Choose \(R\) larger than the absolute value of every point used so far
by either player. We also take \(R\) beyond every candidate interval
discarded in an earlier attempt. By the definition of \(c_n(H)\), there
is a relative interval \(J\) of \(K\) such that
\begin{equation}
\label{eq:upper-choose-J}
        |J|\ge n, \ 
        J\cap[-R,R]=\emptyset,\text{ and }
        p(H,J)\le c_n(H)+\varepsilon.
\end{equation}
In particular, every point of \(J\) is available when this attempt
begins.

The adversary now races the algorithm on
$J\cap L_d$.
As long as this set has an available point, the adversary inputs one
such point.

If, during this race, the algorithm outputs any point outside \(L_d\),
we call the attempt \emph{bad at the leaf}. The interval \(J\) is
discarded, and the adversary starts a new attempt at the same leaf,
including another refill step and a fresh interval farther out.

The leaf cannot have infinitely many bad attempts. Indeed, suppose
that it did. Every adversarial input would belong to \(L_d\), and the
refill steps would ensure that every point of \(L_d\) was eventually
revealed. The resulting sequence would therefore be a full enumeration
of \(L_d\). But every bad attempt contains an output outside \(L_d\),
so the algorithm would output outside \(L_d\) infinitely often. This
contradicts eventual validity on \(L_d\).

Consequently, some leaf attempt is good. By Lemma~\ref{lem:upper-race}, on this attempt the algorithm occupies at
most $\left\lfloor\frac{|J\cap L_d|}{2}\right\rfloor$ points of the bottom layer \(L_d\cap J\). At the end of the round, every point of this layer has been used.

\paragraph{Moving one level upward.}
We prove termination by well-founded induction on the obstruction
tree: each recursive child call terminates by the induction
hypothesis, and infinitely many unsuccessful attempts at the
current vertex contradict eventual validity as follows.

Now suppose that \(v\) is not a leaf. Start a new attempt at \(v\) with
a refill step. Let \(F\) be the finite set of all inputs revealed up
to this point.

By the child condition in the definition of an admissible obstruction
tree, \(v\) has a child \(w\) such that
\[
                         F\subseteq L(w)\subsetneq L(v).
\]
The adversary now carries out the procedure recursively below \(w\).
Since the obstruction tree has no infinite path, this recursion is
well founded. It eventually produces a root-to-leaf path \(H\)
passing through the edge \(v w\), together with a relative interval
\(J\), and it has already completed the races on every layer of \(H\)
strictly below \(v\).

The adversary next races the algorithm on the new layer
$J\cap\bigl(L(v)\setminus L(w)\bigr)$.
As long as this layer contains an available point, the adversary inputs
one such point.

If the algorithm outputs a point outside \(L(v)\) before this race
ends, we call the attempt \emph{bad at \(v\)}. The interval \(J\) and
the path chosen below \(v\) are discarded. The adversary begins a new
attempt at \(v\), performs another refill step, and then chooses a
possibly different child containing the enlarged input history.

Again, \(v\) cannot have infinitely many bad attempts. If it did, all
adversarial inputs made from the first call to \(v\) onward would lie
in \(L(v)\). The refill at the start of each attempt would ensure that
every point of \(L(v)\) eventually appeared. Thus the resulting input
sequence would be a full enumeration of \(L(v)\). Every bad attempt,
however, gives an output outside \(L(v)\), contradicting eventual
validity on \(L(v)\).

It follows that some attempt is good at \(v\). On that attempt,
Lemma~\ref{lem:upper-race} gives
\[
 \left|O\cap J\cap\bigl(L(v)\setminus L(w)\bigr)\right|
 \le
 \left\lfloor
 \frac{\left|J\cap\bigl(L(v)\setminus L(w)\bigr)\right|}{2}
 \right\rfloor.
\]

\paragraph{Why the layers do not interfere.}
Consider the final successful attempt returned by this recursive
procedure. The candidate interval \(J\) was chosen only after all
refill steps on its root-to-leaf path had occurred. Hence no refill
input lies in \(J\).

Moreover, the returned interval $J$ is chosen
after all earlier discarded attempts and refill
steps, and avoids every point used in those steps.
During the final successful construction below
a child $w$, every output in $J$ lies in $L(w)$.
Consequently no point of
$J\cap(L(v)\setminus L(w))$ has been output
before the race on that layer begins.

Similarly, during the race at \(v\), an output in any layer above
\(L(v)\) would lie outside \(L(v)\) and would make the attempt bad.
Hence, on the final successful attempt, no later layer is touched
prematurely.

We have therefore exposed the layers of
\[
              H=(K=L_0\supsetneq L_1\supsetneq\cdots
                                      \supsetneq L_d)
\]
in the order
           $L_d, L_{d-1}\setminus L_d,\ldots,
           L_0\setminus L_1$.
For every layer \(E\) of \(H\), the algorithm occupies at most
$\left\lfloor
\frac{|E\cap J|}{2}\right\rfloor$
points. Therefore
\begin{equation}
\label{eq:upper-path-count}
\begin{split}
                   |O\cap J|\le
     \sum_{E\text{ a layer of }H}
\left\lfloor\frac{|E\cap J|}{2}\right\rfloor=p(H,J)|J|.
\end{split}
\end{equation}
By the choice of \(J\) in \eqref{eq:upper-choose-J},
\[
                   \frac{|O\cap J|}{|J|}
                   \le p(H,J)
                   \le c_n(H)+\varepsilon.
\]

At the end of the mega step, every point of \(J\) has been used by
one of the two players. Since the generator always outputs available
points, no later round can add another output to \(J\). Thus the
density obtained in \eqref{eq:upper-path-count} is permanently sealed.

\subsection{Choosing a bad tree}

We now choose the obstruction tree used in a mega step.

Fix \(n\) and \(\varepsilon>0\). By the definition of \(b_n(K)\), there
is an admissible obstruction tree \(T\), rooted at \(K\), such that
\begin{equation}
\label{eq:upper-near-optimal-tree}
       \sup_{H\text{ a root-to-leaf path of }T}c_n(H)
       \le b_n(K)+\frac{\varepsilon}{2}.
\end{equation}
Run the \(T\)-mega step above, using error \(\varepsilon/2\) when
choosing the interval at the leaf.

The adversary does not need to choose a particular root-to-leaf path
in advance. Its successive child choices are determined by the finite
input history, and the requirement that the obstruction tree have no
infinite path guarantees that some leaf is eventually reached. Let
\(H\) be the path produced by the successful attempt.

Every root-to-leaf path of \(T\) obeys
\eqref{eq:upper-near-optimal-tree}. Hence the interval \(J\) produced
by the mega step satisfies
\begin{equation}
\label{eq:upper-one-stage}
        \frac{|O\cap J|}{|J|}
        \le c_n(H)+\frac{\varepsilon}{2}
        \le b_n(K)+\varepsilon.
\end{equation}

Notice also the consistency property that will be needed when the
construction is repeated: at every downward move, the adversary
chooses a child containing the entire finite input history. Therefore
all languages on the newly selected path contain every input that had
been revealed before the adversary began moving upward along that
path.

\subsection{Repeating the mega step}
The one-mega-step construction gives one permanently sparse interval.
To obtain the lower Banach-density upper bound, we repeat it with
interval sizes tending to infinity and errors tending to zero.

\begin{proof}[Proof of Theorem~\ref{thm:upper}]
Take \(n_s=s\) and \(\varepsilon_s=2^{-s}\).

At mega step \(s\), choose an admissible obstruction tree \(T_s\)
satisfying
\[
       \sup_{H\text{ a root-to-leaf path of }T_s}c_{n_s}(H)
       \le b_{n_s}(K)+\frac{\varepsilon_s}{2}.
\]
Run the corresponding mega step beyond every point and every candidate
interval used in the preceding stages.

Let \(H_s\) be the root-to-leaf path selected during the successful
attempt, and let \(J_s\) be the resulting relative interval. Then $|J_s|\ge n_s$
and
\begin{equation}
\label{eq:upper-stage-final}
       \frac{|O\cap J_s|}{|J_s|}
       \le p(H_s,J_s)
       \le c_{n_s}(H_s)+\frac{\varepsilon_s}{2}
       \le b_{n_s}(K)+\varepsilon_s.
\end{equation}

The input sequence produced by the infinitely many mega steps is a full enumeration of \(K\). Indeed, at the beginning of every root attempt,
the refill rule reveals the first point of \(K\) not previously used
as an input. There is at least one root attempt in every mega step, so
every point of \(K\) is eventually revealed.

Since \(|J_s|\to\infty\), the definition of lower Banach density gives
\[
\lbanach(O,K)\le
     \liminf_{s\to\infty}
\frac{|O\cap J_s|}{|J_s|}\le
     \liminf_{s\to\infty} \bigl(b_{n_s}(K)+\varepsilon_s\bigr)=b(K).
\]
Here the last equality follows from $b_n(K)\nearrow b(K)$. This proves the theorem.
\end{proof}

\section{Achieving the structural lower bound}

We now turn to the constructive direction.  The challenge is stronger
than proving a target-by-target lower bound: we must show there is a single generator  attaining
the structural guarantee simultaneously for every target in the
family.  We first explain the architecture of the construction and
then develop the protected forest, localized board states, and online
generator that realize it.

\subsection{Proof architecture and relation to the earlier tree arguments}
\label{sec:architecture}

The proof combines two features that appeared separately in our earlier
work.  The asymptotic-density argument used an unbounded hierarchy of
hypotheses and repeatedly returned to common ancestors
\cite[Section~6]{kleinberg2026density}. The finite-rank lower-Banach-density
argument used a static inclusion tree whose height was bounded by the
Cantor--Bendixson rank
\cite[Section~4.2]{KWBanach}.

The present construction is again static, but its paths may have
arbitrary finite length. Its new feature is that every finite path
carries its own quantitative pairing-score guarantee.

Two tree objects play complementary roles.

An admissible obstruction tree is an adversarial certificate: it
retains a refinement compatible with every finite set of positive
evidence and attempts to force all terminal histories to have small
score. The generator instead uses a protected forest whose realized
histories remain above the required score.

The closure duality supplies the bridge between them. When no
obstruction tree can force the score below a threshold, a finite set of
positive examples protects the threshold against every compatible next
refinement. We combine these finite protected sets into one static
forest supporting all targets and all relevant ancestor requirements.

The forest determines the language from which the generator currently
draws outputs. We then localize the construction to interval boards.
Each board maintains its own pullback state, and the density analysis on
that board reduces to the pairing score of the corresponding refinement
history. 

\subsection{A simpler fixed-error theorem}\label{sec:structural-number}

We first prove a fixed-error version of the lower bound (Theorem \ref{thm:fixed-epsilon}).  This contains
the main conceptual ingredients of the achievability proof: protected
histories, the static protected forest, localized board states, and the
online argument establishing validity and density.  Allowing a fixed
error \(\varepsilon>0\) keeps the construction substantially simpler,
because each language carries only one density requirement and appears
only once in the forest.  In particular, this version presents the core
mechanism in a much cleaner form, without the additional bookkeeping
created by multiple formal copies of the same language.

Passing from \(b(K)-\varepsilon\) to the exact value \(b(K)\) requires
one additional conceptual ingredient: for each language, we introduce
countably many formal copies carrying successively stronger requirements
whose errors tend to zero.  Executing this idea is technically more
involved, because these copies must coexist consistently in the same
forest and in the same online run.  Apart from this accuracy-copy
mechanism, however, the structural and algorithmic ideas are the same as
in the fixed-error proof.  We therefore present the cleaner fixed-error
construction first, and defer the exactification to
Subsection~\ref{sec:exact-bound}.

\begin{thm}
\label{thm:fixed-epsilon}
For every $\varepsilon>0$ there is one generator $\mathcal A_\varepsilon$
such that, on every full enumeration of every $K\in\X$,
\begin{enumerate}
\item all but finitely many outputs belong to $K$; and
\item the output set satisfies
\[
 \lbanach(O,K)\ge (b(K)-\varepsilon)_+.
\]
\end{enumerate}
\end{thm}

Here \(x_+=\max\{x,0\}\).  We now prove
Theorem~\ref{thm:fixed-epsilon}; the exactification to remove $\varepsilon$ will be in
Subsection~\ref{sec:exact-bound}. It will reuse this construction with
multiple formal copies of each language.

\subsection{Safe histories, topology, and protected sets}
\label{sec:protected-sets}
To build the protected forest, we need finite positive evidence that
prevents a good refinement history from being extended into a bad one.
We formalize this through \emph{safe histories}, identify safety as an
open condition in the finite-evidence history topology, and then extract
finite protected sets witnessing this openness.  These protected sets
will be the local certificates carried by the forest.

\subsubsection{Safety relative to one starting requirement}

Fix a language \(A\in\X\), an integer \(n\ge1\), and a number
\(q<b_n(A)\). These parameters remain fixed throughout this
subsection.

A \emph{finite refinement history} starting at \(A\) is a strictly
descending chain
\(H=(A=M_0\supsetneq M_1\supsetneq\cdots\supsetneq M_r)\).
We write \(\operatorname{end}(H):=M_r\).

Suppose that \(T\) is an admissible obstruction tree whose root is
labelled by \(\operatorname{end}(H)\). Every root-to-leaf path in
\(T\) determines an extension of \(H\). When the path is appended to
\(H\), the common language \(\operatorname{end}(H)\) is written only
once. In particular, if \(T\) consists only of its root, the resulting
history is \(H\) itself.

\begin{defn}\label{def:safe-unsafe-history}
The history \(H\) is \emph{\((A,n,q)\)-unsafe} if there is an
admissible obstruction tree rooted at \(\operatorname{end}(H)\) such
that every root-to-leaf path, when appended to \(H\), produces a
history \(H'\) satisfying \(c_n(H')\le q\).

Otherwise, \(H\) is \emph{\((A,n,q)\)-safe}. When the parameters are
clear, we simply say that \(H\) is safe or unsafe.
\end{defn}

Thus an unsafe history is one below which the adversary can preserve
finite-evidence ambiguity while forcing every terminal continuation
to have score at most \(q\). A safe history is one from which no such
obstruction is possible.

The purpose of the next subsection is to show that this notion of
safety is naturally topological.

\subsubsection{Topology on finite histories}
\label{sec:history-topology}

Our earlier work uses a topology on the language class itself
\cite{kleinberg2025density,KWBanach}. For a language \(L\in\X\) and a finite set
\(F\subseteq L\), the basic finite-evidence neighborhood is
\[U_{L,F}:=\{M\in\X:F\subseteq M\subseteq L\}.\] It records the
languages that remain possible refinements of \(L\) after the positive
examples in \(F\) have appeared.

The endpoint language alone is no longer sufficient here. The score
\(c_n(H)\) depends on the layers created along the entire history
\(H\). Two histories can end at the same language but have different
layer decompositions, different scores, and different safety
properties.

We therefore extend the earlier language topology to a topology on finite histories.

\paragraph{Basic open sets in the topology on finite histories.}

Let \(\mathcal H\) be the set of all finite strictly descending chains
of languages in \(\X\), with arbitrary starting language. If
\(\Gamma\in\mathcal H\) and
\(M\subsetneq\operatorname{end}(\Gamma)\), write
\(\Gamma^\frown M\) for the history obtained by appending \(M\).
We say that a history \(\Gamma\) \emph{extends} a history \(H\) if
\(H\) is an initial segment of \(\Gamma\).

We first define the basic open sets. 
\begin{defn}\label{opensets}
Fix a history \(H\). One open set  of \(H\) is constructed
recursively as follows.

Start by including \(H\). For each history \(\Gamma\) that has been
included, choose one finite set
\(F_\Gamma\subseteq\operatorname{end}(\Gamma)\). Having made this
choice, include \emph{every} one-step extension
\(\Gamma^\frown M\) such that
\(F_\Gamma\subseteq M\subsetneq\operatorname{end}(\Gamma)\).

The same rule is then applied to every newly included history: for
each such history \(\Gamma'\), choose one finite set
\(F_{\Gamma'}\subseteq\operatorname{end}(\Gamma')\), and include every
proper one-step refinement that contains \(F_{\Gamma'}\). Continuing
through all finite depths produces a collection of finite histories,
which we call a \emph{basic neighborhood} of \(H\).

The finite set is chosen separately for each history. In particular,
if two different histories \(\Gamma\) and \(\Gamma'\) have the same
endpoint language, there is no requirement that
\(F_\Gamma=F_{\Gamma'}\). Thus the finite test may depend on the entire
refinement history, not only on its endpoint.

We take all collections obtained in this way, for all starting
histories \(H\) and all possible choices of the finite sets
\(F_\Gamma\), as basic open sets. The topology they generate on
\(\mathcal H\) is called the \emph{finite-evidence history topology}.
\end{defn}

An open set above therefore makes one finite choice at each history
that it contains. If the chosen set at a history \(\Gamma\) is
\(F_\Gamma\), then \emph{all} proper refinements \(M\) with
\(F_\Gamma\subseteq M\subsetneq\operatorname{end}(\Gamma)\) are
included. At each of the resulting histories, a new finite set is
chosen, and the same rule is repeated.

These choices are made separately at different histories. In
particular, if two different histories have the same endpoint
language, the finite sets chosen for them need not be the same. Thus a
basic neighborhood remembers not only the current language but also
the refinement history by which it was reached.

\begin{lem}
\label{lem:history-topology}
The open sets defined above form a basis for a topology on
\(\mathcal H\).

Moreover, a set \(U\subseteq\mathcal H\) is open  in this topology if and only
if it has the following property: for every history
\(H\in U\), there is a finite set
\(F_H\subseteq\operatorname{end}(H)\) such that every language \(M\in\X\) with
\(F_H\subseteq M\subsetneq\operatorname{end}(H)\)
satisfies \(H^\frown M\in\mathcal U\).
\end{lem}

\begin{proof}
Every history $H$ belongs to a basic open set built from $H$, so it remains only to check intersections.

Suppose that a history \(\Gamma\) belongs to two basic neighborhoods
\(\mathcal V\) and \(\mathcal W\). At \(\Gamma\), the first
neighborhood has chosen some finite set \(F_\Gamma^{\mathcal V}\),
while the second has chosen some finite set
\(F_\Gamma^{\mathcal W}\). For a smaller neighborhood inside both,
choose instead their union
\(F_\Gamma^{\mathcal V}\cup F_\Gamma^{\mathcal W}\).

Every proper refinement containing this union passes the test used by
both \(\mathcal V\) and \(\mathcal W\), and hence belongs to both.
At each such extended history, repeat the same procedure: take the
union of the two finite sets chosen there by \(\mathcal V\) and
\(\mathcal W\), and include every proper refinement containing that
union. Continuing recursively gives a basic open set of
\(\Gamma\) contained in \(\mathcal V\cap\mathcal W\). Thus the basic open sets form a basis.

Now let \(\mathcal U\) be open and let \(H\in\mathcal U\). There is a
basic open set of \(H\) contained in \(\mathcal U\). Let
\(F_H\) be the finite set chosen at its first step. By the definition of a basic open set, every proper refinement \(M\) satisfying
\(F_H\subseteq M\subsetneq\operatorname{end}(H)\) gives a history
\(H^\frown M\) lying in that basic open set, and therefore in
\(\mathcal U\).

Conversely, suppose that \(\mathcal U\) has the stated property.
Starting from any \(H\in\mathcal U\), choose such a finite set \(F_H\).
Include every proper refinement that contains \(F_H\); by assumption, all of the resulting histories still lie in \(\mathcal U\). At each of those histories, choose a finite set supplied by the same assumption, and again include every proper refinement containing it.
Continuing recursively produces a basic open set of \(H\) lying
entirely inside \(\mathcal U\). Hence \(\mathcal U\) is open.
\end{proof}

\paragraph{Connection with the topology in our earlier work.}

The new topology uses exactly the same finite-evidence convergence as
the earlier topology, but applies it separately at every history.

Suppose that \(H\) ends at \(L\), and let
\(M_1,M_2,\ldots\) be proper sublanguages of \(L\). Assume that every
finite subset of \(L\) is contained in all sufficiently late \(M_j\).
In the earlier topology on languages, this says that \(M_j\) converges
to \(L\).

In the history topology, we will show that the corresponding statement is
\(H^\frown M_j\to H\). 

For example, consider
\(\X=\{\mathbb N\}\cup
\{\mathbb N\setminus\{j\}:j\ge1\}\).
In the earlier topology,
\(\mathbb N\setminus\{j\}\to\mathbb N\), because every fixed finite
subset of \(\mathbb N\) is contained in
\(\mathbb N\setminus\{j\}\) for all sufficiently large \(j\).
Correspondingly, in the history topology,
\((\mathbb N,\mathbb N\setminus\{j\})\to(\mathbb N)\).

More generally, if \(H\) is any history ending at \(\mathbb N\), then
\(H^\frown(\mathbb N\setminus\{j\})\to H\). If \(H'\) is a different
history with the same endpoint, then the corresponding extensions
converge to \(H'\), not to \(H\). The endpoint convergence is the same,
but the new topology retains the preceding refinement history.

We now make these heuristics precise, and show that the history topology is not introduced in order to make the later
proof work. It is the natural topology obtained by lifting the earlier
finite-evidence convergence to every possible refinement history.

\begin{lem}
\label{lem:finest-history-topology}
The finite-evidence history topology is the finest topology on
\(\mathcal H\) with the following property.

Let \(H\) be a history ending at \(L\), and let
\(M_1,M_2,\ldots\) be proper refinements of \(L\). If
\(M_j\to L\) in the finite-evidence topology on languages, then
\(H^\frown M_j\to H\) in the history topology.
\end{lem}

\begin{proof}
We first check that the finite-evidence history topology has the stated
property. Let \(\mathcal U\) be an open neighborhood of \(H\).
By Lemma~\ref{lem:history-topology}, there is a finite set
\(F_H\subseteq L\) such that every proper refinement \(M\) containing
\(F_H\) satisfies \(H^\frown M\in\mathcal U\).

Since \(M_j\to L\) in the finite-evidence topology on languages, all
sufficiently large \(M_j\) contain \(F_H\). Hence
\(H^\frown M_j\in\mathcal U\) for all sufficiently large \(j\).
Therefore \(H^\frown M_j\to H\).

Now let \(\sigma\) be any other topology on \(\mathcal H\) with this
property. We show that every \(\sigma\)-open set is also open in the
finite-evidence history topology.

Let \(\mathcal U\) be an open set in  \(\sigma\) and let a chain \(H\in\mathcal U\), with
\(L=\operatorname{end}(H)\). We claim that there is a finite set
\(F_H\subseteq L\) such that every proper refinement \(M\) with
\(F_H\subseteq M\subsetneq L\) satisfies
\(H^\frown M\in\mathcal U\).

Suppose not. Since \(L\subseteq\mathbb Z\) is countable, choose finite
sets \(F_1\subseteq F_2\subseteq\cdots\) whose union is \(L\).
By our assumption, for every \(j\) there is a proper refinement
\(M_j\subsetneq L\) such that \(F_j\subseteq M_j\), but
\(H^\frown M_j\notin\mathcal U\).

The sequence \(M_j\) converges to \(L\) in the finite-evidence topology.
Indeed, if \(F\subseteq L\) is finite, then \(F\subseteq F_j\) for all
sufficiently large \(j\), and hence \(F\subseteq M_j\) for all such
\(j\).

By the assumed property of \(\sigma\), it follows that
\(H^\frown M_j\to H\) in \(\sigma\). But \(\mathcal U\) is a
\(\sigma\)-open neighborhood of \(H\), so all sufficiently large
\(H^\frown M_j\) must belong to \(\mathcal U\). This contradicts the
choice of the \(M_j\).

Thus the claimed finite set \(F_H\) exists. Since this holds at every
\(H\in\mathcal U\), Lemma~\ref{lem:history-topology} shows that
\(\mathcal U\) is open in the finite-evidence history topology.
Therefore every \(\sigma\)-open set is open in the history topology,
so the latter is finer than \(\sigma\).
\end{proof}

In conclusion, the history topology is the finest topology that preserves the same
finite-evidence convergence after one passes from languages to finite refinement histories.

\subsubsection{Obstruction trees and topological closure in the history topology}

For a language \(A\in\X\), let \(\mathcal H_A\) be the set of
histories beginning at \(A\). Continuation never changes the first
language of a history, so \(\mathcal H_A\) is both open and closed in
\((\mathcal H,\tau)\).

When studying \(b_n(A)\) defined in \eqref{eq:bn}, we work in this component and evaluate the
relevant topological quantity at the one-language history \((A)\).
All closures and interiors below are taken in \(\mathcal H_A\), with
the subspace topology inherited from \((\mathcal H,\tau)\).

Let \(H\in\mathcal H_A\). If \(T\) is an admissible obstruction tree
whose root is labelled by \(\operatorname{end}(H)\), every
root-to-leaf path in \(T\) can be appended to \(H\), with the common
endpoint written only once.

The following lemma gives the basic connection between the history
topology and the admissible obstruction trees already used in the
definition of \(b_n(A)\).

\begin{lem}
\label{lem:obstruction-closure}
Let \(\mathcal D\subseteq\mathcal H_A\), and a chain
\(H\in\mathcal H_A\). Then \(H\) belongs to the closure of
\(\mathcal D\) if and only if there is an admissible obstruction tree
rooted at \(\operatorname{end}(H)\) such that every root-to-leaf path,
when appended to \(H\), produces a history in \(\mathcal D\).
\end{lem}

\begin{proof}
Suppose first that such an obstruction tree \(T\) exists, and let
\(\mathcal U\) be an open neighborhood of \(H\).

Starting at the root of \(T\), follow a path while keeping the
corresponding extended history inside \(\mathcal U\). Suppose that the
current history is \(\Gamma\in\mathcal U\), ending at \(L\). If the
current tree vertex is a leaf, stop. Otherwise,
Lemma~\ref{lem:history-topology} gives a finite set
\(F_\Gamma\subseteq L\) such that every proper refinement \(M\)
containing \(F_\Gamma\) has
\(\Gamma^\frown M\in\mathcal U\).

The child condition for the obstruction tree supplies a child labelled
by such an \(M\). Move to that child and continue. Since \(T\) has no
infinite path, this process eventually reaches a leaf. The resulting
history belongs both to \(\mathcal U\) and to \(\mathcal D\).

Thus every neighborhood of \(H\) meets \(\mathcal D\), so \(H\) lies
in the closure of \(\mathcal D\).

Conversely, let \(\mathcal C\) be the collection of histories for
which such an obstruction tree exists. Clearly
\(\mathcal D\subseteq\mathcal C\), since a history in \(\mathcal D\)
may use the single-vertex obstruction tree.

We show that \(\mathcal C\) is closed. Suppose that
\(H\notin\mathcal C\), and let \(L=\operatorname{end}(H)\).
We claim that there is a finite set \(F_H\subseteq L\) such that every
proper refinement \(M\) containing \(F_H\) satisfies
\(H^\frown M\notin\mathcal C\).
Otherwise, for every finite \(F\subseteq L\), choose a proper
refinement \(M_F\) containing \(F\) with
\(H^\frown M_F\in\mathcal C\). For each \(F\), choose an admissible
obstruction tree witnessing this membership. Form a new tree rooted
at \(L\), give its root one child labelled \(M_F\) for every finite
\(F\subseteq L\), and attach the corresponding tree below that child.

The resulting tree is countable and satisfies the finite-evidence
child condition. It has no infinite path, since after its first edge
a path remains in one of the attached well-founded trees. Every
root-to-leaf path, when appended to \(H\), produces a history in
\(\mathcal D\). This contradicts \(H\notin\mathcal C\).

The claimed finite set therefore exists. By
Lemma~\ref{lem:history-topology}, the complement of \(\mathcal C\) is
open. Hence \(\mathcal C\) is a closed set containing \(\mathcal D\),
so it contains the closure of \(\mathcal D\). The first part of the
proof gives the reverse inclusion.
\end{proof}

\subsubsection{Topological equivalence of the main characterization theorem}\label{subsec:topo1D}

We now apply the closure lemma to the pairing score.

\begin{cor}
\label{cor:safety-interior}
A history \(H\in\mathcal H_A\) is \((A,n,q)\)-unsafe if and only if
it belongs to the closure of
\(\{G\in\mathcal H_A:c_n(G)\le q\}\).

Equivalently, \(H\) is \((A,n,q)\)-safe if and only if it belongs to
the interior of
\(\{G\in\mathcal H_A:c_n(G)>q\}\).
\end{cor}

\begin{proof}
Apply Lemma~\ref{lem:obstruction-closure} with
\(\mathcal D=\{G\in\mathcal H_A:c_n(G)\le q\}\).
An admissible obstruction tree whose every terminal history belongs to
\(\mathcal D\) is exactly an obstruction tree witnessing that \(H\)
is unsafe. The second statement follows by taking complements.
\end{proof}

Thus safety is stronger than the inequality \(c_n(H)>q\). A safe
history has an open neighborhood consisting entirely of histories
whose scores exceed \(q\). A history may itself have high score but
still be unsafe if low-score histories occur in every one of its
neighborhoods.

The obstruction-tree definition of \(b_n(A)\) now has an equivalent
description in terms of open neighborhoods of \((A)\). Below is the topological characterization of \(b_n(A)\). 

\begin{thm}
\label{thm:topological-bn-restated}
For every \(A\in\X\) and every positive integer \(n\),
\[
 b_n(A)
 =
 \sup_{\substack{\mathcal U\subseteq\mathcal H_A\text{ open}\\
                 (A)\in\mathcal U}}
 \inf_{H\in\mathcal U}c_n(H).
\]
\end{thm}

\begin{proof}
For an admissible obstruction tree \(T\) rooted at \(A\), write
\(s(T)\) for the supremum of \(c_n(H)\) over its root-to-leaf
histories. Thus by definition \(b_n(A)=\inf_T s(T)\).

Let \(T\) be any admissible obstruction tree rooted at \(A\). The main point is that \textbf{for any open set $\mathcal{U}$ containing $(A)$, as in Lemma~\ref{lem:obstruction-closure}, we can always follow a path
through \(T\) while keeping the corresponding history inside
\(\mathcal U\).} Since \(T\) has no infinite path, this procedure
reaches a leaf. Thus \(T\) has some root-to-leaf history
\(H_T\in\mathcal U\).

Consequently,
\(s(T)\ge c_n(H_T)\ge \inf_{H\in\mathcal U}c_n(H)\).
Since this holds for every admissible obstruction tree \(T\) and any open set $\mathcal{U}$ containing $(A)$,
taking the infimum over \(T\) and supremum over $\mathcal{U}$ containing $(A)$, gives
\[
    b_n(A)\ge
    \sup_{\substack{\mathcal U\subseteq\mathcal H_A\text{ open}\\
                    (A)\in\mathcal U}}
    \inf_{H\in\mathcal U}c_n(H).
\]

For the reverse inequality, let \(q<b_n(A)\). We claim that the
initial history \((A)\) is \((A,n,q)\)-safe. Indeed, if it were
unsafe, there would be an admissible obstruction tree rooted at \(A\)
such that every root-to-leaf history \(H\) satisfies \(c_n(H)\le q\).
The supremum of the leaf scores of this tree would then be at most
\(q\). Since \(b_n(A)\) is the infimum of these suprema over all
admissible obstruction trees rooted at \(A\), this would imply
\(b_n(A)\le q\), contradicting \(q<b_n(A)\).

Thus \((A)\) is safe. By
Corollary~\ref{cor:safety-interior}, safety means that \((A)\) has an
open neighborhood \(\mathcal U\subseteq\mathcal H_A\) such that
\(c_n(H)>q\) for every \(H\in\mathcal U\). Therefore $\inf_{H\in\mathcal U}c_n(H)\ge q$.
Since this particular \(\mathcal U\) is one of the open neighborhoods
over which the supremum in
Theorem~\ref{thm:topological-bn} is taken, we obtain
\[
    \sup_{\substack{\mathcal U\subseteq\mathcal H_A\text{ open}\\
                    (A)\in\mathcal U}}
    \inf_{H\in\mathcal U}c_n(H)
    \ge q.
\]

This holds for every \(q<b_n(A)\). Thus we have proved 
the reverse inequality.
\end{proof}

The right-hand side is the best lower bound on \(c_n\) that can be
maintained throughout an open neighborhood of the initial history.
Equivalently, \(b_n(A)\) is the topological lower limit of the pairing
score at \((A)\).

An open neighborhood of \((A)\) specifies a coherent collection of
finite tests, one at every history that may arise. Its score infimum is
the worst score among all histories surviving those tests. Taking the
supremum over neighborhoods asks for the strongest uniform score that
can be protected using finite positive evidence.

Therefore, we can also write our main theorem as the following topological form. 
\begin{thm}[Topological characterization of the optimal guarantee]
\label{thm:topological-main}
For every nonempty countable class \(\X\) of infinite subsets of
\(\mathbb Z\), we have the following optimal lower Banach
density guarantee for generation from positive data.

More precisely:
\begin{enumerate}
\item \emph{Achievability.}
There is one generator such that, on every full enumeration of every
\(K\in\X\), every output is available, all but finitely many outputs
belong to \(K\), and its output set \(O\) satisfies
\[
    \lbanach(O,K)\ge \lim_{n\to\infty}
    \sup_{\substack{\mathcal U\subseteq\mathcal H_K\text{ open}\\
                    (K)\in\mathcal U}}
    \inf_{H\in\mathcal U}c_n(H).
\]

\item \emph{Optimality.}
Let \(\mathcal A\) be any generator that is eventually valid on every
full enumeration of every language in \(\X\). Then, for every
\(K\in\X\), there is an adversarial enumeration  of \(K\) on which the output set
\(O\) of \(\mathcal A\) satisfies
\[
    \lbanach(O,K)\le \lim_{n\to\infty}
    \sup_{\substack{\mathcal U\subseteq\mathcal H_K\text{ open}\\
                    (K)\in\mathcal U}}
    \inf_{H\in\mathcal U}c_n(H).
\]
\end{enumerate}
\end{thm}

\subsubsection{Protected sets}
The topological equivalence Theorem \ref{thm:topological-bn} immediately gives the following.
\begin{cor}[Protected sets]
\label{cor:protected-set}
Fix \(A\in\X\), \(n\ge1\), and \(q<b_n(A)\). Then:
\begin{enumerate}
\item the one-language history \((A)\) is \((A,n,q)\)-safe;
\item every \((A,n,q)\)-safe history \(H\) satisfies \(c_n(H)>q\);
\item if an \((A,n,q)\)-safe history \(H\) ends at \(M\), then there is a finite set \(\Prot(H)\subseteq M\) such that every \(L\in\X\) satisfying $\Prot(H)\subseteq L\subsetneq M$ makes \(H^\frown L\) \((A,n,q)\)-safe.
\end{enumerate}
\end{cor}

\begin{proof}
The first assertion follows from
Theorem~\ref{thm:topological-bn}: since \(q<b_n(A)\), there is an open
neighborhood of \((A)\) on which every history has score greater than
\(q\). Hence \((A)\) is \((A,n,q)\)-safe.

The second assertion follows from
Corollary~\ref{cor:safety-interior}, since every safe history lies in
\(\{H:c_n(H)>q\}\).

For the third assertion, the set of \((A,n,q)\)-safe histories is open by Corollary \ref{cor:safety-interior}.
Therefore, by Lemma~\ref{lem:history-topology}, if \(H\) is safe and
ends at \(M\), there is a finite set \(\Prot(H)\subseteq M\) such that
every proper refinement \(L\) containing \(\Prot(H)\) remains in the
open set of safe histories.
\end{proof}

\begin{remark}\label{rem:safety-local-score}
Safety gives a local score guarantee as follows: since
$c_n(H)>q$, every relative interval $J$ of the starting language
with $|J|\ge n$ and sufficiently far from the origin satisfies
$p(H,J)>q$. The required distance may depend on $H$, so this
conclusion does not give a common radius for all safe histories.
Also, $b_n(L)$ is a min--max over obstruction trees; it is not a
lower bound for the score of every arbitrary chain starting at $L$.
\end{remark}

\begin{defn}\label{def:history-protected-set}
Any finite set \(\Prot(H)\) satisfying Item~3 is called a
\emph{protected set with respect to the history \(H\)}.
\end{defn}

The protected set is not itself an open neighborhood. It is the finite
test at the current history that keeps every permitted next refinement
inside the open set of safe histories. At a later history, a different
protected set may be needed. This is why protected sets are indexed by
histories rather than only by their endpoint languages.

\subsubsection{How the topology guides the protected forest}

At a vertex \(v\) of the static protected forest, each ancestor \(a\)
determines a realized history \(H[a,v]\) and its own safety requirement
\((L_a,N_a,s_a)\), defined below. The forest invariant keeps this history
inside the open set of histories safe for that requirement whenever
\(s_a>0\).

The protected set \(\Prot_a(H[a,v])\) is the finite witness of openness
at this history. The total protected set \(P_v\) contains the union of
these sets over the finitely many ancestors of \(v\), together with
finite board traces specified below. Any proper refinement containing
\(P_v\) passes every ancestor's finite test and therefore preserves all
relevant safe open conditions simultaneously.

The history space contains every possible finite refinement history,
including different histories with the same endpoint. The static
protected forest does not attempt to realize all of them. It chooses
one family of histories and uses finite evidence to keep every realized
ancestor-to-descendant history inside the appropriate high-score
neighborhood.

This gives the topological duality behind the characterization. An
admissible obstruction tree certifies that every neighborhood meets
the low-score histories. A protected set certifies that the current
history lies in an open region avoiding them.

\begin{remark}[Relation to the earlier Cantor--Bendixson argument]\label{rem:history-topology-cb-rank}
The topology on histories is an unfolding of the finite-evidence
topology from \cite{kleinberg2025density,KWBanach}. In the earlier finite-rank
argument, finite evidence could be chosen at each endpoint so that
every proper refinement passing the test had smaller
Cantor--Bendixson rank.

Applying these tests successively gives a neighborhood of \((A)\) in
the history topology in which rank decreases at every strict
extension. If \(A\) has finite point rank \(r\), every history in this
neighborhood has at most \(r+1\) layers. Consequently,
\(c_n(H)\ge 1/2-(r+1)/(2n)\) throughout the neighborhood, and
Theorem~\ref{thm:topological-bn} gives \(b(A)=1/2\).

The present theory replaces bounded history length by a more general
condition: histories in the protected neighborhood may be arbitrarily
long, provided that their pairing scores remain uniformly large. Thus
long refinement histories are not themselves an obstruction; the
obstruction is the inability to isolate them topologically from
low-score histories.
\end{remark}

\subsection{Static Protected Forest}\label{sec:protected-forest}
We now assemble the history-dependent protected sets into one static
forest supporting all target languages simultaneously.  Along every
forest path, the construction must preserve each ancestor's safety
requirement; at the same time, compatible interval boards will provide
the local spatial structure needed by the online generator.

Fix \(\varepsilon>0\) for the proof of
Theorem~\ref{thm:fixed-epsilon}. Enumerate the distinct languages in
\(\X\) as \(L_1,L_2,\ldots\), stopping if \(\X\) is finite, and
identify each vertex with its index. Put
\begin{equation}
 q_i=(b(L_i)-\varepsilon)_+,
 \qquad s_i=(b(L_i)-\varepsilon/2)_+.
 \label{eq:qi}
\end{equation}
Choose a positive integer \(N_i\) such that
\begin{equation}
 \frac1{N_i}\le\frac\varepsilon2,
 \qquad s_i<b_{N_i}(L_i)\quad\text{if }s_i>0.
 \label{eq:Ni}
\end{equation}
These choices exist because \(b_n(L_i)\nearrow b(L_i)\). Thus
\begin{equation}
 \max\{0,s_i-1/N_i\}\ge q_i.
 \label{eq:board-slack}
\end{equation}
Safety and its protected sets will use the threshold \(s_i\).
When \(s_i=0\), no protection is needed for the vertex's own
requirement, although it may still carry protections for its ancestors.

We construct the forest, its board partitions, and its protected sets
in index order. Write \(a\preceq v\) when \(a\) is an ancestor of
\(v\), allowing equality, and let \(H[a,v]\) be the chain of labels
on that path.

\paragraph{Compatible interval boards.}
Each vertex \(v\) will have a partition of \(\mathbb Z\) into
consecutive intervals, called \(v\)-boards, each containing between
\(N_v\) and a finite number \(B_v\) of points of \(L_v\).
Descendant partitions coarsen ancestor partitions. The following
construction requires only the chosen parent's partition.

\begin{figure}
    \centering
    \includegraphics[width=1\linewidth]{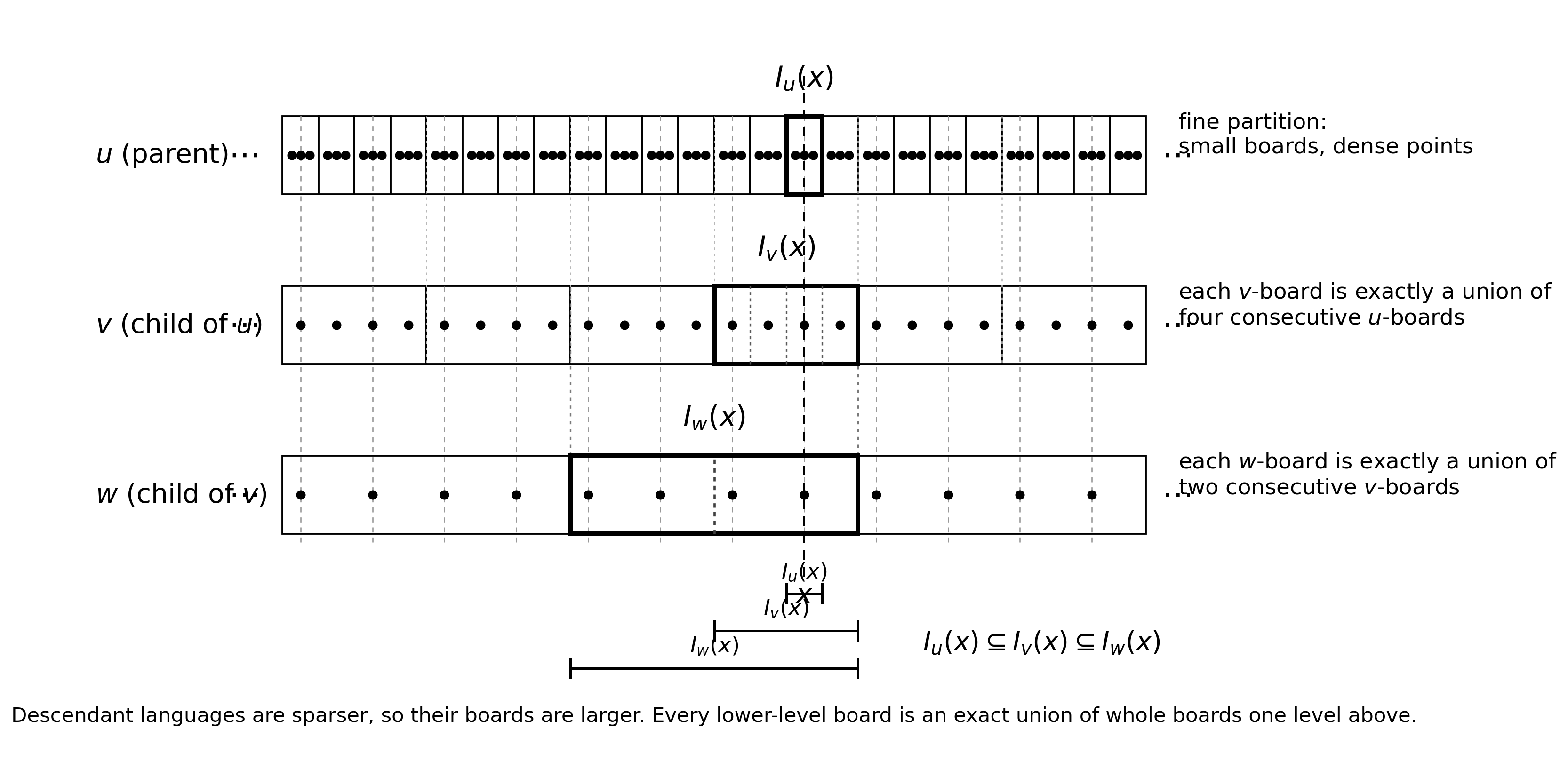}
 \caption{Compatible boards along a forest path \(u\prec v\prec w\).
The language labels satisfy \(L_w\subseteq L_v\subseteq L_u\), with
descendant languages becoming progressively sparser. Each \(v\)-board
is a union of consecutive \(u\)-boards, and each \(w\)-board is a union
of consecutive \(v\)-boards.}
    \label{fig:compatible}
\end{figure}

\begin{lem}
\label{lem:boards}
For every vertex \(v\), there is a partition of \(\mathbb Z\) into
consecutive intervals, called the \(v\)-boards, and a finite integer
\(B_v\), such that every \(v\)-board \(I\) satisfies
\(N_v\le |L_v\cap I|\le B_v\). Moreover, if \(u\) is the parent of
\(v\), then every \(v\)-board is a union of consecutive \(u\)-boards.

The constants may be chosen recursively as follows. If \(v\) is a
root, we may take \(B_v=N_v\). If \(v\) has parent \(u\), we may take
\(B_v=N_v+B_u-1\).
\end{lem}

\begin{proof}
We construct the partitions in the same order as the vertices of the
forest. Since every parent has smaller index than its children, the
partition at the parent is already available when we construct the
partition at a child.

Suppose first that \(v\) is a root. Order the points of \(L_v\) in the
usual order on \(\mathbb Z\), and divide them into consecutive blocks
of \(N_v\) points. If \(L_v\) is unbounded in both directions, choose
an arbitrary cut and form blocks of \(N_v\) points in both directions
from that cut. If \(L_v\) has a first or a last point, begin at that
end and form the blocks in the other direction.

Place the boundaries of the \(v\)-boards between successive blocks.
Any empty unbounded part of \(\mathbb Z\) at one end is included in
the adjacent board. Every resulting board contains exactly \(N_v\)
points of \(L_v\), so we may take \(B_v=N_v\). A board may be
unbounded as an interval of integers, but its intersection with
\(L_v\) is finite.

Now suppose that \(v\) has parent \(u\), and that the \(u\)-boards
have already been constructed. Assign to each \(u\)-board \(I\) the
weight \(|L_v\cap I|\). Since \(L_v\subseteq L_u\), every such weight
is at most \(B_u\). Since \(L_v\) is infinite and each \(u\)-board has
finite weight, there are infinitely many \(L_v\)-points distributed
among the \(u\)-boards.

We form each \(v\)-board by merging consecutive \(u\)-boards. Starting
from one end, continue adding \(u\)-boards until the total weight first
reaches \(N_v\), and then begin a new group. Immediately before the
last \(u\)-board is added, the accumulated weight is at most
\(N_v-1\); the last board contributes at most \(B_u\). Hence the total
weight of the resulting \(v\)-board lies between \(N_v\) and
\(N_v+B_u-1\).

To make this construction precise at the ends, consider the
\(u\)-boards having positive weight. If they have a first member, carry
out the grouping from left to right and include all preceding
zero-weight boards in the first group. If they have a last member,
carry out the symmetric construction from right to left. If they have
neither a first nor a last member, choose an arbitrary boundary between
two \(u\)-boards and group outward in both directions. The positive
weights then extend indefinitely in each direction, so every group
eventually reaches weight \(N_v\). The positive-weight boards cannot
have both a first and a last member, since that would make \(L_v\)
finite.

Assign every intervening zero-weight \(u\)-board to one of its
neighboring groups. The resulting groups partition \(\mathbb Z\), and
each group is a union of consecutive \(u\)-boards. They are therefore
the required \(v\)-boards. Taking \(B_v=N_v+B_u-1\) proves the claim.
\end{proof}

For \(x\in\mathbb Z\), let \(I_v(x)\) be its \(v\)-board. Thus
\(I_u(x)\subseteq I_v(x)\) whenever \(u\preceq v\).

\paragraph{The recursive construction.}
When processing \(v\), perform the following steps in order.
\begin{enumerate}
\item An earlier vertex is a possible parent when
\begin{equation}
 \mathsf{Par}(v)=\{u<v:L_v\subsetneq L_u, P_u\subseteq L_v\}.
 \label{eq:possible-parent}
\end{equation}
If this set is empty, make \(v\) a root. Otherwise scan its members
in increasing index order, retain the first, and replace the retained
\(p\) by \(u\) whenever \(L_u\subseteq L_p\). Attach \(v\) to the
final retained vertex.

\item Construct the \(v\)-boards by Lemma~\ref{lem:boards}.

\item For every \(a\preceq v\) with \(s_a>0\), choose an ordinary
protected set \(\Prot_a(H[a,v])\) for the parameters
\((L_a,N_a,s_a)\), and put
\[
 \mathsf{Bad}(a,v)=\{I:I\text{ is an }a\text{-board and }
                   p(H[a,v],L_a\cap I)<s_a\}.
\]
The purpose of the additional protection is to make the pairing
score uniformly good over all forest paths and boards. As verified
below, each chain $H[a,v]$ with $s_a>0$ is safe. Thus, for this
fixed chain, there is a radius $R=R(a,v)$ such that every $a$-board
$I$ whose trace $L_a\cap I$ avoids $[-R,R]$ has pairing score
greater than $s_a$. Every board in $\mathsf{Bad}(a,v)$ must
therefore contain an $L_a$-point in this finite interval, so there
are only finitely many such boards. However, the radius depends
on the chain, and the forest may contain infinitely many chains
with no common radius. The added points in $P_v$ prevent further
refinements from splitting the remaining bottom layer on any
board in $\mathsf{Bad}(a,v)$. This stops its score from decreasing
further; the uniform bound is proved below.

Define the total protected set by
\begin{equation}
 P_v=\bigcup_{\substack{a\preceq v\\s_a>0}}
 \left(\Prot_a(H[a,v])\ \cup\!
       \bigcup_{I\in\mathsf{Bad}(a,v)}(L_v\cap I)\right).
 \label{eq:total-protected-set}
\end{equation}
\end{enumerate}
The additional traces in \eqref{eq:total-protected-set} freeze a board
once its score falls below \(s_a\).

\begin{lem}
\label{lem:forest-invariant}
The construction is well defined. Every vertex has finitely many
ancestors and a finite set \(P_v\subseteq L_v\). For every
\(a\preceq v\) with \(s_a>0\), the chain \(H[a,v]\) is safe for
\((L_a,N_a,s_a)\).
\end{lem}
\begin{proof}
Parent indices decrease, so ancestry is finite. By
Corollary~\ref{cor:protected-set}, the new one-language chain is safe
for its positive threshold. If \(v\) has parent \(p\), then
\(\Prot_a(H[a,p])\subseteq P_p\subseteq L_v\), so the same corollary
preserves every old ancestor's safety. Hence all ordinary protected
sets in the construction exist.

For a fixed \(a\preceq v\) with \(s_a>0\), safety gives
\(c_{N_a}(H[a,v])>s_a\). Thus every \(a\)-board whose trace avoids
some finite interval \([-R,R]\) has score greater than \(s_a\).
Only finitely many board traces meet this interval, so
\(\mathsf{Bad}(a,v)\) is finite. Each added trace is finite, and
there are finitely many ancestors. This proves that \(P_v\) is finite.
\end{proof}

\begin{lem}[Uniform board bound]
\label{lem:uniform-board}
For every \(a\preceq v\) and every \(a\)-board \(I\),
\begin{equation}
 p(H[a,v],L_a\cap I)\ge\max\{0,s_a-1/N_a\}\ge q_a.
 \label{eq:uniform-board}
\end{equation}
\end{lem}
\begin{proof}
The case \(s_a=0\) is immediate. For the one-language chain, if
\(d=|L_a\cap I|\ge N_a\), then \(b(L_a)\le1/2\) gives
\[
 \frac{\lfloor d/2\rfloor}{d}
 \ge\frac12-\frac1{2N_a}
 \ge b(L_a)-\frac\varepsilon4
 > b(L_a)-\frac\varepsilon2=s_a.
\]
Appending a child \(v\) to a path ending at \(p\) splits only its
terminal layer. Since
\[
 \left\lfloor\frac{h+k}{2}\right\rfloor
 -\left\lfloor\frac h2\right\rfloor
 -\left\lfloor\frac k2\right\rfloor\in\{0,1\},
\]
the normalized score drops by at most \(1/|L_a\cap I|\le1/N_a\).
If the old score is at least \(s_a\), the new score is therefore at
least \(s_a-1/N_a\). If the old score is below \(s_a\), then
\[
 L_p\cap I\subseteq P_p\subseteq L_v\subseteq L_p,
\]
so the terminal trace does not change, and neither does the score.
Induction and \eqref{eq:board-slack} prove the claim.
\end{proof}

\subsection{The online algorithm}
\label{sec:algorithm}
With the protected forest and its compatible board partitions fixed, we
now define the online generator.  The current input history determines
a critical language guess, while each relevant board maintains a local
pullback obtained by lowest-common-ancestor updates.  These localized
records allow distant inputs to affect only their own boards.

Let \(S_t=\{x_1,\ldots,x_t\}\). A vertex \(v\) is consistent if
\(S_t\subseteq L_v\), and is critical if, in addition,
\(L_v\subseteq L_u\) for every earlier consistent \(u<v\).
Let \(Q_t\) be the largest-index critical vertex among the first
\(t\) vertices. If there is none, output any available integer and
leave all records unchanged.

Otherwise the active path runs from its component root to \(Q_t\).
For every active \(v\), update the record on \(I=I_v(x_t)\) by
\begin{equation}
 Z_{v,I}\leftarrow
 \begin{cases}
 Q_t,&\text{if the record is new},\\
 \LCA(Z_{v,I},Q_t),&\text{otherwise}.
 \end{cases}
 \label{eq:raw-lca-update}
\end{equation}
No other record changes. After all updates, scan the active path from
root to leaf and output an available point of the first pod
\(I_v(x_t)\cap L_{Z_{v,I_v(x_t)}}\) containing one. If every pod is
exhausted, output any available point of \(L_{Q_t}\). Such a point
exists because \(L_{Q_t}\) is infinite and only finitely many points
have been used. Fix tie-breaking rules for all choices.

Records belong to vertex--board pairs, even when two vertices use the
same interval. Precisely, after round \(t\), \(Z_{v,I}\) is the LCA
of the nonempty finite history
\[
 \mathcal H^t_{v,I}=\{Q_r:r\le t,\ Q_r\text{ defined},\
                     v\preceq Q_r,\ x_r\in I\}.
\]
An updated record lies on the path from \(v\) to the current \(Q_t\).
After initialization it moves only upward on the finite ancestor path
of its initial value. Thus its pod grows by inclusion, and contains
the input on each round on which that record is updated.

\begin{lem}[History inclusion]
\label{lem:memory}
Let \(u\preceq v\) be active on the current round. If the updated
record on \(I_u(x_t)\) lies in the subtree rooted at \(v\), then
\[
 L_{Z_{u,I_u(x_t)}}\subseteq L_{Z_{v,I_v(x_t)}}.
\]
\end{lem}
\begin{proof}
Every guess in the ancestor-board history descends from its LCA and
hence from \(v\). Therefore \(v\) was active on every contributing
round. Since \(I_u(x_t)\subseteq I_v(x_t)\), all those rounds also
contribute to the current \(v\)-board history. Consequently
\(\mathcal H^t_{u,I_u(x_t)}\subseteq\mathcal H^t_{v,I_v(x_t)}\).
The LCA of the larger history is an ancestor of the LCA of the
smaller history, giving the asserted language inclusion.
\end{proof}

\subsection{Validity guarantee}
\label{sec:validity}

Fix a full enumeration of \(K=L_z\).

\begin{lem}
\label{lem:capture}
There is a finite \(t_0\) such that, for every \(t\ge t_0\), the
guess \(Q_t\) is defined and lies in the subtree rooted at \(z\).
\end{lem}
\begin{proof}
Wait until \(t\ge z\), all of \(P_z\) has appeared, and every earlier
language not containing \(K\) has been contradicted. Thereafter \(z\)
is critical, so \(Q_t\ge z\) and
\begin{equation}
 S_t\subseteq L_{Q_t}\subseteq K.
 \label{eq:guess-below-K}
\end{equation}
For any consistent \(v>z\) with \(L_v\subseteq K\), vertex \(z\)
was a possible parent because \(P_z\subseteq S_t\subseteq L_v\).
Every possible parent before \(z\) contains \(S_t\), hence is
currently consistent and contains \(K\). The parent scan therefore
retains \(z\) when it reaches it. Subsequent replacements remain
inside \(K\), so the actual parent \(p\) satisfies
\(z\le p<v\) and \(S_t\subseteq L_v\subseteq L_p\subseteq K\).
Repeating along strictly decreasing parent indices reaches \(z\).
Apply this to \(v=Q_t\), with \(Q_t=z\) immediate.
\end{proof}

Call a \(z\)-board exceptional if its \(K\)-trace contains an input
or output from before \(t_0\), or if it contains a board at an
ancestor of \(z\), including \(z\), whose record was initialized
before \(t_0\). There are finitely many such boards: each early
round has finite active ancestry, and each ancestor board lies in one
\(z\)-board. Hence
\begin{equation}
 E_z=K\cap\bigcup_{I\text{ exceptional}}I
 \label{eq:exceptional-points}
\end{equation}
is finite. All other \(z\)-boards are regular.

On a round with input in a regular board, every history on a current
subboard at or above \(z\) contains only post-capture guesses.
Its LCA therefore descends from \(z\). Thus every pod examined by
the serving scan lies in \(K\), as does the fallback language.

\begin{lem}
\label{lem:validity}
The generator produces only finitely many outputs outside \(K\).
\end{lem}
\begin{proof}
After \(t_0\), an invalid output can occur only on a round with
input \(x\in E_z\), and only from a strict ancestor \(u\) of \(z\).
Every such output belongs to the finite set
\[
 \bigcup_{x\in E_z}\ \bigcup_{u\prec z}(L_u\cap I_u(x)).
\]
Each trace has at most \(B_u\) points, and outputs never repeat.
Together with the finitely many pre-capture rounds, this proves
 eventual validity.
\end{proof}

\subsection{Proof of the local-density guarantee}
\label{sec:density-properties}

We use the same local pairing mechanism as in \cite{KWBanach}, with
an exact response inside the current visible set.

\begin{lem}[Local pairing game]
\label{lem:pairing-game}
Let a finite board \(B\) be revealed through an increasing filtration
\(\emptyset=Y_0\subseteq Y_1\subseteq\cdots\subseteq Y_m=B\),
and put \(d_i=|Y_i\setminus Y_{i-1}|\). On each turn the adversary
selects a visible point, with repetitions allowed, and the algorithm
selects an unused visible point if one remains. Other events may
also select unused points of \(B\) for the algorithm, whether or not
visible. If every point of \(B\) is eventually selected by the
adversary, the final number of algorithm-selected points is at least
\(\sum_i\lfloor d_i/2\rfloor\).
\end{lem}
\begin{proof}
Call algorithm-selected points white, other adversary-selected points
black, and the rest unused. White points remain white after later
inputs. If \(w\) and \(r\) count visible white and unused points,
consider \(\Phi=w+\lfloor r/2\rfloor\). An input of an unused point
and the required response either replace two unused points by one
white point, or consume the last unused point. Neither decreases
\(\Phi\). Repeated inputs and additional white points cannot decrease
it either.

When visibility adds \(d\) points, of which \(h\) are already white,
all others are unused: any earlier input was already visible.
The increase in \(\Phi\) is
\[
 h+\left\lfloor\frac{r+d-h}{2}\right\rfloor
   -\left\lfloor\frac r2\right\rfloor
 \ge\left\lfloor\frac d2\right\rfloor.
\]
Initially \(\Phi=0\), and at completion \(r=0\). Summing these
increments proves the bound.
\end{proof}

\begin{cor}
\label{cor:regular-board-density}
Every regular \(z\)-board \(I\) satisfies
\( |O\cap K\cap I|\ge q_z|K\cap I|\).
\end{cor}
\begin{proof}
Write \(J=K\cap I\). At its first input round \(t\), the target
record is initialized at \(Q_t\). For \(H_0=H[z,Q_t]\),
Lemma~\ref{lem:uniform-board} gives
\[
 \sum_{D\text{ a layer of }H_0}
       \left\lfloor\frac{|D\cap J|}{2}\right\rfloor
 \ge q_z|J|.
\]
Later target records move upward on \(H_0\). Their pods reveal unions
of consecutive layers of \(H_0\), so visibility increments coarsen
its layer traces. Every point of \(J\) is visible on its own input
round. Completeness therefore makes the final visible set all of
\(J\), with any remaining layer traces empty.

Whenever an input in \(J\) leaves an unused point in the updated
target pod, the serving scan stops at \(z\) or an ancestor \(u\).
In the latter case, regularity puts the ancestor record below \(z\),
and Lemma~\ref{lem:memory} puts its output in the target pod; board
compatibility puts it in \(I\). Thus the local pairing response is
always made. Other outputs in \(J\) only help and are the external
selections allowed in Lemma~\ref{lem:pairing-game}.

Apply that lemma to the local input rounds. Since merging layers
cannot decrease their total rounded-half contribution,
\[
 |O\cap J|\ge\sum_i\left\lfloor\frac{d_i}{2}\right\rfloor
 \ge\sum_{D\text{ a layer of }H_0}
       \left\lfloor\frac{|D\cap J|}{2}\right\rfloor
 \ge q_z|J|.
\]
\end{proof}

\begin{proof}[Completion of the proof of
Theorem~\ref{thm:fixed-epsilon}]
Every output is available, and Lemma~\ref{lem:validity} proves
 eventual validity. Any finite relative interval \(A\) of \(K\)
meets at most two partial boundary \(z\)-boards, accounting for at
most \(2B_z\) points. Removing these points and \(E_z\) leaves whole
regular board traces. Hence Corollary~\ref{cor:regular-board-density}
gives
\[
 |O\cap A|\ge q_z\bigl(|A|-2B_z-|E_z|\bigr).
\]
The constants are independent of \(A\). Taking the lower
Banach-density limit proves
\(\lbanach(O,K)\ge q_z=(b(K)-\varepsilon)_+\).
\end{proof}

\subsection{From \(b(K)-\varepsilon\) to \(b(K)\)}
\label{sec:exact-bound}

We now pass from the fixed-error guarantee to the exact value \(b(K)\).
The only additional conceptual ingredient is an accuracy-copy
construction: for every language \(K\), we introduce formal copies
carrying requirements
$(b(K)-2^{-m})_+$  for  $m=1,2,\ldots$,
so that satisfying every copy forces the limiting guarantee \(b(K)\).

All of the main structural and algorithmic ideas are those of the
fixed-error proof: protected histories, the protected forest,
compatible boards, critical guesses, pullbacks, and the local density
argument. The technical work is to adapt these objects so that
multiple copies of the same underlying language, carrying increasingly
strong requirements, can coexist in one forest and be analyzed within
one online run.

The actual target remains a single language \(K\in\X\), and the
adversary still supplies one full enumeration of \(K\). The copies
introduced below are only formal vertices used by the construction;
they do not represent different possible targets.
\subsubsection{Formal copies and vertex-dependent requirements}

For every language \(L\in\X\) and every positive integer \(m\), create
a formal vertex \(a=(L,m)\).  Its language label is \(L_a:=L\), and
its density requirement is
\(q_a:=q_{L,m}:=(b(L)-2^{-m})_+\). Put
\(s_a=(b(L)-2^{-m-1})_+\). Choose \(N_a:=N_{L,m}\) such that
\(1/N_a\le2^{-m-1}\) and, if \(s_a>0\),
\(s_a<b_{N_a}(L_a)\). Thus \(\max\{0,s_a-1/N_a\}\ge q_a\).

Enumerate all formal vertices in one countable sequence.  We choose
the enumeration so that, for each fixed language \(L\), the vertex
\((L,m)\) appears before \((L,m+1)\).

The parameters \(q_a\), \(s_a\), and \(N_a\) are attached to the formal vertex
\(a\), not merely to the language \(L_a\).  Thus two formal vertices
with the same language label generally impose different density
requirements.

The safety definition is likewise indexed by the starting formal
vertex.  When we say that a chain is safe for \(a\), we mean that it is
safe with respect to the parameters \((L_a,N_a,s_a)\), using the
original language collection \(\X\) in the definition of continuation
trees.

\subsubsection{Forest paths with repeated language labels}

Distinct formal vertices may have the same language label.  Therefore
language labels along a forest path will be nonincreasing rather than
strictly decreasing.

Let \(a\) be an ancestor of \(v\).  Read the language labels along the
forest path from \(a\) to \(v\), and delete consecutive repetitions of
the same label.  Denote the resulting strictly descending language
chain by \(\overline H[a,v]\).

This deletion concerns only the language layers.  It does not identify
the formal vertices or their requirements.  The chain
\(\overline H[a,v]\) is tested using the parameters attached to its
starting vertex \(a\).  In particular, the same compressed language
chain may be safe for one copy of a language and unsafe for another
copy carrying a stronger requirement.

A protected set for this chain is therefore also indexed by the
starting formal vertex.  We write
\(\Prot_a(\overline H[a,v])\) for a protected set with respect to
\(\overline H[a,v]\), using the parameters \((L_a,N_a,s_a)\).

Equal-label edges create no new language layer.  Hence deleting a
repeated label does not change the pairing score.  Nevertheless, the
new formal vertex contributes a new requirement, and that requirement
must be protected separately.

\subsubsection{The protected forest with formal copies}

We repeat the protected-forest construction with nonstrict language
inclusion, again constructing each vertex's boards before its protected
set.

An earlier formal vertex \(u\) is a possible parent of \(v\) if
\(L_v\subseteq L_u\) and \(P_u\subseteq L_v\). Use the same parent
scan: retain the first possible parent and replace the retained \(p\)
by \(u\) whenever \(L_u\subseteq L_p\). The final retained vertex is
the parent; if there are no possible parents, make \(v\) a root.

We verify that the protected-forest invariant remains valid.  Suppose
that \(v\) is attached to a possible parent \(p\), and let \(a\) be an
ancestor of \(p\) with \(s_a>0\).

If \(L_v\subsetneq L_p\), then
\(\Prot_a(\overline H[a,p])\subseteq P_p\subseteq L_v\).  The defining
property of the protected set implies that appending \(L_v\) preserves
safety for the parameters \((L_a,N_a,s_a)\).

If \(L_v=L_p\), then deleting the repeated endpoint gives
\(\overline H[a,v]=\overline H[a,p]\).  The chain relevant to \(a\)'s
requirement has not changed, and hence remains safe.

The new vertex \(v\) also has its own requirement. Its one-language
chain \((L_v)\) is safe whenever \(s_v>0\), because
\(s_v<b_{N_v}(L_v)\). If \(s_v=0\), no protection is needed for its
own requirement.

Thus, for every ancestor \(a\preceq v\) with \(s_a>0\), the compressed
chain \(\overline H[a,v]\) is safe for \(a\)'s requirement. Choose its
ordinary protected set and define \(P_v\) by
\eqref{eq:total-protected-set}, using \(\overline H[a,v]\) and the
parameters of each ancestor copy. In particular, include \(L_v\cap I\)
for every \(a\)-board whose compressed-chain score is below \(s_a\).
Safety makes this a finite family of boards. Finite ancestry then
makes \(P_v\) finite, exactly as in Lemma~\ref{lem:forest-invariant}.

\begin{lem}
\label{lem:copies-chain}
Fix \(L\in\X\).  For every \(m>1\), the parent of the formal vertex
\((L,m)\) is \((L,m-1)\).  Consequently, all formal copies of \(L\)
lie on one ancestor--descendant chain in the protected forest.
\end{lem}

\begin{proof}
Let \(v=(L,m)\) and \(u=(L,m-1)\). Since \(P_u\subseteq L_u=L_v\),
vertex \(u\) is a possible parent. Every possible parent contains
\(L\), so the scan retains \(u\) when it reaches it. A later possible
parent could replace \(u\) only if its label were also \(L\), but
\(u\) is the most recent earlier copy of \(L\). Hence \(u\) is the
actual parent.
\end{proof}

\subsubsection{The critical chain with formal copies}

The online critical chain is also defined using nonstrict inclusion.

At time \(t\), a formal vertex \(v\) is critical if it is consistent
and \(L_v\subseteq L_u\) for every earlier consistent vertex \(u<v\).
The critical vertices form a weakly descending chain of language
labels.  Consecutive critical vertices may be different formal copies
of the same language.

The current guess \(Q_t\) is the final accessible member of this
critical chain, as before.  The proof only uses the fact that
\(L_{Q_t}\) is contained in every earlier consistent language.  It
does not require strict containment between consecutive critical
vertices.

When this critical chain is used to discuss language layers, repeated
labels may be deleted.  When it is used to choose the current formal
vertex \(Q_t\), the copies remain distinct.

\subsubsection{The actual target and its formal target vertices}

Fix an actual target language \(K\in\X\) and a full enumeration of \(K\).
For every positive integer \(m\), let \(z_m:=(K,m)\).  All these formal
vertices have the same language label \(L_{z_m}=K\), but they have
different requirements \(q_{K,m}\), different scales \(N_{K,m}\), and
generally different board partitions.

There is no single formal vertex that serves as the target vertex for
the entire proof.  Instead, we analyze the same run separately with
respect to each fixed \(z_m\).

For eventual validity, it is enough to use one fixed copy, for example
\(z_1\).  For the density conclusion, we fix an arbitrary \(m\),
obtain the bound \(q_{K,m}\), and then let \(m\) tend to infinity.

\begin{lem}
\label{lem:capture-fixed-copy}
Fix \(m\ge1\).  After some finite time, every current guess \(Q_t\) is
defined and lies in the subtree rooted at \(z_m=(K,m)\).
\end{lem}

\begin{proof}
Wait until every point of \(P_{z_m}\) has appeared in the input, the
time has passed the index of \(z_m\), and every earlier formal vertex
whose language does not contain \(K\) has become inconsistent.

Thereafter every earlier consistent language contains
\(K=L_{z_m}\), so \(z_m\) is a critical candidate.  Hence every later
current guess satisfies \(L_{Q_t}\subseteq K\).

Now let \(v>z_m\) be any consistent formal vertex with
\(L_v\subseteq K\).  Since
\(P_{z_m}\subseteq S_t\subseteq L_v\subseteq L_{z_m}\), the vertex
\(z_m\) is a possible parent of \(v\).

Every possible parent \(u<z_m\) of \(v\) contains \(L_v\), and hence
contains the current input set \(S_t\).  Thus \(u\) is consistent at
time \(t\), so its language contains \(K=L_{z_m}\).  Therefore
the parent scan retains \(z_m\) when it reaches that vertex.

Every later replacement remains inside \(K\). Thus the actual parent
\(p\) satisfies \(z_m\le p<v\) and
\(S_t\subseteq L_v\subseteq L_p\subseteq K\).

We may apply the same argument to \(p\).  Parent indices strictly
decrease but never fall below the index of \(z_m\), so the parent
chain eventually reaches \(z_m\).  Hence \(v\) is a descendant of
\(z_m\).  Applying this to \(v=Q_t\) proves the lemma.
\end{proof}

Strict decrease of language labels is not used in this argument.
Termination follows from strict decrease of the formal vertex indices.

\subsubsection{Boards for equal language labels}

The compatible-board construction remains valid with nonstrict
language inclusion.  If \(v\) has parent \(p\), we group consecutive
\(p\)-boards until their total number of \(L_v\)-points first reaches
\(N_v\).  The proof uses only \(L_v\subseteq L_p\).

When \(L_v=L_p\), the two formal vertices may still have different
values of \(N_v\) and \(N_p\).  Their board partitions may therefore
be different.  Every \(v\)-board is nevertheless a union of
consecutive \(p\)-boards, and it contains between \(N_v\) and a finite
upper bound \(B_v\) points of \(L_v\).

For every fixed formal vertex \(v\), the constant \(B_v\) is finite
because \(v\) has only finitely many ancestors.

\subsubsection{Uniform board bounds and LCA records}

The proof of Lemma~\ref{lem:uniform-board} applies with \(2^{-m}\)
in place of \(\varepsilon\) for the starting copy \(a=(L,m)\).
A strict-label extension loses at most one pair; an equal-label
extension changes no layer. Once a board's score is below \(s_a\),
its terminal trace is frozen by \(P_v\). Hence every \(a\)-board
\(I\) and every descendant \(v\) satisfy
\[
 p(\overline H[a,v],L_a\cap I)\ge\max\{0,s_a-1/N_a\}\ge q_a.
\]

Run the same update \eqref{eq:raw-lca-update} on formal vertices,
using the raw current guess \(Q_t\). Each record is the LCA of its
vertex--board history. Equal-label moves may leave a pod unchanged,
which is harmless. Lemma~\ref{lem:memory} uses only vertex ancestry
and board containment, so its history-inclusion proof is unchanged.
In the visibility filtration, omit repetitions of the same pod.

\subsubsection{Validity}

Apply Lemma~\ref{lem:capture-fixed-copy} with the fixed copy
\(z_1=(K,1)\).  After its capture time, every current guess lies in the
subtree rooted at \(z_1\), and hence has language contained in \(K\).

The validity proof from the fixed-error construction uses only forest
ancestry, language containment, finite ancestry, and the local
exceptional-board argument.  All of these remain valid with nonstrict
language inclusion.  Therefore the formal-copy algorithm is eventually
valid on the full enumeration of \(K\).

The choice of \(z_1\) is only for convenience.  Any fixed copy of \(K\)
would give the same eventual-validity conclusion.

\subsubsection{Density for one fixed target copy}

We now fix an arbitrary \(m\) and analyze the same run with respect to
\(z_m=(K,m)\).

The regular boards for \(z_m\) exclude only finitely many boards
containing pre-capture inputs, outputs, or initialized ancestor
records. Each ancestor copy retains its own boards and parameters.
On a regular board, every updated ancestor record lies below \(z_m\),
and history inclusion supplies the same local response as before.

The first chain on the board has score at least \(q_{K,m}\) by the
uniform board bound. Delete equal-label edges when forming its layers.
Subsequent LCA moves merge these layers, and equal-label moves add no
new visible points. Corollary~\ref{cor:regular-board-density} therefore
applies with this fixed copy and gives the following.

\begin{lem}
\label{lem:fixed-copy-density}
Fix \(m\ge1\).  The output set \(O\) of the formal-copy algorithm
satisfies
\(\lbanach(O,K)\ge q_{K,m}=(b(K)-2^{-m})_+\).
\end{lem}

\begin{proof}
Apply the eventual-capture, regular-board, history-inclusion,
and local-game arguments with the chosen target vertex \(z=z_m\), using
the parameters attached to \(z_m\).  Every regular \(z_m\)-board has
output density at least \(q_{K,m}\).

The corresponding board bound \(B_{z_m}\) and exceptional set are
finite.  They may depend on \(m\), but they are fixed once \(m\) is
fixed.  The passage from regular boards to arbitrary large relative
intervals therefore yields
\(\lbanach(O,K)\ge q_{K,m}\).
\end{proof}

No capture time, board bound, or exceptional set is required to be
uniform in \(m\).  For each fixed \(m\), the large-interval limit in
the definition of lower Banach density is taken first.

\subsubsection{Taking the limit over the copies}

The protected forest, board partitions, online rule, and output set
\(O\) are constructed once from the full countable family of formal
vertices.  They do not depend on the copy \(z_m\) chosen afterward for
the analysis.

Lemma~\ref{lem:fixed-copy-density} holds for every \(m\).  Hence
\(\lbanach(O,K)\ge (b(K)-2^{-m})_+\) for every positive integer \(m\).
Letting \(m\) tend to infinity gives
\(\lbanach(O,K)\ge b(K)\).

Thus equal-label copies are compressed only when language layers and
pairing scores are formed.  Their requirements, protected sets, board
partitions, capture times, and exceptional sets remain distinct
throughout the proof.

\begin{thm}\label{thm:oneD-lower-conclusion}
For every nonempty countable class \(\X\) of infinite subsets of
\(\mathbb Z\), there is one generator that is eventually valid on
every full enumeration and whose output set \(O\) satisfies
\(\lbanach(O,K)\ge b(K)\) for every target \(K\in\X\).
\end{thm}

\part{Higher Dimensions}
The higher-dimensional problem has a qualitatively different structure.
We first exhibit an explicit family in which the target-wise optimal
guarantees cannot be attained simultaneously.  Motivated by this
failure, we then characterize the full region of density profiles
achievable by one generator for filtered lower Banach density.  Finally,
we apply the characterization to finite Cantor--Bendixson point rank
and show that the exact \(1/2\) endpoint can still fail even for one
known target.
\section{A Gap Between Per-Language Optima and Uniform Guarantees Across a Language Family}
In one dimension, a single generator can attain the optimal density
simultaneously for every language in the collection. We show that this
can fail in two dimensions: the family admits strong per-language guarantees that cannot be achieved simultaneously.

Write \(\mathbb N=\{0,1,2,\ldots\}\).  Throughout this section, fix a
finite-valued nondecreasing threshold
$h:\mathbb N_{\ge1}\longrightarrow(0,\infty)$
satisfying 
$\sup_{N\ge1}\frac{N}{h(N)}=\infty$. We call such a threshold \(h\) \emph{admissible}.
We use the filtered density \(\delta_h\) from
Definition~\ref{def:filtered-banach-h}.

For a finite nonempty axis-parallel lattice rectangle \(R\), recall 
$V(R)=|R|$ and 
$m_L(R)=|L\cap R|$.
We call \(R\) a \emph{level-\(\lambda\) test for \(L\)} if
\[
m_L(R)\ge\lambda h(V(R)).
\]
Thus \(\delta_h(O,L)\) is the limiting worst output fraction over
level-\(\lambda\) tests as \(\lambda\to\infty\).

For a generator \(G\), let \(\beta_{G,h}(L)\) be the largest
\(\alpha\in[0,1]\) such that \(G\) guarantees
$\delta_h(O,L)\ge\alpha$
on every full enumeration of \(L\) if $L$ is the target.  Define
\[
\beta_h(L)=\sup_G\beta_{G,h}(L),
\]
where the supremum is over generators eventually valid on the entire
candidate family \(\X\).


\begin{thm}[Failure of simultaneous optimality in dimension \(d\ge2\)]
\label{thm:general}
For every \(d\ge2\), there exist a countable family
\(\X\) of infinite languages in \(\mathbb N^d\) and a constant
\(c=1/16\) such that, for every admissible threshold \(h\) and every
generator \(G\) eventually valid on \(\X\), there is a language
\(L\in\X\) satisfying
\[
\beta_{G,h}(L)\le\beta_h(L)-c.
\]
\end{thm}

\begin{thm}
\label{thm:main-not-simul}
For every $d\ge2$, there exist a single countable
family $\X$ of infinite languages in $\mathbb N^d$
and a distinguished language $L_0\in\X$
such that the following hold for every threshold
function $h$ satisfying the conditions above:

\begin{enumerate}
\item $\beta_h(L_0)=1/4$ and $\beta_h(L)=1/2$
for every $L\ne L_0$.

\item There is one generator eventually valid on all of \(\X\) that
guarantees
$\delta_h(O,L)\ge\frac12$
simultaneously for every \(L\in\X\setminus\{L_0\}\).

\item If one eventually valid generator guarantees
density $x$ at $L_0$ and density $y$ at every
other language, then necessarily
$x+y\le\frac12$.
\end{enumerate}

Consequently, no eventually valid generator
guarantees $\delta_h(O,L)\ge\beta_h(L)-\varepsilon$ at every
$L\in\X$ for any fixed $0<\varepsilon<1/8$.
\end{thm}
Importantly, the family $\X$ and distinguished
language $L_0$ are independent of $h$.
Thus the same language family witnesses the incompatibility for
ordinary lower Banach density, corresponding to \(h\equiv1\), and for
every admissible filtered threshold \(h\).

Theorem~\ref{thm:main-not-simul} implies
Theorem~\ref{thm:general} with $c=1/16$,
uniformly over all the thresholds $h$
under consideration.
Indeed, fix any such $h$.
If \(\beta_{G,h}(L)>\beta_h(L)-1/16\) for every \(L\), then $G$ guarantees density $3/16$ at
$L_0$ and density $7/16$ at every
$L\ne L_0$, contradicting $x+y\le1/2$.

\subsection{The idea: vertical coverage leaves horizontal gaps}\label{subsec:heuristic_not_simul_optimal}

A generator sees the points of an unknown set one at a time. After each
input, it must propose a point that has not yet been input. We want its
outputs to cover a large fraction of the target inside every sufficiently
populated rectangle.

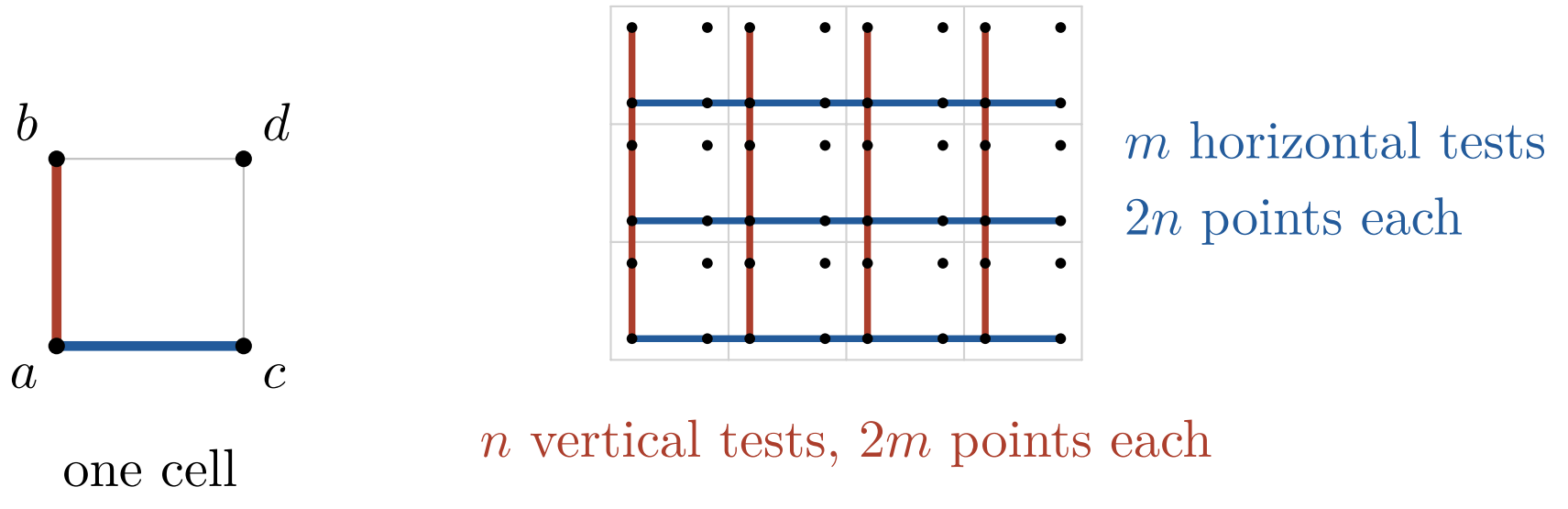
\begin{figure}[ht]
\centering
\begin{tikzpicture}[line cap=round]
  \begin{scope}[x=1.35cm,y=1.35cm]
    \draw[gray!50] (0,0) rectangle (1,1);
    \draw[hc,line width=2pt] (0,0)--(1,0);
    \draw[vc,line width=2pt] (0,0)--(0,1);
    \foreach \x/\y/\lab/\pos in
      {0/0/a/below left,0/1/b/above left,1/0/c/below right,1/1/d/above right}
      {\fill (\x,\y) circle (1.7pt);
       \node[\pos] at (\x,\y) {$\lab$};}
    \node at (.5,-.65) {one cell};
  \end{scope}
  \begin{scope}[shift={(4cm,-.1cm)},x=.85cm,y=.85cm]
    \draw[step=1,gray!35] (0,0) grid (4,3);
    \foreach \j in {0,1,2}
      \draw[hc,line width=1.4pt] (.18,\j+.18)--(3.82,\j+.18);
    \foreach \i in {0,1,2,3}
      \draw[vc,line width=1.4pt] (\i+.18,.18)--(\i+.18,2.82);
    \foreach \i in {0,1,2,3} \foreach \j in {0,1,2}
      \foreach \x/\y in {.18/.18,.18/.82,.82/.18,.82/.82}
        \fill (\i+\x,\j+\y) circle (1.1pt);
    \node[hc,right,align=left] at (4.2,1.5)
      {$m$ horizontal tests\\$2n$ points each};
    \node[vc,below,align=center] at (2,-.35)
      {$n$ vertical tests, $2m$ points each};
  \end{scope}
\end{tikzpicture}
\caption{Horizontal tests use $a,c$; vertical tests use $a,b$.
They intersect only at the $a$ points, which the adversary inputs.}
\label{fig:budget}
\end{figure}

\paragraph{One useful output per cell.}
Our construction uses four-point cells, labelled as in
Figure~\ref{fig:budget}. For now, assume outputs never repeat or use an
already input point; we justify this convention below. In an untouched
cell, input $a$ first. If the generator outputs $b$, immediately input
$c$; if it outputs $c$, immediately input $b$. If it outputs neither,
input $b$ and then $c$ unless $c$ has just been output. Thus $a$ is an
input, all three points $a,b,c$ are finished, and at most one of $b,c$
is an output.

The point $c$ helps a horizontal test, whereas $b$ helps a vertical test.
The fourth point $d$ helps neither.

The cell construction creates a competition between horizontal and
vertical coverage. An output at \(c\) contributes to a horizontal test,
whereas an output at \(b\) contributes to a vertical test; the point
\(d\) contributes to neither. The calculation below shows that one
generator cannot simultaneously attain the individual optimum in both
directions.

\paragraph{Vertical coverage forces a sparse horizontal window.}
Suppose the game is played in every cell of an \(n\times m\) block and
all \(a\)-points are adversarial inputs. There are \(m\) horizontal
tests, each containing \(2n\) points, and \(n\) vertical tests, each
containing \(2m\) points. Their intersections consist only of the
\(a\)-points, so no output is counted in both directions.

Let \(N_{\mathrm h}\) and \(N_{\mathrm v}\) be the total numbers of
outputs in the horizontal and vertical tests. Since each cell
contributes at most one useful output,
\[
N_{\mathrm h}+N_{\mathrm v}\le nm.
\]
By averaging, some horizontal test has output density at most
\(N_{\mathrm h}/(2nm)\), and some vertical test has output density at
most \(N_{\mathrm v}/(2nm)\). Hence these two densities sum to at most
\(1/2\).

\paragraph{Heuristic for the individual density guarantees.}
If the generator only needs to perform well in the horizontal direction, it can always respond to an input \(a\) by outputting \(c\). On a target consisting of infinitely long horizontal rows of cells, this horizontal pairing can achieve density \(1/2\). Similarly, vertical pairing can achieve density \(1/2\) on a target consisting of infinitely tall columns of cells.

Thus, for each orientation separately, density \(1/2\) is achievable. The calculation above suggests the source of a possible conflict: horizontal and vertical coverage compete for the same one useful output per cell. However, this heuristic alone does not yet prove that the two guarantees are incompatible for one generator; in fact, as we explain below, a generator may sometimes learn which orientation it is facing.

There is also a technical obstacle. The cell argument assumes that the adversary reaches a cell before the generator outputs anything in it. To enforce this, our language family includes targets that omit selected future cells. Eventual validity will then prevent the generator from entering those cells too early. This is the role of the block-forcing argument below.

\begin{remark}\label{sec:orientation}
Based on the heuristics above, it is tempting to use only horizontal strips and vertical strips as the set of languages to show the separation. 

Consider a candidate collection whose languages are horizontal and
vertical capped strips, including their allowed whole-cell deletion
variants but not the full grid. Preserve the vertical axis in vertical
targets and the horizontal axis in horizontal targets. A single generator
can achieve density $1/2$ on every language in this collection.

Let $X_t,Y_t$ be the largest first and second coordinates among the
inputs seen by time $t$. Use vertical pairing when $Y_t>X_t$ and
horizontal pairing otherwise. Output the input's partner within its
cell if that point has been neither input nor output; otherwise output
a fresh point on the axis of the chosen orientation.

On a vertical target, $X_t$ is bounded while $Y_t\to\infty$, so the
generator eventually always chooses vertical pairing. Likewise, on a
horizontal target it eventually always chooses horizontal pairing.
Outputs are therefore eventually valid. The pairing argument gives
density $1/2$: only finitely many pairs are affected by early choices,
and every rectangle cuts only a bounded number of pairs.

The lesson is that compatibility with both orientations at every finite
history does not prevent an orientation guess from eventually becoming
permanently correct. Eventual validity does not require the generator
to know when its guess has become correct.
\end{remark}

\paragraph{Where the conflict occurs.}
The actual conflict is between horizontal windows in the full grid \(K\) and vertical windows in a capped vertical strip. A fixed strip by itself is not enough: the horizontal
windows it produces have bounded length, so they do not occur at arbitrarily high filter levels and hence do not affect the filtered lower Banach density.

The idea is therefore to use wider and wider vertical strips inside \(K\). For a fixed strip, the adversary keeps trying taller blocks until a sparse horizontal window appears. If this never happens, the corresponding vertical tests would contradict the strip's density guarantee. Once such a horizontal window is found, the adversary moves to a wider strip and repeats. Since every strip is contained in \(K\), all these inputs still form part of one enumeration of \(K\), while the resulting horizontal windows become longer and occur at arbitrarily high levels.

This is how the vertical density requirement on the strips forces sparse horizontal windows in the full grid.

\paragraph{The remaining obstacle.}
The generator might output in a cell before we visit it. We prevent this
by including possible targets that omit selected cell positions in every
sufficiently distant block. Eventual validity lets us protect unvisited
cells. This is the role of the block-forcing lemma. The density promises
will be used only to show that the strip experiments terminate.

\subsection{Construction of the language family} \label{sec:family}
We now turn the preceding 4-point cell heuristic into a single countable
language family.  The full grid will provide the target on which we
force sparse horizontal windows, while capped vertical strips provide
the competing vertical requirements.  Periodic cell deletions supply
the auxiliary targets needed to keep sufficiently distant blocks
untouched until the adversary chooses to visit them.

Divide the nonnegative lattice into disjoint 4-point \emph{cells} above, each containing
four points $a,b,c,d$ arranged in two rows and two columns. The cell $C_{i,j}$
has lower-left point $(2i,2j)$, and thus 
\[ C_{i,j}=(2i,2j)+\{0,1\}^2\quad(i,j\in\mathbb{N})
\]
and the full grid $K = \mathbb{N} \times \mathbb{N}$ contains every cell:
\[
K=\bigcup_{i,j\in\mathbb{N}}C_{i,j}=\mathbb{N}^2.
\] Thus $i$ specifies the cell-column and $j$ the cell-row. Within each cell,
label the lower-left, upper-left, lower-right, and upper-right points
$a,b,c,d$, respectively as before.

For $1\le k\le\ell$, start with the first $k+1$ cell-columns, extending
upward forever. Then add a finite square of $(\ell+1)\times(\ell+1)$
cells at the lower-left corner. The resulting \emph{capped strip} is
\[
 V_{k,\ell}=
 \underbrace{\bigl(\{0,\ldots,2k+1\}\times\mathbb{N}\bigr)}
             _{\text{infinite vertical strip}}
 \ \cup\
 \underbrace{\{0,\ldots,2\ell+1\}^2}
             _{\text{finite square cap}}.
\]
See the left figure in Figure \ref{fig:family}. 
Above the cap, only the first $k+1$ cell-columns remain. The cap allows
the strip to contain any prescribed finite subset of $K$ by choosing
$\ell$ large enough, without changing the width of its infinite tail.

The language family is built from these capped strips and the full grid \(\mathbb{N}\times\mathbb{N}\) by deleting whole four-point cells in simple arithmetic patterns. We never delete cell-column \(0\), which serves as a common safe region. In any other cell-column \(1\le i\le k\), we may choose a common difference \(m\ge1\) and a starting position \(r\), and delete cells
\[
C_{i,r},\ C_{i,r+m},\ C_{i,r+2m},\ldots,
\]
or equivalently, \(C_{i,tm+r}\) for every \(t\ge0\). Thus, starting from position \(r\), we delete every \(m\)-th cell in column \(i\).

The family $\X $consists of the full grid $K$, together with all capped strips obtained by applying finitely many such deletion rules. In particular, a single cell-column may contain several deleted arithmetic progressions, possibly with different starting positions and common differences. Applying no deletion rules gives the original capped strip.

\begin{figure}[ht]
\centering
\begin{tikzpicture}[x=.39cm,y=.39cm]
  \fill[capcol!30] (0,0) rectangle (6,6);
  \fill[orange!20] (0,0) rectangle (3,8);
  \fill[safe!35] (0,0) rectangle (1,8);
  \draw[step=1,gray!50,thin] (0,0) grid (6,8);
  \draw[thick] (0,8)--(0,0)--(6,0)--(6,6)--(3,6)--(3,8);
  \foreach \i in {0,1,2}
    \draw[-{Latex[length=2mm]}] (\i+.5,7.7)--(\i+.5,8.7);
  \node[rotate=90,font=\small] at (.5,4) {never deleted};
  \node[font=\small,align=center] at (4.5,3) {finite\\cap};
  \node[below,align=center] at (3,-.4)
    {a capped strip\\(squares represent cells)};
  \begin{scope}[shift={(12,0)}]
    \fill[safe!35] (0,0) rectangle (1,9);
    \draw[step=1,gray!50,thin] (0,0) grid (4,9);
    \foreach \t in {0,1,2} {
      \draw[thick] (1,3*\t) rectangle (4,3*\t+3);
      \node[right,font=\small] at (4.2,3*\t+1.5) {block $\t$};
    }
    \foreach \t in {1,2} {
      \fill[vc!15] (2,3*\t+1) rectangle (3,3*\t+2);
      \draw[vc,thick] (2.2,3*\t+1.2)--(2.8,3*\t+1.8);
      \draw[vc,thick] (2.2,3*\t+1.8)--(2.8,3*\t+1.2);
    }
    \foreach \i in {0,1,2,3}
      \draw[-{Latex[length=2mm]}] (\i+.5,8.8)--(\i+.5,9.7);
    \node[below,align=center] at (2,-.4)
      {one deletion rule\\($m=3$, starting row 4)};
  \end{scope}
\end{tikzpicture}
\caption{The cap accommodates any finite input history. Deletions repeat
indefinitely; deleting only finitely many cells would allow a generator
to violate that target finitely often without losing eventual validity.}
\label{fig:family}
\end{figure}

So  $\X$ is countable. Every member is a union of whole cells and contains the zero-th cell column. This column provides an unlimited supply of ``safe'' fallback outputs. Note that each cell is either completely in a language or completely gone from a language. 

Every member other than $K$ has bounded width. For any fixed $k$, choosing
$\ell$ large enough makes $V_{k,\ell}$ contain any prescribed finite
subset of $K$.

\subsection{Two per-language density guarantees}\label{sec:pairing}

We construct two generators. The first guarantees density \(1/2\)
simultaneously on every \(L\neq K\); the second guarantees density
\(1/4\) on \(K\). Both generators remain eventually valid on every
language in \(\X\). 
The key point is that the generator attaining the optimum at \(K\) need not be the same generator that attains the optima at all targets \(L\ne K\).; there need not be a single generator that achieves all of these guarantees simultaneously. Indeed, in the next subsection, we show that the generators attaining the different individual optima cannot all be the same generator.

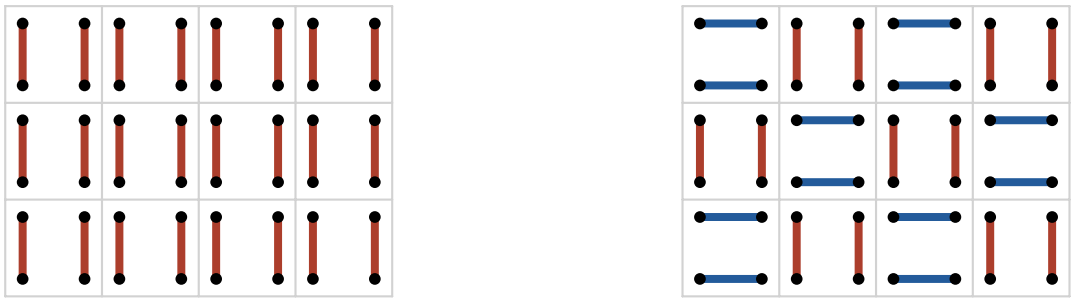
\begin{figure}[ht]
\centering
\begin{tikzpicture}[x=.64cm,y=.64cm,line cap=round]
  \foreach \s in {0,7} {
    \begin{scope}[shift={(\s,0)}]
      \draw[step=1,gray!35] (0,0) grid (4,3);
      \foreach \i in {0,1,2,3} \foreach \j in {0,1,2} {
        \pgfmathtruncatemacro{\horizontal}{(\s>0)*mod(\i+\j+1,2)}
        \ifnum\horizontal=1
          \foreach \y in {.18,.82}
            \draw[hc,line width=1.5pt] (\i+.18,\j+\y)--(\i+.82,\j+\y);
        \else
          \foreach \x in {.18,.82}
            \draw[vc,line width=1.5pt] (\i+\x,\j+.18)--(\i+\x,\j+.82);
        \fi
        \foreach \x/\y in {.18/.18,.18/.82,.82/.18,.82/.82}
          \fill (\i+\x,\j+\y) circle (1.1pt);
      }
    \end{scope}
  }
  \node[below,align=center] at (2,-.25) {vertical pairing\\for bounded-width targets};
  \node[below,align=center] at (9,-.25) {checkerboard pairing\\for the full grid};
\end{tikzpicture}
\caption{Every complete pair supplies an output. Alternating the pairing
direction protects both horizontal and vertical thin rectangles.}
\label{fig:pairing}
\end{figure}

The two constructions use one of the two pairing strategies shown in Figure~\ref{fig:pairing}. Fix one of these strategies. Whenever a point
is input, output its paired partner if that partner has not previously
appeared. Otherwise, output an unused point of the zero-th cell column. In either strategy, the four points of each cell are partitioned into two pairs.

This output is always valid. An input from a cell certifies that the
whole cell belongs to the target, while the zero-th cell column belongs to every language.
Moreover, every present pair eventually contains an output: when its
first point is input, either its partner has already been output or the
generator outputs it immediately.

We first show that the generator using vertical pairs guarantees lower
Banach density at least $1/2$ if the target language is not the full grid. 
Indeed, such a target language $L$ has bounded number of cell-columns.  An axis-parallel rectangle $R$ cuts at most two pairs in each column, at its top and
bottom. So 
\[
                  |{O\cap L\cap R}|\ge\frac{m_L(R)}2-\text{width of } L.
\]
Thus this single generator guarantees density $1/2$ at every $L\ne K$.

We now show that the checkerboard pairing guarantees density at least
\(1/4\) when the target is the full grid \(K\).
Let \(\mathcal M\) denote the matching defined by the checkerboard
pairing in Figure~\ref{fig:pairing}. Every edge of \(\mathcal M\)
eventually contains an output.
Fix an axis-parallel rectangle \(R\), and consider \(K\cap R\), which
is again a lattice rectangle when nonempty. Partition it according to
the \(2\times2\) cells. A complete cell contains two edges of
\(\mathcal M\). Every noncomplete cell intersects \(K\cap R\) in either
one point or in two adjacent points. The two-point intersections form
at most four consecutive runs, one along each side of the rectangle.
Along each such run the checkerboard orientation alternates, so among
any \(m\) consecutive two-point intersections at least
\(\lfloor m/2\rfloor\) contain an edge of \(\mathcal M\).
There are at most four one-point corner intersections.

It follows that the number of matching edges contained entirely in
\(K\cap R\) is at least
$\frac{|K\cap R|}{4}-3$.
Since every such matching edge contains an output,
$|O\cap K\cap R|
\ge
\frac{|K\cap R|}{4}-3$.
The additive constant is negligible along the density filter, so the
generator guarantees density at least \(1/4\) on \(K\).

\subsection{Upper bounds for a single generator}

We now show that no single generator can simultaneously achieve the optimal density on \(K\) and on all languages \(L\neq K\).

The heuristic from Subsection~\ref{subsec:heuristic_not_simul_optimal} is essentially the whole proof. The main
technical point is to force a large block in which the adversary is
the first to touch every cell, and does so at the point \(a\).

We say that a rectangle \(R\) is \emph{finished for \(L\)} when every point of \(L\cap R\) has appeared as either an input or an output.
Because an output can never repeat a point that has already appeared,
the number of outputs in a finished rectangle can never increase.

The crucial observation is the following.

\begin{lem}\label{lem:validity-commitment}
Fix a generator that is eventually valid on \(L\). Suppose the adversary has already supplied a finite sequence of distinct inputs \(x_1,\ldots,x_t\in L\). Then the adversary can choose finitely many additional distinct inputs \(x_{t+1},\ldots,x_s\in L\) as an immediate continuation such that the following holds: for every continuation \(x_{s+1},x_{s+2},\ldots\) consisting of distinct points of \(L\) not among \(x_1,\ldots,x_s\), every generator output produced after \(x_s\) belongs to \(L\).
\end{lem}
Here the generator is fixed throughout, so once the adversary's input sequence is specified, the corresponding output sequence is completely determined.
\begin{proof}
Suppose not. Then, after every finite extension by the adversary inputs from \(L\), some further finite continuation of the adversary input inside \(L\) leads to a generator output outside \(L\).

Fix an enumeration of \(L\). Repeatedly input the first point in this
enumeration that has not yet appeared as an input, and then choose a finite
continuation inside \(L\) that produces another output outside \(L\).
The resulting inputs form a full enumeration of \(L\), while the
generator produces infinitely many outputs outside \(L\). This
contradicts eventual validity.
\end{proof}

We use this lemma to protect future cells: temporarily keep the input
history inside a possible target that omits them. To visit those cells
later, we will leave the smaller target but remain inside the base strip.

\begin{lem}\label{lem:block}
Fix a generator that is eventually valid on \(\X\), a capped strip \(V_{k,\ell}\), and any finite sequence of distinct adversary inputs contained in \(V_{k,\ell}\). Then, for every \(1\le n\le k\) and every \(m\ge1\), the adversary can supply finitely many additional inputs from \(V_{k,\ell}\) so as to produce an \(n\times m\) block of cells in cell-columns \(1,\ldots,n\) with the following properties:
\begin{enumerate}
\item in every cell, the adversary is the first to touch the cell, by inputting its point \(a\);
\item every \(a,b,c\) point in the block is eventually either an input or an output;
\item each cell contains at most one output among \(a,b,c\).
\end{enumerate}
\end{lem}

\begin{proof}
Our goal is to find a single \(n\times m\) block in which every cell is first touched by the adversary, at its point \(a\). The idea is to construct nested possible target languages
\[
J_{nm}\subset J_{nm-1}\subset\cdots\subset J_1\subset J_0=V_{k,\ell}
\]
such that, when restricted to one sufficiently distant \(n\times m\) block, each step going down the chain removes exactly one  cell. We then visit these cells in the reverse order.

For \(t\ge0\), let \(B_t\) be the \(n\times m\) block in cell-columns
\(1,\ldots,n\) and cell-rows \(tm,\ldots,tm+m-1\). Thus 
\[
B_t=\bigcup_{\substack{1\le i\le n\\0\le j<m}} C_{i,tm+j}.
\]

 Label the \(nm\) relative positions in such a block by
$e_1,\ldots,e_{nm}$ in an arbitrary order. 
Thus if \(e_p\) has column \(i_p\) and row-position \(j_p\in\{0,\ldots,m-1\}\), then its copy in \(B_t\) is \(C_{i_p,tm+j_p}\).

We now construct \(J_1,\ldots,J_{nm}\). Suppose \(J_{p-1}\) has already been constructed and the adversary has supplied only finitely many inputs. Choose \(T_p\) sufficiently large that none of the inputs used so far lies in any block \(B_t\) with \(t\ge T_p\). Define \(J_p\) from \(J_{p-1}\) by deleting the copy of \(e_p\) from every block \(B_t\) with \(t\ge T_p\); that is, delete cells
$C_{i_p,tm+j_p}$ for every $t\ge T_p$.
Then \(J_p\in\X\), and all adversary inputs supplied so far still belong to \(J_p\). By Lemma~\ref{lem:validity-commitment}, the adversary can supply finitely many additional inputs from \(J_p\) until the generator becomes committed to \(J_p\).

Repeat this for \(p=1,\ldots,nm\).

After all \(nm\) commitments have been made, only finitely many points have been touched. We may therefore choose $T\ge \max\{T_1,\ldots,T_{nm}\}$ so large that the entire block \(B_T\) has never been touched by either the adversary or the generator.

This is the single block on which all the languages \(J_p\) line up as desired. Indeed, for every \(p\),
\[
J_p\cap B_T
=
B_T\setminus\{e_1,\ldots,e_p\},
\]
where we identify \(e_i\) with its copy in \(B_T\). Thus, restricted to \(B_T\), moving from \(J_p\) to \(J_{p-1}\) adds exactly the one cell \(e_p\).

Now the adversary visits the cells of \(B_T\) in reverse order, $e_{nm},e_{nm-1},\ldots,e_1$.
We claim that \(e_q\) is still completely untouched when its turn arrives. After the generator was committed to \(J_q\), all inputs used to construct the later commitments \(J_{q+1},\ldots,J_{nm}\) belonged to smaller languages and hence also to \(J_q\). Moreover, the cells already visited in \(B_T\) are \(e_{nm},\ldots,e_{q+1}\), all of which belong to \(J_q\). Therefore every adversary input since the commitment to \(J_q\), up until \(e_q\) is visited, has belonged to \(J_q\). The validity commitment by Lemma \ref{lem:validity-commitment} then forces every corresponding generator output to lie in \(J_q\). But \(e_q\) is absent from \(J_q\), so the generator cannot have touched \(e_q\).

Thus, when we reach \(e_q\), the adversary is the first to touch that cell and can begin by inputting its point \(a\). We then use the three-point strategy from Subsection~\ref{subsec:heuristic_not_simul_optimal}: finish \(a,b,c\) while allowing at most one of them to be an output.

Repeating this for \(q=nm,nm-1,\ldots,1\) gives the desired \(n\times m\) block.
\end{proof}

Now we can finish the proof of Theorem \ref{thm:main-not-simul}. The remaining part is essentially the heuristic in Subsection \ref{subsec:heuristic_not_simul_optimal}. 
\begin{lem}\label{prop:tradeoff}
Suppose a generator is eventually valid on \(\X\), guarantees density \(x\) at \(L_0=K\), and density \(y\) at every \(L\neq L_0\). Then
$x+y\le \frac12$.
\end{lem}

\begin{proof}
Fix \(v<y\). We will construct a full enumeration of \(L_0\) whose density generated by the generator is at most \(1/2-v\).

Consider an \(n\times m\) block given by Lemma~\ref{lem:block}. In each cell-row, take the \(2n\) points rectangle window consisting of the $(a,c)$-points of its $n$ cells; we call this a \emph{horizontal test}. Similarly, in each cell-column, take the $2m$ points consisting of the $(a,b)$-points of its $m$ cells; we call this a \emph{vertical test}.

We first prove a claim. 
\begin{claim}Fix \(n\) and a capped strip \(V_{n,\ell}\), and suppose the adversary has already supplied finitely many inputs from \(V_{n,\ell}\). Then, after finitely many additional inputs from \(V_{n,\ell}\), the adversary can produce a finished horizontal test of length \(2n\) and density at most \(1/2-v\).
\end{claim}
\begin{proof}
Suppose not. Repeatedly apply Lemma~\ref{lem:block} to \(n\times m\) blocks, choosing $m$ larger and larger. In one such block, let \(N_h\) and \(N_v\) be the total numbers of outputs in the horizontal and vertical tests.
By Lemma~\ref{lem:block}, each cell contains at most one output among \(a,b,c\), and the horizontal and vertical tests intersect only at the \(a\)-points, which are adversary inputs. Hence
$N_h+N_v\le nm$.
Since, by assumption, every horizontal test has density greater than \(1/2-v\), then we have 
$N_h>2nm\left(\frac12-v\right)$,
and therefore $N_v<2nmv$.
There are \(n\) vertical tests, each containing \(2m\) points, so at least one of them has density less than \(v\).

Now repeat this construction with \(m\) chosen increasingly large so that these vertical tests have arbitrarily high levels. Between successive applications of Lemma~\ref{lem:block}, input the first point in a fixed enumeration of \(V_{n,\ell}\)
that has not yet appeared as an input. If the desired horizontal test never appeared, this would produce a full enumeration of \(V_{n,\ell}\) with finished vertical tests of density below \(v\) at arbitrarily high levels. This contradicts the assumed density-\(y\) guarantee on \(V_{n,\ell}\), since \(v<y\). The claim follows.
\end{proof}

We now use the claim repeatedly to enumerate \(L_0\). Fix an enumeration of \(L_0\). For each \(j=1,2,\ldots\), first input its next point that has not yet been used as an adversary input. Choose \(n_j\) large enough that $2n_j\ge jh(2n_j)$, and then choose \(\ell_j\) large enough that all adversary inputs  so far lie in \(V_{n_j,\ell_j}\).

By the claim, after finitely many additional inputs inside \(V_{n_j,\ell_j}\), we obtain a finished horizontal test \(R_j\) of length \(2n_j\) and density at most \(1/2-v\). Since \(V_{n_j,\ell_j}\subseteq L_0\), all these inputs also belong to \(L_0\). Moreover,
\[
|R_j|=|L_0\cap R_j|=2n_j\ge jh(2n_j),
\]
so \(R_j\) is a level-\(j\) test for \(L_0\).

Repeating this for every \(j\) gives a full enumeration of \(L_0\) containing finished tests of density at most \(1/2-v\) at arbitrarily high levels. Hence $\delta_h(O,L_0)\le \frac12-v$.
Since this holds for every \(v<y\), letting \(v\uparrow y\) gives 
$x+y\le\frac12$, as desired.
\end{proof}

\subsection{The individual upper bounds}
The trade-off argument shows that these guarantees cannot all be
achieved by one generator.  To complete the determination of the
individual optima in Theorem~\ref{thm:main-not-simul}, it remains to
show that no generator can exceed \(1/2\) on the bounded-width targets
or \(1/4\) on the full grid.
\paragraph{No target has optimum above $1/2$.}
Fix $L\in\X$. At stage $j$, after the listing input, choose $N\ge jh(N)$
and an untouched $N$-point segment of the common line \(\{0\}\times\mathbb N\). Keep inputting
segment points that have been neither input nor output until the segment
is finished. If it has $a$ inputs and $w$ outputs, then $a+w=N$ and
$w\le a$, since there is only one output per round. Its output fraction
is therefore at most $1/2$.

The segment has volume and target population both equal to $N$, so it
is a level-$j$ test. These permanently sparse tests show that the resulting
full enumeration has density at most $1/2$. Together with vertical
pairing, this proves $\beta_h(L)=1/2$ for every $L\ne K$.

\paragraph{The full-grid optimum is at most $1/4$.}
At stage $j$ of an enumeration of $K$, after the listing input, choose
$n$ with $2n\ge jh(2n)$. Choose $\ell\ge n$ so that the current input
history lies in $V_{n,\ell}$, and force an $n\times n$ block there by
Lemma~\ref{lem:block}.

The block has $n$ horizontal tests and $n$ vertical tests, each with
$2n$ points. No output is counted twice, and together the tests contain
at most $n^2$ outputs. Their average output fraction is at most
$\frac{n^2}{(2n)(2n)}=\frac14$.
Some test therefore has fraction at most $1/4$. It is finished and is a
level-$j$ test for $K$. Repeating gives a full enumeration of $K$ with
density at most $1/4$. Together with checkerboard pairing, this proves
$\beta_h(K)=\frac14$.

For $d\ge2$, embed each language as $L\times\{0\}^{d-2}$.
The upper-bound tests have singleton extra sides and unchanged volume.
Every higher-dimensional rectangle meeting the embedded language projects
to a planar rectangle with the same target points and no larger volume; monotonicity of $h$
then transfers the lower bounds as well.

\section{Characterization theorem for the generation frontier}
The preceding section shows that separate target-wise optima do not
describe the higher-dimensional problem: the best guarantees at
different targets may be incompatible.  We therefore turn to the full
region of density profiles that one generator can achieve
simultaneously.

The refinement structure remains the same as in one dimension, but the
local quantitative certificate changes.  The scalar pairing score is
replaced by a finite geometric stopping game that tests whether one
causal strategy can satisfy all relevant rectangle requirements along a
refinement history.

Throughout this section, fix \(d\ge2\) and a nonempty countable family \(\collec\) of infinite subsets of \(\mathbb Z^d\). We use the model
conventions from Part~I.

A rectangle is a nonempty finite axis-parallel integer box; singleton
coordinate intervals are allowed.  Let 
$V(R)=|R|$ and $m_K(R)=|K\cap R|$.

For the profile characterization, fix a threshold of the form
\[
h(V)=u(V)\log(2+V),
\]
where \(u:[1,\infty)\to (0,\infty)\) is nondecreasing.  Equivalently,
\(h(V)/\log(2+V)\) is nondecreasing.  We use the filtered density
\(\delta_h\) from Definition~\ref{def:filtered-banach-h}. 

A simultaneous target profile is a function
\[
\alpha:\collec\longrightarrow[0,1].
\]
If \(K\) is \(h\)-vacuous, then
$\delta_h(O,K)=1$
for every output set \(O\), so the requirement
\(\delta_h(O,K)\ge\alpha(K)\) is automatic.  Thus vacuous targets impose
no restriction on the achievable-profile problem.  

The main proof
first treats an arbitrary fixed quota profile $q:\collec\to[0,1]$.  For a
prescribed profile $\alpha$ and $\varepsilon>0$, we use
$q_\varepsilon(K)=(\alpha(K)-\varepsilon)_+$.  Thus the fixed-error argument
has one requirement per language and no accuracy-copy indices.

\begin{thm}
\label{thm:fixed-quota}
\begin{enumerate}
\item Fix a quota profile \(q:\collec\to[0,1]\). If every $K\in\collec$ has a $q$-protected tree in the sense of
Definition~\ref{def:protected-tree}, then one globally fresh, universally
eventually valid generator satisfies $\delta_h(O,K)\ge q(K)$ on every full
enumeration of every $K\in\collec$.

\item Conversely, let \(\alpha\) be an achievable profile.
Then, for every \(\varepsilon>0\), every \(K\) has a
\(q_\varepsilon\)-protected tree.
\end{enumerate}
\end{thm}

The two directions of this theorem correspond
to lower and upper bounds.
The first direction is constructive:
protected trees, equivalently decorated open
neighborhoods, yield a generator satisfying
the prescribed density requirements.
The second direction gives the matching
obstruction by contraposition.
If, for some $K$ and $\varepsilon>0$,
no $q_\varepsilon$-protected tree exists,
then no eventually valid generator can
achieve the profile $\alpha$.
By Corollary~\ref{cor:hd-obstruction},
the failure of a protected tree is equivalent
to the following adversarial condition:
for every consistent geometric assignment
$\Theta$, there is an admissible obstruction
tree rooted at $K$ whose every terminal
history fails the corresponding geometric
stopping games.

The slack $\varepsilon$ arises because a limiting
density guarantee need not be witnessed exactly at
any finite stage. To obtain exact attainment, we
coordinate all accuracy requirements within one
protected construction; separate generators
achieving every fixed-error approximation would
not suffice.

\begin{thm}[Exact simultaneous profile characterization]
\label{thm:main}
A profile $\alpha$ is achieved by one globally fresh, universally eventually
valid generator if and only if every accuracy copy $(K,k)$ has a protected
tree on copies in the sense of Definition~\ref{def:copy-protected-tree}.
\end{thm}
Although a protected tree may itself be infinite, every local condition
in the structural criterion is finite: it is tested on a finite
refinement history, finite positive evidence, and a finite geometric
stopping game. In particular, the criterion contains no quantification
over infinite input enumerations or online generators. Thus the theorem
replaces the original infinite online interaction by static refinement
certificates whose quantitative conditions are all finite.

The trade-off theorem above gives an explicit nontrivial slice of this
profile frontier, while the finite-rank theorem below identifies a broad
regime in which the frontier contains the simultaneous near-half
profile.

Two consequences explain the geometric content of the characterization.
\begin{thm}[Finite rank and the endpoint]\label{thm:overview}
\begin{enumerate}
\item Let $\tau >0$ and define $h_\tau(V)=\log^{1+\tau}(2+V)$.
 For every $0<\varepsilon<1/2$, there is an eventually valid generator
that guarantees
$\delta_{h_\tau}(O,K)\ge1/2-\varepsilon$ at {\it every} target $K$ of finite
point rank in the finite-evidence topology on $\collec$ from Definition \ref{def:CBrank}.  The
finite ranks need not have a common bound.
\item Let 
$h:\mathbb N_{\ge1}\to(0,\infty)$
be nondecreasing and satisfy
$\sup_{n\ge1}\frac{n}{h(n)}=\infty$. No eventually valid generator can guarantee
$\delta_h(O,\mathbb Z^d)\ge1/2$ on every full enumeration of $\mathbb Z^d$.
This already fails for the rank-zero singleton class $\{\mathbb Z^d\}$.
\end{enumerate}
\end{thm}

\subsection{Proof architecture and the one-dimensional analogy}
\label{sec:architecture-higher-dim}

The proof follows the order of the one-dimensional lower-Banach-density
argument.  We first fix the quotas.  We then define the local quantitative
condition, identify safe chains and protected sets, build one static protected
forest, define the online generator, and prove validity and density.  Only
then do we introduce accuracy copies to reach the exact profile.

The new issue is geometric.  In one dimension a refinement chain is
summarized by a numerical layer-pairing score.  For rectangles, many
possibly overlapping tests must be served by the same responses, and the
adversary may stop after revealing any language on the chain.  The numerical
score is therefore replaced by a finite stopping game.

\begin{center}
\begin{tabularx}{\textwidth}{@{}p{.26\textwidth}XX@{}}
\toprule
 & One-dimensional proof & Present proof in $d\ge2$ \\
\midrule
Unknown-target structure
& Finite descending chains
& The same finite descending chains \\
Local quantitative object
& Rounded-half chain score $c_n(H)$
& Feasibility of a finite stopping game \\
Finite evidence
& Protected set for a safe chain
& Protected set for a safe chain \\
Static object
& Protected forest
& Protected forest \\
Local spatial state
& Board, pullback, and pod
& Board, pullback, and pod \\
Exact endpoint
& Accuracy copies introduced only at the end
& Accuracy copies introduced only at the end \\
\bottomrule
\end{tabularx}
\end{center}

Thus we retain the one-dimensional terminology whenever the logical role is
the same.  The finite stopping game is the only new structural term in the
main proof.

\subsection{The finite stopping game}\label{sec:game}

Fix a quota profile $q$.  Let
$H=(L_0\supsetneq L_1\supsetneq\cdots\supsetneq L_r)$ be a finite chain. 

A \emph{finite rectangle packet for \(H\)} is a tuple
\(\mathbf R=(\mathcal R_0,\ldots,\mathcal R_r)\), where each
\(\mathcal R_i\) is a finite, possibly empty, family of rectangles
assigned to the language \(L_i\). Its finite support is
\[
U=\bigcup_{i=0}^r\ \bigcup_{R\in\mathcal R_i}(L_i\cap R),
\]
the union of all strings $L_i\cap R$ with $R\in\cR_i$.

Initially the level is \(r\), and the visible set is \(U\cap L_r\).

At level \(i\), Black may make either of the following moves:
\begin{enumerate}
\item claim an unused visible point, after which White may claim at most
one unused visible point;
\item if \(i>0\), move to level \(i-1\), thereby enlarging the visible
set to \(U\cap L_{i-1}\). This move gives White no immediate response.
\end{enumerate}

Black may stop at level \(i\) only after every point currently visible
at that level has been claimed. At level \(0\), once the visible set is
exhausted, the game ends.

At a stop at level $i$, White wins when every $R\in\cR_i$ satisfies
\[|W\cap L_i\cap R|\ge q(L_i)|L_i\cap R|.\]  
The packet is \emph{feasible} if White has a single strategy such that, no matter how Black plays, whenever Black stops at a level \(i\) after all points currently visible at that level have been claimed, White satisfies all the rectangle requirements assigned to \(L_i\).

\begin{remark}
  This finite stopping game is the abstract version of the local density game used in the finite-rank proof in \cite{KWBanach}. There, finite Cantor--Bendixson rank bounded the number of times the visible language could change, so the local problem involved only a finite chain of visible sets. In higher dimensions, a size-sensitive discrepancy theorem proved in \cite{KWBanach} provides a single successful White strategy. 

Here we do not have the finite rank condition. For an arbitrary finite refinement chain, the stopping game simply asks whether such a common White strategy exists. Finite rank will enter only later, when we verify this condition using the discrepancy theorem.  
\end{remark}

\begin{defn}
    For \(a\ge0\), let \(Q_a=[-a,a]^d\cap\mathbb Z^d\). Given cutoffs
\(s_0,\ldots,s_r\ge1\), we say that \(H\) is
\emph{geometrically \(q\)-feasible at these cutoffs outside \(Q_a\)}
if every finite rectangle packet
\(\mathbf R=(\mathcal R_0,\ldots,\mathcal R_r)\) is feasible whenever
each \(R\in\mathcal R_i\) is disjoint from \(Q_a\) and satisfies
\(m_{L_i}(R)\ge s_i h(V(R))\).
\end{defn}

\paragraph{Comparison with the one-dimensional score.}
\begin{center}
\begin{tabularx}{\textwidth}{@{}p{.25\textwidth}XX@{}}
\toprule
 & One dimension & Higher dimensions \\
\midrule
Chain
& $L_0\supsetneq\cdots\supsetneq L_r$
& The same \\
Density test windows
& Relative intervals
& Finite packets of rectangles \\
Local quantity
& The numerical score $c_n(H)$
& The value of the finite stopping game \\
Threshold condition
& score $c_n(H)>q$
& One strategy meets all quotas at every stop \\
Reason for the difference
& Order makes layer pairing additive
& Overlapping rectangles may require incompatible replies \\
\bottomrule
\end{tabularx}
\end{center}

\begin{remark}
The one-dimensional case is more explicit: White has a canonical
pairing strategy. Consider an interval
\(J\) and a descending chain
\(L_0\supsetneq L_1\supsetneq\cdots\supsetneq L_r\). The chain partitions
\(L_0\cap J\) into layers: the innermost layer \(L_r\cap J\), followed
by the differences \((L_{r-1}\setminus L_r)\cap J,\ldots,
(L_0\setminus L_1)\cap J\).
White uses a very simple strategy: after each Black move, if an unused
visible point remains, White claims any such point. Applied to the game on \(J\), the aggregate invariant from
Lemma~\ref{lem:pairing-game} shows that, if Black stops when
\(L_i\) is visible, White has at least
$\sum_{j=i}^{r}\left\lfloor\frac{|E_j\cap J|}{2}\right\rfloor$
points in \(L_i\cap J\), where \(E_i,\ldots,E_r\) are the visible
layers. Thus, if
$H_i=(L_i\supsetneq L_{i+1}\supsetneq\cdots\supsetneq L_r)$,
the guaranteed density at that stop is
$\frac{
\sum_{j=i}^{r}
\left\lfloor |E_j\cap J|/2\right\rfloor}
{|L_i\cap J|}
=
p(H_i,L_i\cap J)$.
Taking the worst-case value over the relevant relative
intervals gives the numerical chain score.

In higher dimensions, there is no comparably simple formula. Different
rectangles can overlap and compete for the same White responses, so
having enough points for each rectangle separately does not imply that
one strategy can satisfy them all. We therefore have to use this abstract finite stopping game to test whether the required quotas are jointly feasible.
\end{remark}

The essential compatibility requirement is that White must use one
strategy before Black decides how far to enlarge visibility and at
which language to stop. Strategies that work separately for individual
rectangles or stopping levels need not be mutually compatible, because
earlier White choices cannot be revised when a larger language becomes
visible. This is the main challenge in higher dimensions. In one dimension, the numerical pairing score encodes this
compatibility automatically.

\subsection{Sufficiency of Theorem~\ref{thm:fixed-quota}: from protected trees to a generator}
\label{sec:topology}

We first prove the sufficiency direction of
Theorem~\ref{thm:fixed-quota}.
Assuming that every language has a $q$-protected
tree, we construct a single eventually valid
generator achieving the prescribed density
guarantees.

We begin by relating protected trees to the
finite-evidence history topology. We then build
a static protected forest and use the finite
geometric stopping games to define an online
generator. Finally, we verify its validity
and density guarantees.

For a finite strictly descending chain $H$, recall $\EndLang(H)$ is its final language, and again write
$H^\frown M$ for the chain obtained by appending a proper refinement
$M\subsetneq\EndLang(H)$. 

We use the same finite-evidence history topology as defined in Subsection \ref{sec:history-topology}. In short, a finite-test rule assigns to every chain $H$ a finite set
$\mathfrak f(H)\subseteq\EndLang(H)$.  Starting from $H$, retain every
extension $\Gamma^\frown M$ whenever $\Gamma$ has already been retained and
$\mathfrak f(\Gamma)\subseteq M\subsetneq\EndLang(\Gamma)$.  The collections
obtained in this way form a basis for the continuation topology on $\Hist$.

\subsubsection{Protected trees and their topological formulation}
\label{subsec:protected-trees}

Fix a quota profile
\[
q:\mathcal X\longrightarrow[0,1].
\]
Thus \(q(L)\) is the density requirement imposed when the true language
is \(L\). For example, for a desired profile \(\alpha\) and a fixed
\(\varepsilon>0\), we take
\[
q(L)=(\alpha(L)-\varepsilon)_+.
\]

The finite-evidence structure is the same as in the one-dimensional
proof. The difference is only in the local quantitative condition. In
one dimension, a chain has an explicit numerical score. Here, a chain
must pass the finite stopping game for every eligible rectangle packet.

\paragraph{Combinatorial formulation.}

\begin{defn}[\(q\)-protected tree]
\label{def:protected-tree}
Let \(K\in\mathcal X\). A \emph{\(q\)-protected tree rooted at \(K\)}
is a countable rooted tree whose vertices carry the following data:
\begin{itemize}
\item a language label \(L_v\in\mathcal X\);
\item a finite set \(P_v\subseteq L_v\), called its
      \emph{protected set};
\item a cutoff \(s_v\ge1\); and
\item a radius \(a_v\ge0\).
\end{itemize}

The root is labelled \(K\). If \(w\) is a child of \(v\), then
\(L_w\subsetneq L_v\) and \(a_w>a_v\). We require:
\begin{enumerate}
\item for every vertex \(v\) and every \(M\in\mathcal X\) satisfying
\[
P_v\subseteq M\subsetneq L_v,
\]
the vertex \(v\) has a child labelled \(M\);

\item for every finite root-to-vertex path
\(v_0,\ldots,v_r\), the chain
\[
L_{v_0}\supsetneq L_{v_1}\supsetneq\cdots\supsetneq L_{v_r}
\]
is geometrically \(q\)-feasible at the cutoffs
\(s_{v_0},\ldots,s_{v_r}\) outside $[-a_{v_r},a_{v_r}]^d\cap\mathbb Z^d$
\end{enumerate}
\end{defn}

The first condition gives the protected set the same meaning as in one
dimension. Once the finite evidence \(P_v\) has appeared, every proper
refinement that still contains this evidence must remain represented.
The tree is not allowed to retain only a convenient refinement.

A vertex may be a leaf. This happens when its protected set isolates its
language, meaning that there is no \(M\in\mathcal X\) with
\(P_v\subseteq M\subsetneq L_v\). If compatible proper refinements
remain, then the tree must continue through all of them.

The same language may label several vertices on different branches.
These are different occurrences, reached through different refinement
histories, and they may carry different protected sets, cutoffs, and
radii. Strict containment prevents the same language from occurring
twice on one branch.

For a given quota profile \(q\) and root language \(K\), such a tree
need not exist. One tries to construct it recursively by choosing the
protected set, cutoff, and radius at each vertex and then continuing
through every compatible proper refinement. The construction fails if
these choices cannot preserve geometric \(q\)-feasibility along every
finite path.

\paragraph{Comparison with the one-dimensional construction.}

\begin{center}
\begin{tabularx}{\textwidth}{@{}p{.22\textwidth}XX@{}}
\toprule
& One-dimensional construction
& Higher-dimensional construction \\
\midrule

Refinement histories
& Finite descending chains of languages
& Same \\

Finite positive evidence
& A protected set in the current language
& Same \\

Role of the protected set
& Every compatible proper refinement must admit a protected continuation
& Same \\

Terminal case
& A protected set may isolate the current language
& Same \\

History topology
& The finite-evidence history topology
& Same \\

Local quantitative condition
& The explicit rounded-half chain score exceeds the quota
& Every eligible rectangle packet is feasible in the finite stopping game \\

White strategy
& Claim any unused visible point; the resulting guarantee follows
  directly from the layer sizes
& One causal strategy must satisfy all overlapping rectangle
  requirements and all possible stopping levels \\

Additional geometric data
& A minimum interval population; the distant-region conclusion follows
  from the chain score
& A cutoff at every occurrence and an increasing excluded core \\

\bottomrule
\end{tabularx}
\end{center}

In one dimension, the local condition can be computed explicitly.
Consider the game restricted to a single interval \(J\), and suppose
that it stops at \(L_i\). Let \(w\) be the total number of White
points and let \(u\) be the number of currently visible unused
points in \(J\). Consider
$\Phi=w+\left\lfloor\frac{u}{2}\right\rfloor$.
Each Black claim followed by a greedy White response leaves
\(\Phi\) unchanged, including the case in which Black claims the
last visible unused point. Revealing \(s\) additional points
increases \(\Phi\) by at least \(\lfloor s/2\rfloor\).
Initially,
\(\Phi=\lfloor |E_r\cap J|/2\rfloor\), and at a legal stop
\(u=0\). Hence White obtains at least
\[
\sum_{j=i}^{r}
\left\lfloor\frac{|E_j\cap J|}{2}\right\rfloor
\]
points in \(L_i\cap J\). After division by \(|L_i\cap J|\),
this gives the one-dimensional chain score.

There is no analogous scalar formula in higher dimensions. Different
rectangles can overlap and compete for the same White responses. Even
if each rectangle can be served well separately, the corresponding
strategies may be incompatible. The finite stopping game therefore
tests directly whether one causal strategy satisfies all requirements
at every possible stopping language.

\paragraph{Topological formulation.}

We use the finite-evidence history topology introduced in the
one-dimensional part; see Definition~\ref{opensets}. This topology does
not depend on the dimension.

For \(K\in\mathcal X\), let \(\mathcal H_K\) be the space of all finite
strictly descending chains beginning at \(K\). If
\[
H=(K=L_0\supsetneq L_1\supsetneq\cdots\supsetneq L_r),
\]
write \(H_i=(K=L_0\supsetneq\cdots\supsetneq L_i)\) for its \(i\)-th
prefix.

\begin{defn}[\(q\)-good geometric decoration]
\label{def:q-good-decoration}
Let \(\mathcal U\subseteq\mathcal H_K\) be an open neighborhood of
\((K)\) that is closed under taking prefixes. A \emph{geometric decoration} of \(\mathcal U\) assigns to
each history \(H\in\mathcal U\) a cutoff \(s_H\ge1\) and a radius
\(a_H\ge0\), such that
$a_{H^\frown M}>a_H$
whenever \(H^\frown M\in\mathcal U\).

The decoration is \emph{\(q\)-good} if, for every
\[
H=(K=L_0\supsetneq\cdots\supsetneq L_r)\in\mathcal U,
\]
the chain \(H\) is geometrically \(q\)-feasible at the cutoffs
$s_{H_0},s_{H_1},\ldots,s_{H_r}$
outside  $Q_{a_H}=[-a_H,a_H]^d\cap\mathbb Z^d$.
\end{defn}

The cutoffs assigned to the prefixes are used because they are the
cutoffs carried by the successive vertices on the corresponding tree
path. The radius belongs to the final history and increases whenever the
history is extended.

\begin{thm}[Topological characterization]
\label{thm:topological-protected-tree}
A \(q\)-protected tree rooted at \(K\) exists if and only if
\((K)\) has an open neighborhood in \(\mathcal H_K\) that admits a
\(q\)-good geometric decoration.
\end{thm}

\begin{proof}
The proof is the same unfolding argument as in the one-dimensional
case. The only change is that the explicit chain-score condition is
replaced by geometric feasibility of the stopping games.

Suppose first that a \(q\)-protected tree rooted at \(K\) exists.
If a vertex has several children with the same language label, retain
one such child together with its entire descendant subtree. This
pruning preserves all the defining conditions.

Let \(\mathcal U\) be the collection of language histories appearing
on root-to-vertex paths in the pruned tree. A history ending at a
vertex \(v\) has the protected set \(P_v\). For every proper refinement
\(M\) containing \(P_v\), the tree has a child labelled \(M\), and hence
the extended history \(H^\frown M\) also belongs to \(\mathcal U\).
By the local characterization of the finite-evidence history topology,
\(\mathcal U\) is an open neighborhood of \((K)\).

Assign to a history \(H\in\mathcal U\) the cutoff and radius carried by
its final vertex. Radii increase along extensions. The path-feasibility
condition in Definition~\ref{def:protected-tree} says exactly that this
decoration is \(q\)-good.

Conversely, suppose that an open neighborhood
\(\mathcal U\) of \((K)\) has a \(q\)-good geometric decoration. Choose
a basic open neighborhood
\(\mathcal V\) of \((K)\) with
\(\mathcal V\subseteq\mathcal U\). Recall that the construction of a
basic neighborhood chooses, at every history \(H\in\mathcal V\), a
finite set \(P_H\subseteq\operatorname{end}(H)\), and includes every
one-step refinement \(H^\frown M\) satisfying
$P_H\subseteq M\subsetneq\operatorname{end}(H)$.

Make one tree vertex for every history in \(\mathcal V\). The root is
the one-language history \((K)\), and the parent of \(H^\frown M\) is
\(H\). Label the vertex corresponding to \(H\) by
\(\operatorname{end}(H)\), give it protected set \(P_H\), and assign it
the cutoff \(s_H\) and radius \(a_H\) from the decoration.

The definition of a basic neighborhood gives a child for every
compatible proper refinement. The decoration gives increasing radii,
and its \(q\)-goodness gives geometric feasibility along every finite
root-to-vertex path. Since \(\mathcal X\) is countable, the set of its
finite refinement histories is countable. The resulting tree is
therefore a \(q\)-protected tree rooted at \(K\).
\end{proof}

\begin{cor}[Admissible obstruction-tree characterization]
\label{cor:hd-obstruction}
Fix a quota profile $q$ and a language $K$.

A \emph{global geometric assignment} $\Theta$
specifies a cutoff $s_H\ge1$ and a radius
$a_H\ge0$ for every finite history
$H\in\mathcal H_K$, with
$a_{H^\frown M}>a_H$
for every one-step extension.

For a history
$H=(L_0\supsetneq\cdots\supsetneq L_r)$,
write $H_i=(L_0\supsetneq\cdots\supsetneq L_i)$.
Call $H$ \emph{$\Theta$-good} if it is
geometrically $q$-feasible at cutoffs
$s_{H_0},\ldots,s_{H_r}$ outside $Q_{a_H}$.

Then a $q$-protected tree rooted at $K$
exists if and only if there is a global
geometric assignment $\Theta$ such that
every admissible obstruction tree rooted
at $K$ has at least one terminal history
that is $\Theta$-good.
\end{cor}

\begin{proof}
Fix a global geometric assignment $\Theta$,
and let $\mathcal B_\Theta$ denote the set
of histories that are not $\Theta$-good.

By Lemma~\ref{lem:obstruction-closure},
the root history $(K)$ belongs to
$\overline{\mathcal B_\Theta}$ if and only if
there exists an admissible obstruction tree
rooted at $K$ whose every terminal history
belongs to $\mathcal B_\Theta$.
The lemma applies unchanged in higher
dimensions, since its proof uses only the
finite-evidence history topology.

Consequently, every admissible obstruction
tree has a $\Theta$-good terminal history
if and only if $(K)$ has an open neighborhood
disjoint from $\mathcal B_\Theta$.

In this case, restricting $\Theta$ to a
basic open neighborhood of $(K)$ gives a
$q$-good geometric decoration.
Conversely, any $q$-good decoration of a
basic open neighborhood of $(K)$ can be
extended to a global geometric assignment
by choosing arbitrary cutoffs and
increasing radii outside that neighborhood.

The conclusion now follows from
Theorem~\ref{thm:topological-protected-tree}.
\end{proof}

\begin{cor}
\label{cor:topological-protected-set}
Let \(\mathcal U\) be an open neighborhood of \((K)\) with a
\(q\)-good decoration. For every \(H\in\mathcal U\), there is a finite
set
$P_H\subseteq\operatorname{end}(H)$
such that every proper refinement \(M\) satisfying
$P_H\subseteq M\subsetneq\operatorname{end}(H)$
has \(H^\frown M\in\mathcal U\).
\end{cor}

\begin{proof}
This is exactly the local characterization of open sets in the
finite-evidence history topology.
\end{proof}

Thus the protected sets have precisely the same topological meaning as
in one dimension: they are finite witnesses that all compatible
one-step refinements remain inside the chosen open region.

The topological statements in the two settings are now directly
parallel. In one dimension, for a fixed starting language \(K\), scale
\(n\), and quota \(q\), the condition \(q<b_n(K)\) is equivalent to the
existence of an open neighborhood \(\mathcal U\) of \((K)\) on which
every chain satisfies \(c_n(H)>q\). Here, the corresponding condition
is the existence of an open neighborhood of \((K)\) carrying a
\(q\)-good geometric decoration.

The need for the decoration is the only new topological subtlety.
The continuation topology still records which refinements survive
finite positive evidence. The decoration records the cutoffs and
excluded cores required to make the finite stopping games feasible
coherently along those refinements.

\subsubsection{The static protected forest}
\label{sec:forest}

The forest construction and the capture argument are the same as in the
one-dimensional proof; see Section~\ref{sec:protected-forest} and
Lemma~\ref{lem:capture}. The only difference is that each vertex here
inherits its protected set, cutoff, and radius from an occurrence in a
supplied \(q\)-protected tree.

Enumerate the distinct languages in \(\mathcal X\) as
\(L_1,L_2,\ldots\), and construct the forest in this order. We identify
each forest vertex with its index.

Suppose that vertices \(1,\ldots,j-1\) have already been constructed.
An earlier vertex \(i<j\) is a \emph{possible parent} of \(j\) if
$L_j\subsetneq L_i$
and 
$P_i\subseteq L_j$.
If no possible parent exists, make \(j\) a root and assign to it the
root occurrence of a supplied \(q\)-protected tree rooted at \(L_j\).
Otherwise, scan the possible parents in increasing index order. Retain
the first one, and replace a retained parent \(p\) by a later possible
parent \(i\) whenever
$L_i\subseteq L_p$.
Attach \(j\) to the final retained parent \(p\). Since
\(P_p\subseteq L_j\subsetneq L_p\), the protected-tree condition at the
occurrence assigned to \(p\) supplies a child occurrence labelled
\(L_j\). Assign this occurrence to \(j\), and let \(j\) inherit its
protected set \(P_j\), cutoff \(s_j\), and radius \(a_j\).

It follows inductively that every root-to-vertex path in the forest is
a path in one of the supplied \(q\)-protected trees. Since every parent
has smaller index than its child, every forest vertex has only finitely
many ancestors.

We use the same critical-language rule as in the one-dimensional
construction. To recall, after inputs \(S_t = \{x_1,\ldots,x_t\}\), a language $L_i$ or a vertex \(i\) is \emph{consistent at time \(t\)} if
\(S_t\subseteq L_i\). It is \emph{critical at time \(t\)} if it is
consistent and
$L_i\subseteq L_u$
for every earlier consistent vertex \(u<i\). Let \(Q_t\) be the
largest-index critical vertex among the first \(t\) vertices. If none
exists, leave \(Q_t\) undefined.

We again obtain the same lemma as in the one-dimensional case, where $Q_t$ eventually will be in the subtree rooted at the true target language. 

\begin{lem}
\label{lem:hd-capture}
Fix a true target \(K=L_z\). On every full enumeration of \(K\), all
sufficiently late guesses \(Q_t\) are defined and lie in the subtree
rooted at \(z\), where \(z\) itself is included in its subtree.
\end{lem}

\begin{proof}
This is exactly the proof of Lemma~\ref{lem:capture} in the
one-dimensional part. It uses only the possible-parent rule, the parent
scan, and the eventual appearance of the finite protected set \(P_z\).
The cutoffs, radii, and finite stopping games play no role in the
argument.
\end{proof}
Indeed, after \(P_z\) has appeared and every earlier language not
containing \(K\) has been refuted, \(z\) is critical. The same induction
on vertex indices as in Lemma~\ref{lem:capture} shows that every later
consistent vertex whose label lies inside \(K\) is a descendant of
\(z\).

Thus, the combinatorial role of the forest is unchanged: it places all sufficiently late guesses below the vertex corresponding to the true target and provides a fixed ancestry structure for the later pullback and lowest-common-ancestor (LCA) operations. The additional difficulties in higher dimensions arise only when we construct the spatial boards and their finite-game policies.

\subsubsection{A spatial partition with a logarithmic sewing bound}

The next lemma replaces the one-dimensional fact that an interval has only
two boundary pieces. In the one-dimensional proof, each vertex has its own compatible
partition into interval boards. In higher dimensions, we instead use
one fixed partition \(\mathcal P\) of \(\mathbb Z^d\) into rectangular
boards, independent of the language and of the protected forest.

The role of these boards is the same: they localize the density
argument. The new issue is that a rectangle \(R\) may intersect many
boards, so we need a quantitative bound on the total error created when
the local guarantees are sewn together.

Recall that \(V(R)=|R|\) denotes the number of lattice points in the
rectangle \(R\). Recall that throughout this section the threshold has the form
$h(V)=u(V)\log(2+V)$,
where $u:[1,\infty)\to(0,\infty)$ is nondecreasing.

The following logarithmic partitioning lemma provides a fixed dyadic
partition for which the errors from all intersected boards remain
controlled by the error scale of the original rectangle.

\begin{lem}\label{lem:sewing}
There is a partition $\cP$ of $\mathbb Z^d$ into finite rectangular boards
such that, for every finite rectangle $R$,
\begin{align}
 \sum_{C\in\cP:\,C\cap R\ne\emptyset}
       \log(2+V(C\cap R))
   &\le  (12d^2+4d) \log(2+V(R)),\label{eq:logsew}\\
 \sum_{C\in\cP:\,C\cap R\ne\emptyset}
       h(V(C\cap R))
   &\le (12d^2+4d) h(V(R)).\label{eq:hsew}
\end{align}
The partition is independent of $u$ and of every language.
\end{lem}

\begin{proof}
For this particular $h$, we could use dyadic partitioning. Let $\Q_\ell = [-\ell, \ell]^d$. Put $r_j=2^{2^j}$ for $j\ge0$ and use $\Q_2$ as the central board.  For
$j\ge1$, split the shell $\Q_{r_j}\setminus \Q_{r_{j-1}}$ into $2d$ boxes,
according to the first coordinate outside the smaller cube and its sign.
These boxes partition the shell.

Fix \(R\), and let \(\ell\) be its largest integer side length.
If \(\ell=1\), then \(R\) is a singleton and both estimates are
immediate. Hence assume \(\ell\ge2\).
The boards with outer radius at most \(\ell\) contribute at most
\(12d^2\log \ell\): at radius \(r_j\), there are at most \(2d\)
boards, each contributing at most \(3d\log r_j\), and
\[
\sum_{r_j\le \ell}\log r_j<2\log \ell.
\]
The infinity norm varies by at most \(\ell-1\) across \(R\), so
\(R\) meets at most two shell levels with outer radius greater than
\(\ell\). These contribute at most
\(4d\log(2+V(R))\). This proves \eqref{eq:logsew}.

Finally, since \(u\) is nondecreasing,
\[
u(V(C\cap R))\le u(V(R))
\]
for every board \(C\) meeting \(R\). Therefore
\begin{align*}
\sum_{C\in\cP:\,C\cap R\ne\emptyset}
h(V(C\cap R))
&=
\sum_{C\in\cP:\,C\cap R\ne\emptyset}
u(V(C\cap R))\log(2+V(C\cap R))\\
&\le
u(V(R))
\sum_{C\in\cP:\,C\cap R\ne\emptyset}
\log(2+V(C\cap R))\\
&\le
(12d^2+4d)u(V(R))\log(2+V(R)) =
(12d^2+4d)h(V(R)).
\end{align*}
This proves \eqref{eq:hsew}.
\end{proof}
The same shell-partition idea admits extensions to more general
thresholds, but no such extension is needed for the present results.

\paragraph{Comparison with the one-dimensional boards.}
In the one-dimensional proof, the boards are adapted to the language at
each forest vertex. For a fixed vertex \(v\), every \(v\)-board \(I\)
contains between \(N_v\) and \(B_v\) points of \(L_v\). Thus the
\(L_v\)-population of a board is uniformly bounded, although its ambient
length as an interval of integers need not be bounded. The order on the
line makes this construction possible: one obtains a descendant board
by merging consecutive parent boards until enough points of the
descendant language have been collected. Consequently, a relative
interval of \(L_v\) is a union of whole \(v\)-boards together with at
most two partial boundary boards. 

In higher dimensions, we instead use one spatial partition
\(\mathcal P\), independent of both the language and the protected
forest. Its boards are finite rectangles, but their volumes are not
uniformly bounded, and even for a fixed language \(L\), the populations
\(|L\cap C|\) need not be uniformly bounded over \(C\in\mathcal P\).
This is deliberate. If one merges neighboring rectangular parent
boards until a prescribed number of language points has been collected,
the union need not remain a rectangle. Thus the simple
language-adapted, nested construction from one dimension has no direct
rectangular analogue.

The universal shell partition preserves rectangular geometry, but an
arbitrary rectangle may meet many boards at several scales. The sewing
lemma controls the total filter mass of all these pieces, replacing the
two-boundary-board argument used in one dimension.

The same shell partition could be used in one dimension, but it would
make the proof longer. It would lose both the uniform bound on
the number of language points in a board and the decomposition into
whole boards plus only two boundary pieces. For a sparse language, a
long relative interval could meet many shell boards containing very few
language points each. One would then need a sewing estimate and an
additional rule for discarding poorly populated pieces. The
language-adapted boards are therefore simpler and stronger in one
dimension, while the universal partition is useful in higher dimensions
because it avoids the difficulty of constructing nested rectangular
partitions adapted to each language.

\subsubsection{Boards, pullbacks, pods, and the online algorithm}
\label{sec:hd-algorithm}

The pullback and pod mechanism is the same as in the one-dimensional
construction; see Subsection~\ref{sec:algorithm}. A board stores a
pullback vertex, the corresponding pod grows when the pullback moves
upward by a lowest-common-ancestor update, and outputs are taken from
the current pod whenever possible.

There are two differences. First, we use the fixed spatial partition
\(\mathcal P\) from Lemma~\ref{lem:sewing}, rather than a separate
language-dependent board partition at each forest vertex. Second, the
simple one-dimensional pairing response is replaced by one stored
winning policy for the finite stopping game.

Let \(C\in\mathcal P\). The spatial board \(C\) is initialized when it
receives its first input \(x_t\). It may already contain output points,
but it contains no earlier inputs; such earlier outputs will only help.

We use the following fallback rule. If \(Q_t\) is defined, output any
globally fresh point of \(L_{Q_t}\); if \(Q_t\) is undefined, output any
globally fresh point of \(\mathbb Z^d\).

If \(Q_t\) is undefined when \(C\) is initialized, declare \(C\)
fallback-only and use this rule on the current and every later input in
\(C\). Otherwise, among the ancestors of \(Q_t\), including \(Q_t\)
itself, choose the deepest vertex \(v\) such that
$C\cap\Q_{a_v}=\emptyset$.
If there is no such vertex, again declare \(C\) fallback-only and use
the fallback rule on every input in \(C\).

\paragraph{Simulating the finite stopping game on a board.}
We now bridge the simulated finite game and the actual outputs of a generator. 

Let 
$v_0,v_1,\ldots,v_r=v$
be the root-to-vertex path and the forest, listed from the root $v_0$ to \(v\), and write
\(L_i=L_{v_i}\). For each \(i\), let \(\mathcal R_i\) be the family of
all nonempty axis-parallel rectangles \(R'\subseteq C\) satisfying
$|L_i\cap R'|\ge s_{v_i} h(V(R'))$. 
Let
\[
U=
\bigcup_{i=0}^r\ \bigcup_{R'\in\mathcal R_i}(L_i\cap R')
\]
be the support of this packet. Thus \(U\subseteq C\) consists of all
points that occur in at least one of the rectangle requirements.

Since \(C\cap\Q_{a_v}=\emptyset\), every rectangle in this packet is
disjoint from \(\Q_{a_v}\). Moreover, the retained root-to-\(v\) path
is a path in the supplied \(q\)-protected tree. Hence the geometric
feasibility condition in Definition~\ref{def:protected-tree} implies
that this packet is feasible.

Choose once and for all a winning White strategy for the finite stopping
game associated with this path and packet. The algorithm keeps track of
the current state of this game on the board \(C\). The correspondence
between the simulated game and the actual input-output process is as
follows.

\begin{enumerate}
\item
Initially the pullback is \(v_r=v\). This corresponds to level \(r\) of
the finite game, whose visible set is
$U\cap L_r$.
Thus the initial pod is exactly the initial visible set of the game.

The input \(x_t\) that initializes \(C\) is now processed immediately
according to Items~3--5 below. Item~2 applies only to subsequent inputs
in \(C\).

\item
Suppose the current pullback is \(v_j\), and a later input arrives in
\(C\). Let \(Q_t\) be the current guess. If \(Q_t\) is undefined, stop
the simulation on \(C\) and use the fallback rule on this and all later
inputs in the board. If \(Q_t\) is defined but \(v_j\) and \(Q_t\) lie
in different trees of the forest, stop the simulation on \(C\) and use
the fallback rule on this and all later inputs in the board.

Otherwise, let
$v_i=\operatorname{LCA}(v_j,Q_t)$,
where \(i\le j\). Replace the pullback \(v_j\) by \(v_i\).

The net effect is to move the simulated game from level \(j\) to level
\(i\). Since the stopping game enlarges visibility one level at a time,
we perform the successive moves
$j\longrightarrow j-1\longrightarrow\cdots\longrightarrow i.$
No White response is made during these moves. The visible set thereby
grows from \(U\cap L_j\) to \(U\cap L_i\).

\item
After these visibility moves, consider the actual input \(x_t\). 
\begin{enumerate}
    \item If
\(x_t\in U\) and is unused in the simulated game, use \(x_t\) as
Black's next claim.
This is a legal Black move. Indeed, \(x_t\in L_{Q_t}\), while the
updated pullback \(v_i\) is an ancestor of \(Q_t\). Hence
$L_{Q_t}\subseteq L_i$,
so \(x_t\in U\cap L_i\), the current visible set. \label{item:3a}
\item If \(x_t\notin U\), or if \(x_t\) has already been claimed in the
simulated game, make no Black move in the simulation on this round.
\end{enumerate}

\item
When a virtual Black move is made, follow the response prescribed by
the fixed winning White strategy. If the strategy passes, make no
virtual White claim.

If the strategy chooses a point \(y\), mark \(y\) White in the
simulated game. If \(y\) is globally fresh, output \(y\) in the actual
process. If \(y\) has already been output on an earlier round, then its
required contribution is already present in the actual output set; in
that case output a fallback point for the current round.

\item
On any round on which the simulated game does not provide an available
output (available here means the string has not appeared as an
input by the time and has not previously been output)---because no virtual Black move is made, the stored strategy
passes, or the White point prescribed by the strategy has already been
output---output instead any available point of the current guess
language.

Such fallback outputs are not recorded as moves of the simulated game.
If a fallback output happens to belong to \(U\), it is simply an
additional actual White point and can only improve the final rectangle
counts.
\end{enumerate}

\begin{lem}
\label{lem:hd-help}
For a board \(C\) on which the finite-game simulation continues, the
following hold:
\begin{enumerate}
\item after each actual round, every point of \(U\) that has appeared
as an actual input is no longer unused in the simulated game;
\item every point marked White in the simulated game belongs to the
actual output set \(O\).
\end{enumerate}

Consequently, let \(v_i\) be an ancestor of the current pullback on the
retained path. If every point of \(U\cap L_i\) has appeared as an actual
input, then $|O\cap L_i\cap R|
\ge q(L_i)|L_i\cap R|$ for every $R\in\mathcal R_i$.
\end{lem}

\begin{proof}
The first assertion follows directly from the simulation rule: whenever
an actual input in \(U\) is unused in the simulated game, it is entered
as Black's next move by Item \ref{item:3a}.

Now suppose the stored strategy chooses a virtually unused point \(y\)
as White's response. If \(y\) has not been used by either adversary or generator, the generator outputs
it. Otherwise \(y\) has already appeared in the actual process. It
cannot have appeared only as an input, by the first assertion, since
then it would already be claimed virtually. Hence \(y\) was already
output. This proves the second assertion.

Finally, enlarge visibility to level \(i\). By hypothesis, every point
of \(U\cap L_i\) has appeared as an actual input. By the first assertion,
each such point has therefore already been claimed in the simulated
game. Hence the visible set \(U\cap L_i\) is exhausted, so stopping at
level \(i\) is legal. The winning strategy gives
$|W\cap L_i\cap R|\ge q(L_i)|L_i\cap R|
\qquad\text{for every }R\in\mathcal R_i.$
By the second assertion, every simulated White point belongs to the
actual output set \(O\), proving the claim.
\end{proof}

\paragraph{Comparison with the one-dimensional algorithm.}
Up to the local response rule, this is the same cumulative-pullback
construction as in one dimension. The pullback is updated by an LCA,
the pod grows, and later guesses are interpreted through one fixed
ancestry structure. The proof that these updates remain compatible is
the same as the history-inclusion argument in
Lemma~\ref{lem:memory}.

The differences are local and geometric. In one dimension, any unused
visible point supplies the pairing response, and a long interval loses
only two boundary boards. Here each spatial board stores one finite-game
policy capable of serving all of its overlapping rectangle tests, and
Lemma~\ref{lem:sewing} is needed to combine the guarantees from all
boards met by a rectangle.

\subsubsection{Validity and density and completion of Proof of Theorem~\ref{thm:fixed-quota}}
\label{sec:hd-validity-density}

Fix a true target \(K=L_z\). By Lemma~\ref{lem:hd-capture}, there is a
time \(T\) after which every current guess is a descendant of \(z\).

\paragraph{Validity.}
The proof is the same as the one-dimensional validity argument in
Lemma~\ref{lem:validity}. We only indicate the changes.

Call a board \(C\in\mathcal P\) exceptional if it was touched by an
input or output by time \(T\), or if it meets the core \(\Q_{a_z}\).
There are only finitely many exceptional boards.

A nonexceptional board is initialized after time \(T\), with a current
guess below \(z\), and it misses \(\Q_{a_z}\). Hence \(z\) lies on its
root-to-vertex path. Every later guess is also below \(z\), so every later
pullback, being an LCA of descendants of \(z\), remains below \(z\).
Therefore every local output and every fallback output on this board
belongs to \(K\).

Thus an invalid output after time \(T\) can occur only as a local output
from one of the finitely many exceptional boards. Each such board is
finite and outputs never repeat, so only finitely many invalid outputs
are possible. Hence the generator is eventually valid on \(K\).

\paragraph{Density.}
Let \(C\) be a nonexceptional board. Since \(C\) is finite and the input
is a full enumeration of \(K\), every point of \(K\cap C\) is eventually
input. The vertex \(z\) is an ancestor of the current pullback, so
Lemma~\ref{lem:hd-help}, applied at level \(z\), gives
$|O\cap K\cap R'|
\ge q(K)|K\cap R'|$
for every rectangle \(R'\subseteq C\) satisfying $|K\cap R'|\ge s_z h(V(R'))$.

Let \(B_z\) be the total number of \(K\)-points in exceptional boards,
and fix a finite rectangle \(R\). For every nonexceptional board \(C\),
the intersection \(R\cap C\) is again a rectangle. If
$|K\cap R\cap C|\ge s_z h(V(R\cap C))$,
then the local guarantee applies to \(R\cap C\). The remaining pieces
have total \(K\)-mass at most
\[
s_z\sum_{C\in\mathcal P}h(V(R\cap C))
\le s_z(12d^2+4d)h(V(R))
\]
by Lemma~\ref{lem:sewing}. Therefore
\begin{equation}
\label{eq:finalbound}
|O\cap K\cap R|
\ge
q(K)\bigl[m_K(R)-B_z-s_z(12d^2+4d) h(V(R))\bigr].
\end{equation}

Along the density filter,
\(m_K(R)/h(V(R))\to\infty\), so both error terms are negligible compared
with \(m_K(R)\). Hence
$\delta_h(O,K)\ge q(K)$.
This proves the sufficiency direction of
Theorem~\ref{thm:fixed-quota}.

\subsection{Necessity of Theorem \ref{thm:fixed-quota}: from a generator to protected trees}
\label{sec:necessity}

We now prove the reverse implication of
Theorem~\ref{thm:fixed-quota}.
Suppose that one globally fresh, universally
eventually valid generator achieves a density
profile $\alpha$. Fix $\varepsilon>0$ and let
\[
q(L)=(\alpha(L)-\varepsilon)_+.
\]
We must construct a $q$-protected tree rooted
at every language $K$.

The key is to extract finite commitments from
the assumed generator. Eventual validity
forces a finite commitment to producing only
valid outputs, while the density guarantee
forces an analogous commitment to the
rectangle quotas. These commitments provide
the protected sets and geometric parameters
needed to construct the tree.

\begin{lem}[Finite validity and density commitments]
\label{lem:hd-commitments}
Suppose that a finite input prefix consists only of points of
\(L\in\mathcal X\). The adversary can extend this prefix finitely using
only inputs from \(L\), and can choose a cutoff \(s\ge1\), so that every
further finite continuation inside \(L\) has the following properties:
\begin{enumerate}
\item every output produced after the committing prefix belongs to \(L\);
\item every occupied level-\(s\) test \(R\) for \(L\) contains at least
\(q(L)m_L(R)\) White points.
\end{enumerate}
A trace is \emph{occupied} when each of its points has appeared as an
input or an output.
\end{lem}

\begin{proof}
The first assertion is the same finite-commitment argument used for
eventual validity in one dimension. If no finite extension committed
the generator to \(L\), repeatedly input the next missing point of a
fixed enumeration of \(L\), and then continue finitely inside \(L\)
until the generator outputs a point outside \(L\). This would produce
a full enumeration of \(L\) with infinitely many invalid outputs,
contradicting eventual validity.

Keep this validity commitment fixed. If \(q(L)=0\), the second assertion
is automatic. Otherwise \(q(L)<\alpha(L)\).

Suppose that no further finite extension, together with any cutoff,
committed the density requirement. For \(j=1,2,\ldots\), first input
the next missing point of the fixed enumeration of \(L\). Then extend
finitely inside \(L\) until there is an occupied trace \(L\cap R\) satisfying 
$m_L(R)\ge j\,h(V(R))$
and having White fraction below \(q(L)\).

The resulting input sequence is a full enumeration of \(L\). Once a
finite trace is occupied, global freshness prevents any later output
from changing its White count. We therefore obtain permanently
deficient traces at arbitrarily high filter levels, implying
$\delta_h(O,L)\le q(L)<\alpha(L)$,
a contradiction.
\end{proof}

We now construct a \(q\)-protected tree rooted at an arbitrary
\(K\in\mathcal X\). Apply Lemma~\ref{lem:hd-commitments} starting from
the empty input prefix in \(K\).

At a vertex \(v\), let \(L_v\) be its language label. Let \(P_v\) be
the finite set of all inputs in the committing prefix supplied by the
lemma, let \(s_v\) be its cutoff, and choose \(a_v\) large enough to
contain every input and output in that prefix. At a nonroot vertex,
also require \(a_v\) to be larger than its parent's radius.

For every proper refinement \(M\in\mathcal X\) satisfying
\[
P_v\subseteq M\subsetneq L_v,
\]
the committing prefix at \(v\) is also a legal input prefix for \(M\).
Apply Lemma~\ref{lem:hd-commitments} to \(M\), starting from that same
prefix, and use the resulting data to form a child labelled \(M\).
Continue recursively. Since \(\mathcal X\) is countable, the resulting
tree is countable, and its child condition is exactly the protected-set
condition.

It remains to verify geometric feasibility. Fix a finite path
$v_0,\ldots,v_r$
and an eligible finite rectangle packet for the corresponding language
chain, outside \(Q_{a_{v_r}}\). Its support \(U\) is untouched at the
committing prefix of \(v_r\), because that prefix and all of its outputs
lie inside \(Q_{a_{v_r}}\).

Suppose Black had a strategy forcing a failed stop in the finite game.
Continue the actual generator from the committing prefix of \(v_r\),
using Black's claimed points as inputs. When the game is at the level
corresponding to \(v_i\), all inputs supplied since the committing
prefix at \(v_i\) belong to \(L_{v_i}\). The validity commitment at
\(v_i\) therefore forces every generator output to lie in \(L_{v_i}\).

If such an output lies in \(U\), use it as White's move in the finite
game; it is visible and fresh. If it lies outside \(U\), regard it as
a pass. This defines a legal causal White strategy against Black's
play.

At a failed stop at level \(i\), the visible set is occupied. Hence
every requested trace \(L_{v_i}\cap R\) is occupied. The packet
eligibility condition gives \(m_{L_{v_i}}(R)
\ge
s_{v_i}h(V(R))\), while
the failed quota says that the trace contains fewer than
\(q(L_{v_i})|L_{v_i}\cap R|\) White points. This contradicts the density
commitment at \(v_i\).

Thus every finite path is geometrically
$q$-feasible, and the constructed tree is
$q$-protected.

By Corollary~\ref{cor:hd-obstruction},
this also gives a global geometric assignment
$\Theta$ such that every admissible obstruction
tree rooted at $K$ has a $\Theta$-good
terminal history.
Equivalently, if every geometric assignment
can be defeated by an admissible obstruction
tree whose terminal histories all fail,
then no generator can achieve $\alpha$.
This proves the necessity direction of
Theorem~\ref{thm:fixed-quota}.

\paragraph{Comparison with the one-dimensional protected-set argument.}
The finite-evidence recursion is the same as in
Corollary~\ref{cor:protected-set}: a finite set \(P_v\) is chosen at
the current history, and every proper refinement containing \(P_v\)
must remain in the protected construction.

The validity commitment also uses the same diagonal argument as in one
dimension. The difference is the local quantitative certificate. In one
dimension, the explicit chain score and the protected-set lemma show
that every compatible continuation remains above the desired score.
Here, finite density commitments and the stopping-game simulation show
that every compatible continuation remains feasible for all required
rectangle packets.

\subsection{From fixed error to the exact profile}\label{sec:exact-profile}

The fixed-quota theorem contains the geometric
core of the proof. We now introduce accuracy
copies to enforce all strict approximations within
one generator. The certificates for different
accuracies must be unified: separate achievability of every fixed-error profile does not imply exact attainment.

For every \(K\in\collec\) and \(k\ge1\), introduce a formal copy
\((K,k)\) with quota
\[
q(K,k)=\bigl(\alpha(K)-2^{-k}\bigr)_+.
\]
The underlying language of the copy is still \(K\); the index records
only the required accuracy.

\begin{defn}[Protected tree on accuracy copies]
\label{def:copy-protected-tree}
A protected tree rooted at \((K,k)\) is defined as in
Definition~\ref{def:protected-tree}, except that vertices carry copy
labels \((L_v,k_v)\).  Equal underlying language labels are permitted
along an edge.  A stop at \((L_v,k_v)\) uses quota \(q(L_v,k_v)\).

For every copy \((M,n)\) satisfying
$P_v\subseteq M\subseteq L_v$,
the vertex \(v\) has a child labelled \((M,n)\).  An edge whose
underlying language does not change exposes no new points in the finite
stopping game.
\end{defn}

\paragraph{Necessity.}

Suppose one globally fresh, universally eventually valid generator
achieves the profile \(\alpha\).  Fix a root copy \((K,k)\).
We construct a protected tree recursively.

At a vertex labelled \((L,j)\), apply
Lemma~\ref{lem:hd-commitments} with quota \(q(L,j)\).  When
\(q(L,j)>0\), this quota is strictly below \(\alpha(L)\), so the lemma
provides a finite committing prefix, a finite protected set \(P_v\),
and a cutoff \(s_v\).  When \(q(L,j)=0\), only the validity commitment
is needed, and the density requirement is automatic.  Choose a radius
\(a_v\) containing the committing prefix and its outputs, larger than
the radius at the parent.

For every compatible copy \((M,n)\) with
$P_v\subseteq M\subseteq L$,
the same committing prefix is a legal prefix for \(M\).  Apply the
commitment lemma again and continue recursively below a child labelled
\((M,n)\).  The stopping-game argument from the necessity proof of
Theorem~\ref{thm:fixed-quota} applies unchanged.  If the underlying
language is unchanged, no new points become visible; only the quota
attached to the stopping copy changes.

Thus every root copy \((K,k)\) has a protected tree on copies.

\paragraph{The protected forest on copies.}

Assume conversely that every root copy has a protected tree.  Enumerate
all copies in one sequence, with \((L,k)\) preceding \((L,k+1)\) for
every fixed language \(L\).

Repeat the protected-forest construction from
Section~\ref{sec:forest}, allowing nonstrict language inclusion in the
possible-parent rule:
\[
L_v\subseteq L_u,
 \text{ and }
P_u\subseteq L_v.
\]
Use the same parent scan, replacing the retained parent whenever the
new possible parent's underlying language is contained in the retained
one.  Each forest vertex inherits its protected set, cutoff, radius,
and occurrence in the supplied protected tree on copies.

Every finite forest path is therefore a path in a protected tree on
copies.  Equal-language edges change the formal requirement but expose
no new points in the corresponding stopping game.

\begin{lem}
\label{lem:hd-copy-chain}
For every language \(L\) and every \(k\ge1\), the copy \((L,k+1)\) is
a descendant of \((L,k)\).  Consequently, all copies of one language
lie on one ancestor--descendant chain in the protected forest.
\end{lem}

\begin{proof}
The protected set of \((L,k)\) is contained in \(L\), so \((L,k)\) is
a possible parent of \((L,k+1)\).  Every possible parent of
\((L,k+1)\) has underlying language containing \(L\).  When the parent
scan reaches \((L,k)\), it retains that copy; no later possible parent
can have underlying language strictly contained in \(L\).  Hence the
chosen parent is \((L,k)\), or a later same-language copy lying below
it.  Induction gives the conclusion.
\end{proof}

\begin{lem}
\label{lem:hd-copy-capture}
Fix a target \(K\), a copy \(z_k=(K,k)\), and a full enumeration of
\(K\).  After some finite time, every current guess is defined and lies
in the subtree rooted at \(z_k\).
\end{lem}

\begin{proof}
Use the critical-vertex rule with nonstrict inclusion of underlying
languages.  Wait until the index of \(z_k\) has been reached, all
elements of its protected set have appeared, and every earlier vertex
whose underlying language does not contain \(K\) has become
inconsistent.

Thereafter \(z_k\) is a critical candidate.  If a later consistent
vertex \(v\) has underlying language contained in \(K\), then
\(z_k\) is a possible parent of \(v\).  The parent scan retains a
vertex no earlier than \(z_k\) whose underlying language is contained
in \(K\).  Iterating along strictly decreasing parent indices reaches
\(z_k\).  Hence every sufficiently late current guess is a descendant
of \(z_k\).
\end{proof}

\paragraph{Validity.}

Apply Lemma~\ref{lem:hd-copy-capture} to the first copy \(z_1=(K,1)\).
After its capture time, every current guess has underlying language
contained in \(K\).  The exceptional-board argument from
Section~\ref{sec:hd-validity-density} therefore proves eventual
validity.  Equal-language edges do not affect this argument.

\begin{lem}
\label{lem:hd-fixed-copy-density}
For every \(k\ge1\), the output set of the copy construction satisfies
\[
\delta_h(O,K)\ge q(K,k).
\]
\end{lem}

\begin{proof}
Fix \(k\) and apply Lemma~\ref{lem:hd-copy-capture} to
\(z_k=(K,k)\).  Only finitely many boards are exceptional for this
capture time or meet the core attached to \(z_k\).

On every other board, the current pullback remains below \(z_k\), and
the stored stopping-game policy uses the quota attached to \(z_k\).
Equal-language edges expose no new points, so the simulation lemma,
the local rectangle guarantees, and the sewing argument from
Section~\ref{sec:hd-validity-density} apply without change.  They give
$\delta_h(O,K)\ge q(K,k)$.
The exceptional boards and the core may depend on \(k\), but they are
finite for each fixed \(k\), which is all the density limit requires.
\end{proof}

\begin{proof}[Completion of the proof of
Theorem~\ref{thm:main}]
The necessity direction was established above.

For sufficiency, the forest, spatial partition, board simulations, and
output set are constructed once from the complete countable family of
copies.  Lemma~\ref{lem:hd-fixed-copy-density} gives, for every
\(k\ge1\),
\[
\delta_h(O,K)
\ge
\bigl(\alpha(K)-2^{-k}\bigr)_+.
\]
As \(k\to\infty\) we have
$\delta_h(O,K)\ge\alpha(K)$.
The same generator works for every target \(K\), completing the proof.
\end{proof}

Thus the copies remove the endpoint slack; they are not part of the
geometric core of the characterization.

\section{Finite Cantor--Bendixson rank}
\label{sec:finite-rank}

We now derive the finite-rank consequence of the profile
characterization. The topological ingredient is the same rank-drop
principle used in one dimension: finite positive evidence forces every
proper surviving refinement to have smaller rank.

The geometric ingredient is a  finite-game strengthening of the
size-sensitive discrepancy ideas from \cite{KWBanach}. The finite
stopping game requires one strategy to serve many overlapping
rectangles while remaining valid under transfers of responsibility
between spatial scales. Appendix~\ref{app:geometry} develops a
multiscale potential with a recursive descendant correction for this
purpose.

\begin{lem}
\label{lem:local-rectangle-strategy} Let $d \geq 1$. 
For every \(\tau>0\) and \(0<\eta<1/2\), let
$h_\tau(V)=\log^{1+\tau}(2+V)$.
There is a constant \(C_{d,\eta,\tau}\) with the following property.

Let \(E\subseteq\mathbb Z^d\) be finite, and let \(\mathcal R\)
be a finite family of axis-parallel rectangles. In the finite game
in which Black claims an unused point of \(E\) and White then claims
at most one unused point, White has one strategy such that, when
\(E\) is exhausted, every \(R\in\mathcal R\) satisfying
$|E\cap R|\ge C_{d,\eta,\tau}h_\tau(V(R))$
also satisfies
\[
|W\cap R|
\ge
\left(\frac12-\eta\right)|E\cap R|.
\]
\end{lem}

\begin{thm}[Finite-rank near-half guarantee]
\label{thm:finite-rank-near-half}
Fix \(d\ge2\), \(\tau>0\), and \(0<\varepsilon<1/2\), and let
\(\mathcal X\) be a nonempty countable family of infinite subsets of
\(\mathbb Z^d\). Define
$h_\tau(V)=\log^{1+\tau}(2+V)$.
There is one universally eventually valid generator such that, for
every target \(K\in\mathcal X\) of finite Cantor--Bendixson point rank,
\[
\delta_{h_\tau}(O,K)\ge\frac12-\varepsilon.
\]
No common upper bound on the finite ranks is required.
\end{thm}

\begin{proof}
If \(L\) is \(h_\tau\)-vacuous, then
\(\delta_{h_\tau}(O,L)=1\) for every output set \(O\), so the theorem is
automatic at \(L\).  It therefore suffices for the construction to
enforce the near-half quota on the remaining finite-rank targets.
Define the quota profile
\[
q(L)=
\begin{cases}
\frac12-\varepsilon,
 & \text{if \(L\) has finite point rank and is not
 \(h_\tau\)-vacuous},\\
0,
 & \text{otherwise}.
\end{cases}
\]
By Theorem~\ref{thm:fixed-quota}, it is enough to construct a
\(q\)-protected tree rooted at every \(K\in\mathcal X\).

The protected sets are chosen exactly as in the one-dimensional
finite-rank proof. Suppose that \(L\) has Cantor--Bendixson point rank
\(r\). Then
$L\in\mathcal X^{(r)}\setminus\mathcal X^{(r+1)}$,
so \(L\) is isolated in the subspace \(\mathcal X^{(r)}\). Since the
basic finite-evidence neighborhoods of \(L\) are
$U_{L,F}=\{M\in\mathcal X:F\subseteq M\subseteq L\}$,
there is a finite set \(P_L\subseteq L\) such that $U_{L,P_L}\cap\mathcal X^{(r)}=\{L\}$.
Consequently, every proper refinement \(M\in\mathcal X\) satisfying
\[
P_L\subseteq M\subsetneq L
\]
has point rank strictly smaller than \(r\). For languages without finite point rank, set \(P_L=\emptyset\).
Their quota is zero, so no density estimate is required at those
vertices.

Thus every strict step in a protected tree starting from a finite-rank
language lowers the rank. In particular, a path starting at a rank-\(r\)
language contains at most \(r+1\) finite-rank language layers.

By Lemma \ref{lem:local-rectangle-strategy}, write 
$C=C_{d,\varepsilon/2,\tau}$,
and 
$s_r=
\left\lceil
\max\left\{1,\frac{(r+1)C}{\varepsilon}\right\}
\right\rceil$.
At an occurrence of a rank-\(r\) language, use protected set \(P_L\)
and cutoff \(s_r\). At an occurrence without finite rank, use cutoff
\(1\). Choose any radii that increase strictly down the tree, and add a
child for every compatible proper refinement. Since \(\mathcal X\) is
countable, this gives a countable tree with the required protected-set
coverage.

It remains to verify the finite stopping games. Fix a finite path
$L_0\supsetneq L_1\supsetneq\cdots\supsetneq L_t$
and a finite eligible rectangle packet with support \(U\). Write its
disjoint language layers as
$E_t=U\cap L_t$, and for each $0 \leq j < t$, the layer is  $E_j=U\cap(L_j\setminus L_{j+1})$

For each layer \(E_j\), apply
Lemma~\ref{lem:local-rectangle-strategy} to all rectangles in the packet
whose \(E_j\)-trace contains at least
\(C h_\tau(V(R))\) points. Whenever Black claims a point of \(E_j\),
White follows the strategy chosen for that layer. Since the layers are
disjoint, these strategies can be interleaved into one causal strategy
for the stopping game. This strategy is fixed before Black chooses the
stopping level.

Suppose Black stops at \(L_i\). If \(q(L_i)=0\), there is nothing to
prove. Otherwise \(L_i\) has some finite rank \(r\). Every strict
refinement below \(L_i\) has smaller rank, so at most \(r+1\) language
layers are visible at this stop.

Fix a requested rectangle \(R\) at level \(i\), and write $m=|L_i\cap R|$.
Since the packet is eligible, \(m\ge s_r h_\tau(V(R))\).

Call a visible layer ``small" if it contains fewer than \(C h_\tau(V(R))\)
points. Since at most \(r+1\) layers are visible, the total number of points of the small layers is at most 
\[
(r+1)C h_\tau(V(R))\le\varepsilon s_r h_\tau(V(R))\le\varepsilon m.
\]
On every remaining layer trace,
Lemma~\ref{lem:local-rectangle-strategy} gives White density at least
\(1/2-\varepsilon/2\). Hence
\[
\begin{aligned}
|W\cap L_i\cap R|
&\ge
\left(\frac12-\frac{\varepsilon}{2}\right)(m-\varepsilon m)\ge
\left(\frac12-\varepsilon\right)m
=
q(L_i)m.
\end{aligned}
\]

Thus the same White strategy satisfies every requested rectangle at
every possible stopping level. Every finite path is therefore
geometrically \(q\)-feasible, so the trees constructed above are
\(q\)-protected.

Theorem~\ref{thm:fixed-quota} now gives one universally eventually valid
generator satisfying
$\delta_{h_\tau}(O,K)\ge\frac12-\varepsilon$
for every finite-rank target \(K\) that is not \(h_\tau\)-vacuous.
For an \(h_\tau\)-vacuous target, the same inequality holds
automatically because
$\delta_{h_\tau}(O,K)=1$.
This proves the theorem for every finite-rank target \(K\).
\end{proof}

\section{Why exact half-density fails even for one known target}
\label{sec:endpoint}

Fix \(d\ge2\) and the known target \(K=\mathbb Z^d\). We show that no
causal generator satisfying the freshness condition can guarantee exact
density \(1/2\). Thus the preceding \(1/2-\varepsilon\) guarantee is sharp: the loss
\(\varepsilon\) cannot in general be removed.

The obstruction is already visible in a two-row strip. Black takes the
top point of a fresh column. White must immediately enter that column,
or Black takes its other point next and obtains a completely Black
column. If White always responds inside the attacked column, the top
row becomes Black. The following lemma extends this elementary threat
argument to strips of arbitrary height.

\begin{thm}[Failure of exact half-density]
\label{end:main}
Let \(h:\mathbb N_{\ge1}\to(0,\infty)\) be nondecreasing and satisfy
$\sup_{n\ge1}\frac{n}{h(n)}=\infty$.
For every causal generator satisfying the freshness condition, there
is a full enumeration of \(\mathbb Z^d\) on which
$\delta_h(O,\mathbb Z^d)<\frac12$.
\end{thm}
This theorem will be implied by the following lemma.
\begin{lem}
\label{end:strip}
Fix \(s\ge2\) and \(N\ge1\). In a completely unused
\(s\)-by-\(N\) strip, Black has a strategy that produces
one of the following:
\begin{enumerate}
\item a column of length \(s\) containing no White point;
\item a completely occupied row of length \(N\) whose White fraction is
at most
$\frac12-\gamma_s$
for some constant \(\gamma_s>0\) depending only on \(s\).
\end{enumerate}
White may output anywhere in \(\mathbb Z^d\).
\end{lem}

\begin{proof}
Number the rows from top to bottom and assign them weights
$w_1=2^{s-2}$, and for each $2 \leq i \leq s$,  let decreasing weights be
$w_i=2^{s-2}-2^{i-2}$.
Thus \(w_1>\cdots>w_s=0\), and a direct calculation gives, for
\(1\le k<s\),
\begin{equation}
    \sum_{i=1}^k w_i-(k-1)w_1-w_{k+1}=1. \label{eq:trivial}
\end{equation}
Let \(W_s=\sum_{i=1}^s w_i\).

Call a column protected once it contains a White point. While an
unprotected column remains, Black chooses one and claims its unused points from
top to bottom until either White enters that column or Black fills it.

If Black fills the column, the first outcome holds. Otherwise for this particular column, suppose
White first enters after Black's \(k\)-th move, where \(1\le k<s\).
Black has obtained weights \(w_1,\ldots,w_k\). The first \(k-1\) White
responses have weight at most \(w_1\), and the response entering the
current column has weight at most \(w_{k+1}\). Thus this run consists of \(k\) Black moves and \(k\) White responses,
with the final White response entering the current column. By
\eqref{eq:trivial}, the run increases the weighted
Black-minus-White count by at least \(1\).

Each run has \(k\le s-1\) White responses, and each response can protect
at most one column. Hence one run protects at most \(s-1\) columns. If no Black column appears,
at least \(N/(s-1)\) runs are needed before all columns are protected.

Once all columns are protected,  Black occupies the remaining points of
the strip, always choosing an unused point of greatest weight. Each subsequent White point inside
the strip has no greater weight, so the weighted deficit cannot
decrease. 

When the game ends, if \(D_i\) denotes the number of Black points
minus White points in row \(i\), then the total weighted Black minus White is at least the total such weight difference for all the runs, which is at least $\frac{N}{s-1}$. Equivalently,  
\[
\sum_{i=1}^s w_iD_i\ge \frac{N}{s-1}.
\]
Therefore some row satisfies
\[
\frac{D_i}{N}\ge\frac{1}{(s-1)W_s}.
\]
Since that row is completely occupied, its White fraction is at most
$\frac12-\frac{1}{2(s-1)W_s}$.
Thus the lemma holds with
$\gamma_s=\frac{1}{2(s-1)W_s}>0$.
\end{proof}

\begin{proof}[Proof of Theorem~\ref{end:main}]
Choose strictly increasing integers \(2 \leq s_1 < s_2 < \ldots\) such that
$\frac{s_j}{h(s_j)}\ge j$,
and strictly increasing integers \(N_1,N_2,\ldots\) such that
$\frac{N_t}{h(N_t)}\ge t$.
Such choices are possible because \(n/h(n)\) is unbounded.

Fix an enumeration \(z_1,z_2,\ldots\) of \(\mathbb Z^d\). We construct
the adversary's full enumeration in stages. Maintain a height index
\(j\), initially \(j=1\).

At stage \(t\), first input the first \(z_m\) that has not yet appeared
as an input. After the generator responds, choose a translate of an
\(s_j\)-by-\(N_t\) strip that avoids every point used so far, and apply
Lemma~\ref{end:strip}.

If the lemma produces a completely Black column, replace \(j\) by
\(j+1\). If it produces a sparse row, leave \(j\) unchanged.

Every stage is finite, and the first input of each stage ensures that
the resulting sequence is a full enumeration of \(\mathbb Z^d\).

There are now two cases.

If completely Black columns occur infinitely often, then the \(k\)-th
such column has length \(s_k\), White fraction zero, and 
$\frac{s_k}{h(s_k)}\ge k$.
These columns therefore occur at arbitrarily high filter levels, so
$\delta_h(O,\mathbb Z^d)=0$.

Otherwise, only finitely many completely Black columns occur. The height
index is then eventually fixed at some value \(j_0\), and every
sufficiently late stage produces a row of length \(N_t\) with White
fraction at most
$\frac12-\gamma_{s_{j_0}}$.
Since
$\frac{N_t}{h(N_t)}\ge t\longrightarrow\infty$,
these rows occur at arbitrarily high filter levels. Hence
$\delta_h(O,\mathbb Z^d)
\le
\frac12-\gamma_{s_{j_0}}
<
\frac12$.
This proves the theorem.
\end{proof}

\appendix

\section{A multiscale discrepancy proof of the finite rectangle game}\label{app:geometry}

In this section we prove Lemma~\ref{lem:local-rectangle-strategy}.  The main difficulty
is that White must satisfy many overlapping rectangle requirements from different directions at widely different spatial scales while receiving only one response to
each Black move.

The proof is by combining a
greedy exponential potential with a  multi-scale sparse dyadic hierarchy.
  After each Black move, White claims
the point whose loss would cause the greatest potential increase.  The
new difficulty is choosing the spatial scale at which this response
should be made, since several overlapping rectangle requirements may
become endangered simultaneously.

The proof has three stages. First, we formulate abstract properties
of a potential function that imply a winning strategy. Second, we
distribute rectangle requirements over a sparse tree of dyadic cubes,
passing sufficiently populated pieces to smaller scales. Finally, we
construct an explicit potential satisfying the required properties.

\paragraph{Proof roadmap.}
The formal dependencies are as follows.
\begin{enumerate}[label=(\arabic*),leftmargin=*]
\item Lemma~\ref{lem:new-abstract} shows that any potentials satisfying
the locality conditions and \textup{(P1)--(P4)} yield a White strategy
meeting every local quota.

\item Claims~\ref{claim:discardsmall} and
\ref{claim:quotassuffice} show that meeting the local quotas loses only
an \(\varepsilon\)-fraction of each eligible rectangle.

\item Proposition~\ref{prop:cf-conditional} converts these local
guarantees into the finite rectangle strategy, provided the scale
conditions \textup{(G1)--(G3)} hold.

\item Claim~\ref{claim:cf-potential-existence} constructs explicit
scales, thresholds, and potentials satisfying
\textup{(G1)--(G3)} and \textup{(P1)--(P4)}.

\item Combining these statements proves
Lemma~\ref{lem:local-rectangle-strategy}.
\end{enumerate}

\subsection{The board and the spatial hierarchy}

\paragraph{Spatial tree.}
Let $E\subseteq\mathbb Z^d$ be the finite set on which Black and White
play. In each round Black claims an unused point of $E$ and White may
claim at most one unused point. The game ends once every point of $E$
has been claimed.

We organize $E$ using a finite tree of nested spatial cubes. To picture
the tree, think of the ordinary aligned dyadic cubes, whose side lengths
are $1,2,4,8,\ldots$. We will retain only certain generations, of side
lengths $H_0<H_1<\cdots<H_J$, each a power of two; the exact sizes will
be chosen later. A cube at one retained level is a child of the larger
cube containing it at the next retained level. We discard cubes that do
not meet $E$, and add a dummy root $o$ above the largest retained cubes.

For a nonroot vertex $v$, write $C_v$ for its geometric cube and
\[E_v=E\cap C_v\] for the \emph{game points} in that cube; set $E_o=E$.
Thus $C_v$ is a cube, but $E_v$ need not be a cube or rectangle. Different
children of a vertex have disjoint point sets. For example, in two
dimensions, retaining the toy scales $2,8,32$ makes a $32\times32$ square
the parent of sixteen $8\times8$ squares; each of those has sixteen
$2\times2$ children. A $1\times32$ horizontal rectangle meets four of
the $8\times8$ children. The actual proof uses much more widely separated
scales, but the tree is defined in exactly the same way.

At each vertex $v$, fix a finite family $\mathcal A_v$ of subsets
$Q\subseteq E_v$. Each such $Q$ is a \emph{local requirement} and has a
fixed  quota $a_Q\ge0$: White must claim at least $a_Q$ points
of $Q$. The families and quotas are determined before the game begins.
A set $Q$ need not be a rectangle intersecting $E$. Later, $Q$ will be the points of a
dyadic rectangle that are left at this cube after its sufficiently
populated child intersections have been handed to smaller cubes.
We reserve $R$ for geometric rectangles and $Q$ for local requirements.

\paragraph{Cube activation.}
Throughout the procedure, we would like to keep the potential below $1$. 

Initially only the dummy root is \emph{active}. When Black claims $x$,
White's response is chosen by the deepest active vertex whose set
$E_v$ contains $x$. A smaller region can become active when its own
requirements become endangered (in the sense of Definition \ref{def:activate}). Once it becomes active, this transfer
of responsibility is permanent, even if its requirements subsequently
become safe. The precise activation threshold will be $1/8$ for the
potential introduced below. This is a bookkeeping rule, not a
restriction on Black: Black can always claim any unused point of $E$.

When a descendant $u$ becomes active, every ancestor $v$ permanently
stops \emph{counting on future White points} inside $E_u$. The points
have not left the game; they now belong to the smaller region's
accounting. Accordingly, the portion of the board still counted at $v$
is
\begin{equation}\label{eq:new-effective}
E_v^*=E_v\setminus\bigcup_{\substack{u\text{ active}\\u\text{ proper descendant of }v}}E_u.
\end{equation}
 Each local requirement \(Q\in\mathcal A_v\) comes with
\begin{itemize}
\item a quota \(a_Q\ge0\), meaning that White must eventually claim at least \(a_Q\) points of \(Q\). The value $a_Q$ is fixed before the algorithm.
\item let $r_Q(t)$ be its quota minus
the number of White claims credited to it so far, and
\item  let $n_Q(t)$ be the
number of unused points in $Q\cap E_v^*$. 
\end{itemize}
A White claim is credited to
$Q$ precisely when the point is in $Q\cap E_v^*$ at the time of the
claim; earlier credits are never withdrawn. We allow $r_Q<0$ when the
quota has been exceeded.

If a descendant becomes active, then $n_Q$ may suddenly decrease while
$r_Q$ stays unchanged. The ancestor still has the same quota, but may
rely on fewer points to achieve it. \emph{This is the only complication
that our potential will need to pay for.}

\subsection{Requirements for the potential function.}

Suppose each vertex $v$ has a nonnegative potential $P_v$,
which is a function of the current game state. For
now $P_v$ is \emph{unspecified}: think of it as a measure of how close
the requirements in this region are to being impossible. We want to
keep every $P_v$ well below $1$. 
In particular,
$P_v$ depends on the current values of the local quantities
$r_Q,n_Q$, on which descendants are active, and, recursively,
on the potentials of the descendants still monitored by $v$.

Once $P_v$ is fixed, the score of an unused point $x$ is
{defined from the potential}: keeping the active vertices
unchanged, let
\begin{equation}
  s_v(x)
=
P_v(\text{state after Black claims }x)
-
P_v(\text{current state}).   \label{eq:svx}
\end{equation}
Since the current game state changes after every move, the
scores $s_v(x)$ are recomputed after every move as well.
When $x\notin E_v^*$, set $s_v(x)=0$.

Write $M_v$ for the largest current score $s_v(x)$ over
unused $x\in E_v^*$, or $0$ if no such point remains, and set
\[
I_v=P_v+M_v.
\]
The extra $M_v$ is a reserve for the next Black move:
we must allow for Black to act before White can respond.

\begin{defn}[Activation]\label{def:activate}
    After Black actually claims a point, first calculate all potentials
\emph{without yet changing activation}. If an inactive vertex now has
potential at least $1/8$, activate the deepest such vertex. 
\end{defn}
Later we
will prove that this one activation also settles any endangered
ancestors. No vertex is deactivated. 

For this plan to work, we ask the potentials to have four properties.
The numerical constants just leave convenient room in the inequalities.

We also assume the following locality properties. With the active
vertices fixed, a Black or White claim at a point
\(x\notin E_v^*\) leaves \(P_v\) unchanged. Moreover, activating
a vertex \(u\), without any further claim, can change potentials
only at strict ancestors of \(u\). In particular, activation
does not change \(P_u\) itself.

\medskip
\noindent\textbf{(P1) Small initial potentials and detection of
unmet quotas.}
Initially, every vertex has potential at most \(1/16\).

More importantly, the potential must detect failure of any local
requirement. Recall that each \(Q\in\mathcal A_v\) has a prescribed
quota \(a_Q\), and that \(r_Q\) is the number of additional White
points still needed to meet this quota. We require that, at the
end of the game, if \(r_Q>0\) for any \(Q\in\mathcal A_v\),
then \(P_v\ge1\).

The key point is that keeping every potential strictly below \(1\) throughout
the game is sufficient to ensure that all local quotas are met.

\medskip
\noindent\textbf{(P2) One move has a small effect on the potential.}
There is a fixed constant $0<\sigma\le1/16$, to be
chosen only at the end of the proof, with the following
property. Suppose that all potentials in the subtree
rooted at $v$ are at most $1/4$. Then every hypothetical
Black claim at an unused point $x\in E_v^*$ satisfies
\begin{equation}\label{eq:new-small}
0\le s_v(x)\le \sigma P_v.
\end{equation}
A White claim never increases any potential.

Thus a single Black move cannot make the danger of a
region jump by a large factor. In particular,
$M_v\le\sigma P_v$.

\medskip

\noindent\textbf{(P3) White's advantage exceeds the possible
growth of the remaining scores.}
Keep the active vertices fixed, and suppose again that
all relevant potentials are at most $1/4$.

If White claims an unused point $y\in E_v^*$, then
White's claim decreases $P_v$ by at least
$(1+8\sigma)s_v(y)$.

Moreover, after White claims $y$ and the scores are
recomputed in the new state, the Black score of every
other remaining point $x\in E_v^*$ is at most
$(1+2\sigma)s_v(x)$,
where on the right-hand side $s_v(x)$ denotes its score
immediately before White's claim.

Thus White's move may slightly increase some remaining
scores, but White's gain at the chosen point is strictly
larger than this possible increase. This is enough for
the greedy strategy below.

This suggests a simple \textbf{White strategy}: after Black's move, White chooses
the remaining point \(y\) maximizing \(s_v(y)\). In other words,
White claims the point that would be most damaging if Black took it next.

\medskip
\noindent\textbf{(P4) Delegation protects the ancestor's reserve.}
Suppose that an inactive descendant $u$ has pre-move
potential at least $1/12$, and Black next claims an unused
point of $E_u^*$. Compare the state before this Black claim
with the state after the claim and the activation of $u$,
but before White's response.

For every ancestor $v$ of $u$, we require
\[
P_v'\le P_v
\qquad\text{and}\qquad
I_v'\le I_v,
\]
where $I_v=P_v+M_v$ is the reserve defined above.
Ancestors already separated from $u$ by an active vertex
are unaffected.

Thus activation need not decrease every surviving score
individually. It is enough that it does not increase the
total reserve available for the next Black move.

\paragraph{White's strategy.}
We summarize White's strategy here. 

Initially only the root is active. Each round proceeds as follows.

\begin{enumerate}
\item After Black claims an unused point \(x\), update the
potentials while keeping the active vertices unchanged.

\item If an inactive vertex now has potential at least \(1/8\),
activate the deepest such vertex \(u\). Its ancestors then
exclude \(E_u\) from their effective boards, as in
\eqref{eq:new-effective}. Their outstanding quotas remain
unchanged, and earlier White credits are retained.
Recompute the potentials and response scores in this updated
state.

\item Let \(v\) be the deepest active vertex with \(x\in E_v\).
White claims an unused point \(y\in E_v^*\) maximizing
\(s_v(y)\). If no unused point remains in \(E_v^*\), White
passes. Break ties using a fixed ordering of \(E\).

\item Update the accounting for White's claim: every local
requirement containing \(y\) receives a credit if its vertex
still counts \(y\). Update the remaining populations,
outstanding quotas, and potentials accordingly.
\end{enumerate}

A vertex never becomes inactive again. Its potential may
subsequently decrease, but it continues to handle Black moves
in its region unless a deeper active descendant handles them.
We prove below that the vertices crossing the activation
threshold lie on one ancestor chain, and that activating the
deepest one is sufficient.

We now show that any potential satisfying the locality assumptions and
\textup{(P1)--(P4)} yields a winning strategy.

\begin{lem}\label{lem:new-abstract}
If the potentials satisfy the locality properties above and
\textup{(P1)--(P4)}, then White has one causal strategy meeting
every local quota.
\end{lem}

\begin{proof}
After Black claims $x$, first compute the potentials without changing
which vertices are active. If an inactive vertex has reached $1/8$,
activate the deepest such vertex $u$. Next let $v$ be the deepest
active vertex containing $x$. White claims an unused point of $E_v^*$
with largest $s_v(y)$, if such a point exists, and otherwise passes.
Fix an arbitrary ordering to break ties. All these decisions depend
only on the current state.

We show that after every completed round, we would keep
\begin{equation}\label{eq:new-invariant}
I_v\le\frac14\quad\text{for active }v, \text{ and }
P_v<\frac18\quad\text{for inactive }v.
\end{equation}

For a fixed round, we use no superscript for quantities just before
Black's move, a superscript \(B\) for quantities after Black's move
but before any activation or White response, and a superscript \(W\)
for quantities after White's response. Thus, if Black claims \(x\),
\[
P_v^B=P_v+s_v(x).
\]
We write \(M_v^B\) and \(M_v^W\) for the maximum Black scores
in the post-Black and post-White states, respectively, and set
$I_v^B=P_v^B+M_v^B, 
I_v^W=P_v^W+M_v^W.$

Initially \eqref{eq:new-invariant} follows from (P1)--(P2).
Moreover, the two bounds imply \(I_v\le1/4\) for every vertex.
Indeed, for an inactive vertex,
$I_v=P_v+M_v\le(1+\sigma)P_v
<\frac{1+\sigma}{8}<\frac14$,
while this is part of the invariant for active vertices.
Hence the next Black claim keeps every potential at most \(1/4\):
with the active vertices fixed,
$P_v^B=P_v+s_v(x)\le P_v+M_v=I_v\le\frac14$.
Thus (P2)--(P4) are applicable when White responds.

Suppose first that no region activates. Let \(v\) be the responsible
active vertex. Let \(M_v^B\) be the largest score of a point remaining
unused after Black's claim. If an unused point remains in \(E_v^*\), White takes a maximizer
\(y\), whether or not \(M_v^B=0\).
By (P3),
$P_v^W\le P_v^B-(1+8\sigma)M_v^B$.
Moreover, every point remaining after White's claim had score at most
\(M_v^B\) immediately before that claim, so the second part of (P3)
gives
$M_v^W\le(1+2\sigma)M_v^B$.
Consequently,
$I_v^W =P_v^W+M_v^W \le P_v+s_v(x)
 -(1+8\sigma)M_v^B+(1+2\sigma)M_v^B =P_v+s_v(x)-6\sigma M_v^B\le P_v+M_v=I_v$,
since \(s_v(x)\le M_v\), where \(M_v\) is the maximum score before
Black's move. If no point remains in \(E_v^*\), then
\(M_v^{B}=0\) and the same conclusion follows with a pass.
All other active reserves are unchanged, while a White claim cannot
increase the potential of an inactive vertex. Thus
\eqref{eq:new-invariant} is preserved when no activation occurs.

Now suppose that a region activates. Only vertices on the ancestor
chain of Black's point can have increased. Choose the deepest inactive
vertex \(u\) whose potential crossed \(1/8\). Before Black's move,
\(P_u<1/8\), while by (P2), we have 
$P_u^{B}\le(1+\sigma)P_u$.
Since \(P_u^{B}\ge1/8\), we have 
$P_u\ge\frac{1}{8(1+\sigma)}>\frac1{12}$.
Thus (P4) applies. Activating \(u\) returns the potentials and reserves
of all its relevant ancestors to values no larger than they had before
Black's move. In particular, every ancestor that crossed the threshold
is safe again, so no second activation is needed.

It remains to check the newly active vertex \(u\). Activation itself
does not change \(P_u\). Moreover,
$P_u^{B}<\frac{1+\sigma}{8}$.
Applying (P2) in the post-Black state gives
$M_u^{B}\le\sigma P_u^{B}$,
and hence
$I_u^{B}\le(1+\sigma)P_u^{B}
<\frac{(1+\sigma)^2}{8}<\frac14$.
After activation \(u\) is the responsible active vertex for this Black
move. The same calculation from (P3) used in the no-activation case
shows that White's response cannot increase \(I_u\). All remaining
vertices satisfy the invariant as well. This proves
\eqref{eq:new-invariant} after every round.

At exhaustion every potential is at most $1/4$. An unmet quota would
make its vertex's potential at least $1$ by (P1), a contradiction.
Thus all quotas are met.
\end{proof}

We are left to construct the local requirements and
then build a potential with the four desired properties.

\subsection{Constructing the local requirements}

We now construct local requirements from axis-parallel lattice
rectangles. Only the spatial cubes are required to be dyadic; the
rectangles to be protected need not be dyadically aligned.

Fix $0<\varepsilon \le 1/4$. Use the retained scales $H_j=2^{h_j}$,
whose sizes will be chosen at the end of the proof. Take $J$ large
enough that the longest side of every requested rectangle is at most
$H_J$.

\paragraph{Passing populated pieces downward.}
At each scale $j$, choose thresholds $0<\alpha_j<\gamma_j$.
For every scale-$j$ cube $C_v$, consider \emph{all} nonempty
axis-parallel lattice rectangles $R\subseteq C_v$.
The pair $(v,R)$ is \emph{eligible} if $|E\cap R|\ge\gamma_j$.
We consider all such pairs, not just the rectangles in the original
packet; the same geometric rectangle at different scales is treated
as a different label.

For an eligible rectangle $R\subseteq C_v$, search downward
through the descendants of $v$. Start with the children of
$v$. If a cube $C_u$ lies at level $l(u)$, compare its
intersection with the threshold belonging to that level:
\[
|E\cap R\cap C_u|\ge\gamma_{l(u)}.
\]

If this inequality holds, pass $R\cap C_u$ to $u$ and
stop searching below $u$ at this step. If it fails, continue
the search among the children of $u$. Cubes disjoint from
$R$ are ignored.

Thus a piece may be passed directly to a descendant several
levels below $v$. The selected cubes are pairwise disjoint,
because along any branch we select only the first cube that
meets its own threshold. The points not belonging to any
selected cube remain at $v$ and form the local remainder
$Q_R$.

This is similar to the dense/frozen dyadic partition
procedure of Fox and Wei~\cite{FoxWei2018, FoxWeiSODA2017}: sufficiently
populated regions continue to be refined, while sparse
regions stop. The important difference from the simpler
immediate-child procedure is that failure at one scale does
not end the search. A smaller descendant may still meet the
threshold appropriate to its own level.

\begin{figure}[ht]
\centering
\begin{tikzpicture}[
    x=1.07cm,y=1.07cm,
    line cap=round,line join=round,
    frame/.style={draw=cubeedge,line width=0.9pt},
    coarsegrid/.style={draw=cubeedge!32,line width=0.48pt},
    finegrid/.style={draw=cubeedge!28,line width=0.28pt},
    rectborder/.style={draw=rectedge,line width=1.25pt},
    chosen/.style={draw=passedge,line width=1.35pt,densely dashed},
    nextarrow/.style={-{Latex[length=2mm]},draw=passedge,line width=1.1pt}
]

\def\fixedpoints{%
  0.28/1.36,0.72/1.75,%
  1.06/1.05,1.19/1.18,1.56/1.30,1.69/1.45,%
  1.31/1.81,1.44/1.94,1.81/1.56,1.94/1.68,%
  1.38/1.38,1.87/1.12,%
  2.45/1.37,%
  3.30/1.54,3.44/1.67,3.80/1.05,3.94/1.18,%
  3.12/1.36,3.63/1.88}
\newcommand{\drawE}{%
  \foreach \xp/\yp in \fixedpoints
    \fill[ink] (\xp,\yp) circle[radius=1.50pt];%
}

\node[font=\bfseries\small,text=ink] at (2,4.74)
  {Level $j$};
\node[font=\scriptsize,text=cubeedge] at (2,4.36)
  {eligible rectangle};

\node[font=\bfseries\small,text=ink] at (7.45,4.74)
  {Level $j-1$};
\node[font=\scriptsize,text=cubeedge] at (7.45,4.36)
  {test threshold $\gamma_{j-1}=8$};

\node[font=\bfseries\small,text=ink] at (12.90,4.74)
  {Level $j-2$};
\node[font=\scriptsize,text=cubeedge] at (12.90,4.36)
  {search inside the failed pieces};

\begin{scope}[shift={(0,0)}]
  \fill[pending] (0,1) rectangle (4,2);

  \draw[frame] (0,0) rectangle (4,4);
  \draw[rectborder] (0,1) rectangle (4,2);

  \node[font=\scriptsize,text=rectedge] at (2,2.23) {$R$};
  \node[font=\scriptsize,text=cubeedge] at (2,3.48) {$C_j$};

  \drawE
\end{scope}

%
\begin{scope}[shift={(5.45,0)}]

  \fill[pending] (0,1) rectangle (1,2);
  \fill[pending] (2,1) rectangle (3,2);
  \fill[pending] (3,1) rectangle (4,2);

  \fill[passed] (1,1) rectangle (2,2);

  \draw[coarsegrid,step=1] (0,0) grid (4,4);
  \draw[frame] (0,0) rectangle (4,4);
  \draw[rectborder] (0,1) rectangle (4,2);
  \draw[chosen] (1,1) rectangle (2,2);

  \node[font=\scriptsize,text=rectedge] at (2,2.23) {$R$};
  \node[font=\scriptsize,text=cubeedge] at (2,3.48) {$C_j$};

  \node[font=\tiny,text=cubeedge] at (0.5,0.72) {$2<8$};
  \node[font=\tiny,text=passedge] at (1.5,0.72) {$10\ge8$};
  \node[font=\tiny,text=cubeedge] at (2.5,0.72) {$1<8$};
  \node[font=\tiny,text=cubeedge] at (3.5,0.72) {$6<8$};

  \drawE
\end{scope}

%
\begin{scope}[shift={(10.90,0)}]

  \fill[pending] (0,1) rectangle (1,2);
  \fill[pending] (2,1) rectangle (3,2);
  \fill[pending] (3,1) rectangle (4,2);

  \fill[passed] (1,1) rectangle (2,2);

  \draw[finegrid,step=0.25] (0,1) grid (1,2);
  \draw[finegrid,step=0.25] (2,1) grid (3,2);
  \draw[finegrid,step=0.25] (3,1) grid (4,2);

  \fill[retained] (0.25,1.25) rectangle (0.50,1.50);
  \fill[retained] (0.50,1.75) rectangle (0.75,2.00);

  \fill[retained] (2.25,1.25) rectangle (2.50,1.50);

  \fill[passed] (3.25,1.50) rectangle (3.50,1.75);
  \fill[passed] (3.75,1.00) rectangle (4.00,1.25);

  \fill[retained] (3.00,1.25) rectangle (3.25,1.50);
  \fill[retained] (3.50,1.75) rectangle (3.75,2.00);

  \draw[coarsegrid,step=1] (0,0) grid (4,4);
  \draw[finegrid,step=0.25] (0,1) grid (1,2);
  \draw[finegrid,step=0.25] (2,1) grid (3,2);
  \draw[finegrid,step=0.25] (3,1) grid (4,2);

  \draw[frame] (0,0) rectangle (4,4);
  \draw[rectborder] (0,1) rectangle (4,2);

  \draw[chosen] (1,1) rectangle (2,2);
  \draw[chosen] (3.25,1.50) rectangle (3.50,1.75);
  \draw[chosen] (3.75,1.00) rectangle (4.00,1.25);

  \node[font=\scriptsize,text=rectedge] at (2,2.23) {$R$};
  \node[font=\scriptsize,text=cubeedge] at (2,3.48) {$C_j$};

  \drawE
\end{scope}

\draw[nextarrow] (4.13,3.06) -- (5.30,3.06);
\node[font=\scriptsize,text=passedge] at (4.715,3.36)
  {test $\gamma_{j-1}$};
\node[font=\tiny,text=cubeedge] at (4.715,2.78)
  {select or continue};

\draw[nextarrow] (9.58,3.06) -- (10.75,3.06);
\node[font=\scriptsize,text=passedge] at (10.165,3.36)
  {test $\gamma_{j-2}$};
\node[font=\tiny,text=cubeedge] at (10.165,2.78)
  {inside failed pieces};

\node[font=\scriptsize,text=cubeedge] at (2,-0.43)
  {19 points in $R$};

\node[font=\scriptsize,text=cubeedge] at (7.45,-0.43)
  {10 selected; 9 searched further};

\node[font=\scriptsize,text=cubeedge] at (12.90,-0.43)
  {4 more selected; 5 remain in $Q_R$};

\fill[pending] (0.10,-1.19) rectangle (0.38,-0.93);
\draw[frame] (0.10,-1.19) rectangle (0.38,-0.93);
\node[anchor=west,font=\scriptsize,text=ink] at (0.52,-1.06)
  {search continues};

\fill[passed] (4.35,-1.19) rectangle (4.63,-0.93);
\draw[frame] (4.35,-1.19) rectangle (4.63,-0.93);
\node[anchor=west,font=\scriptsize,text=ink] at (4.77,-1.06)
  {selected and passed down};

\fill[retained] (8.75,-1.19) rectangle (9.03,-0.93);
\draw[frame] (8.75,-1.19) rectangle (9.03,-0.93);
\node[anchor=west,font=\scriptsize,text=ink] at (9.17,-1.06)
  {left in $Q_R$};

\fill[ink] (13.48,-1.06) circle[radius=1.7pt];
\node[anchor=west,font=\scriptsize,text=ink] at (13.64,-1.06)
  {$E$};

\end{tikzpicture}

\caption{The downward search for an eligible rectangle.
In this toy example, the four level-$(j-1)$ pieces contain
$2,10,1,$ and $6$ points, and the threshold is
$\gamma_{j-1}=8$. The 10-point piece is selected immediately,
so the search stops below it. The other three pieces fail
and are searched further. At level $j-2$, where the
illustrative threshold is $\gamma_{j-2}=2$, two descendants
of the 6-point piece become eligible and are selected.
Thus a piece can be passed directly to a descendant several
retained levels below its parent. For illustration,
level $j-2$ is taken to be the bottom retained level, so
the five points not contained in any selected cube form
the local remainder $Q_R$.}

\label{fig:dyadicpartition}
\end{figure}
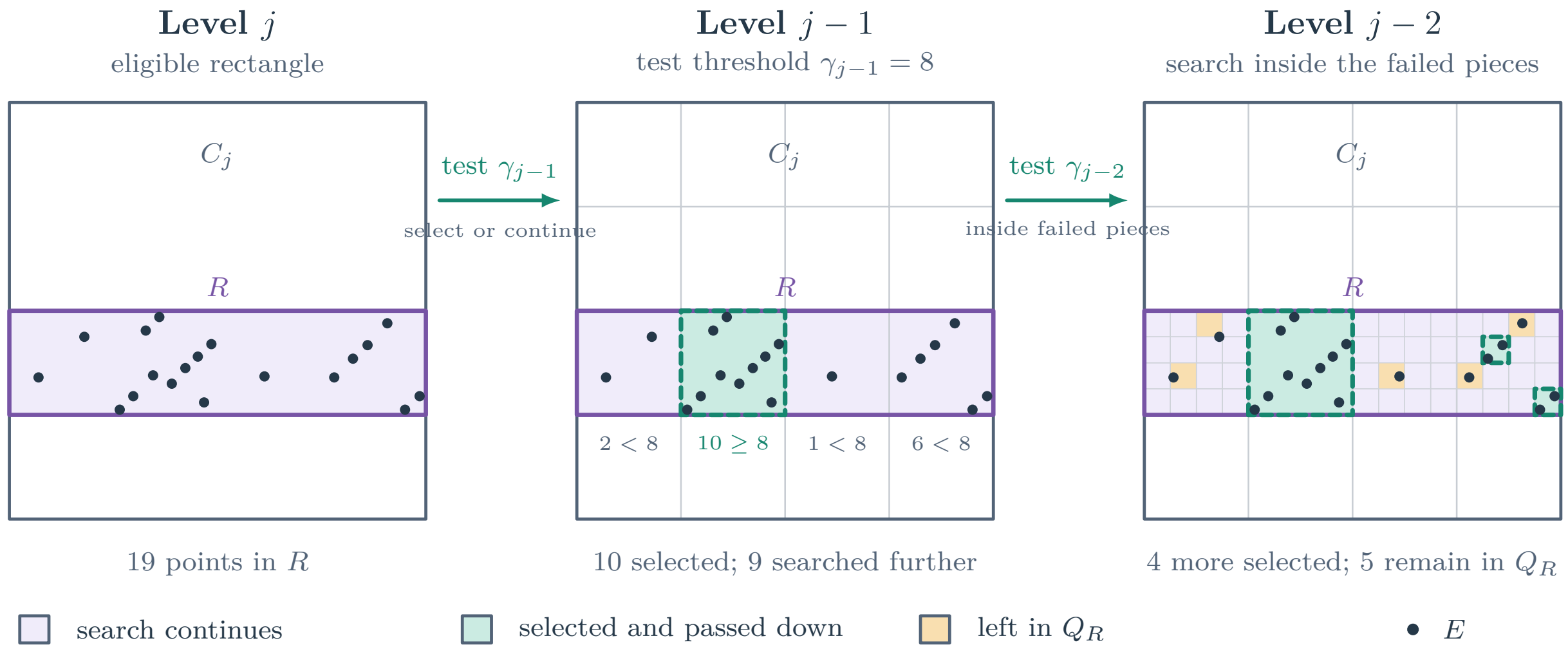

Figure~\ref{fig:dyadicpartition} illustrates the downward search,
including pieces passed directly across multiple retained levels.

\paragraph{Local requirements and discarded sets.} Let \(Q_R\) be the union of the points remaining at the current
level. If \begin{equation}
    |Q_R|\ge\alpha_j, \label{eq:alphaj}
\end{equation}
we add \(Q_R\) to the local
requirement family \(\mathcal A_v\) of the cube containing \(R\).
Otherwise, we {\it discard} this remainder. At level zero, there
are no smaller cubes, so \(Q_R=E\cap R\). This quota makes sense, since we want White to cover roughly half of the points in every axis-parallel non-degenerate rectangle. 

This construction has two useful properties. First, if
$Q$ is a local requirement and $u$ is any proper descendant
lying at level $l(u)$, then
\begin{equation}\label{eq:allscale-light}
|Q\cap E_u|<\gamma_{l(u)}.
\end{equation}
Indeed, if the intersection contained at least $\gamma_{l(u)}$
points, the downward search would have selected either
$u$ or an earlier qualifying cube on the path to $u$;
those points therefore could not remain in $Q$.

Second, every piece passed to a descendant is an
axis-parallel rectangle eligible at the receiving level,
so the recursion is well defined.

\paragraph{Requirements on \(\alpha_j\) and \(\gamma_j\).} We now identify the conditions required of the thresholds \(\alpha_j\) and \(\gamma_j\); their explicit values will be chosen in the final verification. Recall that \(\alpha_j\) determines whether a local remainder
is large enough to be protected, whereas \(\gamma_j\) determines whether a rectangle is eligible at level \(j\).

First, we must ensure that discarding small remainders
loses only a small fraction of the points of each rectangle. We claim it suffices to choose small $\delta_j$ such that $\alpha_j / \gamma_j \leq \delta_j$ and $\sum_j \delta_j \leq \varepsilon / 4$.  
\begin{claim}\label{claim:discardsmall}
Suppose $\alpha_j/\gamma_j\le\delta_j$ for every $j$,
where $\sum_j\delta_j\le\varepsilon/4$.
Then, for every eligible rectangle $R$ within a spatial cube, the total
number of points in all discarded remainders $Q_{R'}$,
where $R'$ ranges over $R$ and all its descendants in the
recursive decomposition, is at most
$\frac{\varepsilon}{4}|E\cap R|$.
\end{claim}
\begin{proof}
Indeed, fix one eligible 
rectangle \(R\) within a spatial cube and follow its recursive decomposition.
At any scale \(j\), there may be many eligible rectangles
\(R_1,\ldots,R_m\) being processed, each one we can compute \(|Q_{R_i}|\). Notice the \(R_i\) are pairwise
disjoint subsets of the original rectangle \(R\), since
they arise by passing down intersections with disjoint
child cubes.
If \(Q_{R_i}\) is discarded, then
\[
\sum_{i: \ Q_{R_i} \text{ discarded }}|Q_{R_i}|<\sum_{i: \ Q_{R_i} \text{ discarded }} \alpha_j
\le \sum_{i: \ Q_{R_i} \text{ discarded }}\delta_j\gamma_j
\le\delta_j \sum_{i: \ Q_{R_i} \text{ discarded }}|E\cap R_i| \leq \delta_j |E \cap R|.
\]
Summing over all scales $j$ gives a total discarded population
of at most \(\varepsilon|E\cap R|/4\).
\end{proof}

Second, assign each retained local requirement \(Q\) the quota
\begin{equation}
    a_Q=\left(\frac12-\frac{\varepsilon}{2}\right)|Q|. \label{eq:localrequirementaQ}
\end{equation}

\begin{claim}\label{claim:quotassuffice}
Suppose $\alpha_j/\gamma_j\le\delta_j$ for every $j$,
where $\sum_j\delta_j\le\varepsilon/4$. And suppose every retained local requirement $Q$ has quota
$a_Q$ defined in \eqref{eq:localrequirementaQ}. If White meets all local quotas,
then every eligible rectangle $R$ within a spatial cube satisfies
\begin{equation}\label{eq:new-dyadic-goal}
|W\cap R|\ge
\left(\frac12-\varepsilon\right)|E\cap R|.
\end{equation}
\end{claim}

\begin{proof}
Fix an eligible rectangle $R$. The recursive
decomposition partitions $E\cap R$ into retained and
discarded remainders, all pairwise disjoint.
By Claim~\ref{claim:discardsmall}, at most an $\varepsilon/4$
fraction of $E\cap R$ is discarded. Thus the retained
remainders contain at least $(1-\varepsilon/4)|E\cap R|$ points.
Since White meets the quota of every retained remainder,
and these remainders are disjoint, we obtain
$|W\cap R|
\ge \left(\frac12-\frac{\varepsilon}{2}\right)
     \left(1-\frac{\varepsilon}{4}\right)|E\cap R|
\ge \left(\frac12-\varepsilon\right)|E\cap R|.$
\end{proof}

\paragraph{From local eligibility to arbitrary rectangles.}
The preceding two claims protect every eligible rectangle contained
in a spatial cube. The passage to an arbitrary rectangle needs only
the following scale bound, without a dyadic decomposition of the
rectangle into many pieces.

\begin{prop}\label{prop:cf-conditional}
Suppose that for every $0<\varepsilon\le1/4$ we can choose retained
spatial scales and thresholds $\alpha_j,\gamma_j$ such that:
\begin{enumerate}
\item[(G1)] $\sum_{j\ge0}\frac{\alpha_j}{\gamma_j}
\le\frac{\varepsilon}{4}$.
  \item[(G2)] there is a constant
$c_{d,\varepsilon,\tau}$, independent of $E$ and the
packet, such that
\[
\gamma_0\le c_{d,\varepsilon,\tau},
\qquad
\gamma_j\le
c_{d,\varepsilon,\tau}h_{j-1}^{1+\tau}
\quad(j\ge1);\]
\item[(G3)] the resulting local requirements, with quotas
  \eqref{eq:localrequirementaQ}, admit potentials satisfying
  \textup{(P1)--(P4)}.
\end{enumerate}
Then Lemma~\ref{lem:local-rectangle-strategy} holds for every finite
packet of arbitrary axis-parallel rectangles, using a single White
strategy for the packet.
\end{prop}

\begin{proof}
By (G3) and Lemma~\ref{lem:new-abstract}, White meets every local
quota. By (G1) and Claims~\ref{claim:discardsmall} and
\ref{claim:quotassuffice}, White therefore gets at least a
$1/2-\varepsilon$ fraction in every eligible rectangle contained
in a spatial cube.

Fix a desired accuracy $0<\eta\le1/4$ and apply this with
$\varepsilon=\eta/2$. Let $R$ be any rectangle in the packet,
and choose the smallest $j$ for which its longest side is at most
$H_j$. Since every side of $R$ has length at most $H_j$, it meets
at most $2^d$ scale-$j$ cubes. Each nonempty intersection
$S_v=R\cap C_v$ is itself an axis-parallel rectangle contained
in $C_v$.

If $|E\cap S_v|\ge\gamma_j$, this intersection is eligible and
receives White fraction at least $1/2-\eta/2$. The other
intersections contain fewer than $2^d\gamma_j$ points altogether.
For $j\ge1$, the longest side of $R$ exceeds $H_{j-1}$, so
$\log_2V(R)>h_{j-1}$. By (G2), and also for $j=0$,
\[
\gamma_j\le c_{d,\eta/2, \tau}(1+\log_2V(R))^{1 + \tau}.
\]
Thus, if $|E\cap R|\ge C_{d,\eta, \tau}(1+\log(2+V(R)))^{1 + \tau}$
for sufficiently large $C_{d,\eta, \tau}$, the ineligible pieces
together contain at most $\eta|E\cap R|$ points. Summing the
guarantees on the other pieces gives
\begin{equation}\label{eq:cf-final}
|W\cap R|\ge\left(\frac12-\frac\eta2\right)
(1-\eta)|E\cap R|\ge\left(\frac12-\eta\right)|E\cap R|.
\end{equation}
The same strategy works for every rectangle in the packet.
For $\eta>1/4$, apply the result with accuracy $1/4$.
\end{proof}

\emph{Thus the complete finite rectangle game follows from
(G1)--(G3).} We have not yet chosen the potentials or any of the
thresholds. We do that now, and verify the assumptions in the order
suggested by their roles.

\paragraph{Relation to earlier methods.}
Before constructing the potential, we explain how the argument combines
two ideas that appeared separately in earlier work.

The first is the potential-function method for discrepancy
games.  In classical balancing games, going back to Spencer
\cite{SpencerBalancingGames} and in particular to the
discrepancy-game framework of Alon, Krivelevich, Spencer,
and Szabó \cite{AKSSDiscrepancyGames}, one controls a family
of competing requirements by an exponential or
hyperbolic-cosine potential.  The potential determines which
available points are currently dangerous: after the potential
has been chosen, the danger of a point is measured by the
increase that would result if the adversary claimed it.
The balancing player then chooses a response that offsets the
largest potential increase.  Our basic exponential weight and
the rule of responding at a most dangerous point follow this
same philosophy.  As in these classical arguments, we first
formulate the one-move inequalities that the potential should
satisfy and only afterward choose an exponential function
realizing them.

The second ingredient is multiscale rather than game
theoretic.  The downward decomposition is closer to the
dyadic partitioning and freezing procedure used by Fox and
Wei \cite{FoxWei2018}.  A rectangle is examined on a hierarchy
of dyadic spatial scales.  If a piece is sufficiently
populated at its current scale, it is passed to that scale;
otherwise the search continues to smaller cubes.  Thus the
decision to stop is made using the threshold appropriate to
the scale where the piece is encountered.  In particular, a
piece may fail the test at one level but have a descendant
that passes the test at a smaller level.

The additional feature needed here is the interaction between
these two mechanisms.  A standard discrepancy-game potential
has a fixed collection of requirements and does not need to
transfer responsibility between different spatial scales.
Here a smaller cube may become dangerous during the game and
take responsibility for its region.  Its ancestors then stop
counting on the unused points in that cube.  Consequently,
the potential at a coarse scale cannot depend only on its own
local deficit: it must also anticipate the possible future
activation of its descendants.  This is the reason for the
recursive descendant correction in our potential.  The
correction acts as an allowance that is accumulated while a
descendant is being monitored and is spent when responsibility
is transferred to that descendant.

Thus the proof may be viewed as coupling a classical
discrepancy-game potential with a dyadic freezing hierarchy.
The former supplies the greedy ``most dangerous point''
response, while the latter determines the spatial scale at
which responsibility should be assigned.  The recursive
allowance is what makes these two parts compatible.

\subsection{Constructing the potential and choosing the thresholds}

We now construct potentials satisfying (P1)--(P4). Rather than
guessing their formula, we work backward from the properties
needed by White's strategy.

\paragraph{Deriving the potential from the required inequalities.}
We seek a potential that is a sum of nonnegative weights, one
for each local requirement:
\[
P_v=\sum_{Q\in\mathcal A_v}p_Q.
\]
For the dummy root \(o\), set \(\mathcal A_o=\emptyset\)
and \(P_o=0\).

We first write down the inequalities that these weights
should satisfy, by the conditions in Proposition \ref{prop:cf-conditional}.
\begin{enumerate}
    \item 
\emph{The Black--White comparison (P3).}
Fix the current game state and an unused point $x\in E_v^*$,
keeping the active vertices unchanged. 

We regard $p_Q$ as a function of $r_Q$, $n_Q$,
and the overlap data $m_{Qw}=|Q\cap E_w|$ and
potentials $P_w$ of all descendants $w$ still
monitored by $v$. Let $\mathcal D_v$ be the set of proper descendants
$w$ of $v$ such that no vertex on the path from
$v$ to $w$, other than $v$, is active.
So
\[
p_Q=p_Q\bigl(r_Q,n_Q;
      (m_{Qw},P_w)_{w\in\mathcal D_v}\bigr).
\]

Under a hypothetical Black claim at $x$, the outstanding
quota $r_Q$ is unchanged, while $n_Q$ decreases by $1_Q(x)$.
Under a hypothetical White claim, both $r_Q$ and $n_Q$
decrease by $1_Q(x)$. The descendant potentials may also change.
Thus
\[
\begin{aligned}
p_Q^{B_x}
&=p_Q\bigl(r_Q,n_Q-1_Q(x);
      (m_{Qw},P_w^{B_x})_{w\in\mathcal D_v}\bigr),\\
p_Q^{W_x}
&=p_Q\bigl(r_Q-1_Q(x),n_Q-1_Q(x);
      (m_{Qw},P_w^{W_x})_{w\in\mathcal D_v}\bigr),
\end{aligned}
\]
where $P_w^{B_x}$ and $P_w^{W_x}$ are the descendant
potentials after the respective hypothetical claims,
each starting from the same current state.

Since $P_v=\sum_{Q\in\mathcal A_v}p_Q$, the basic
Black--White comparison asks that the decrease under a
White claim dominate the increase under the corresponding
Black claim. We first seek the stronger termwise comparison
\begin{equation}\label{eq:design-P3}
p_Q^{B_x}+p_Q^{W_x}\le2p_Q.
\end{equation}
This will determine the basic form of the weight.
After adding the descendant correction, we will return
to the full (P3), including the strict White advantage
and the controlled growth of the remaining scores.

\item \emph{The delegation condition (P4).}

Suppose Black claims an unused point $x$, triggering
the activation of a descendant $u$. Let $k_Q$ be
the number of usable points removed from $Q$ by
the Black claim and activation. Then $r_Q$ is unchanged
and $n_Q$ decreases by $k_Q$.

Let $\mathcal D_v'$ be the descendants still monitored
after activation, and let $P_w'$ denote their new potentials.
A sufficient condition for the potential part of (P4) is
\begin{equation}\label{eq:design-P4}
p_Q\bigl(r_Q,n_Q-k_Q;
    (m_{Qw},P_w')_{w\in\mathcal D_v'}\bigr)
\le
p_Q\bigl(r_Q,n_Q;
    (m_{Qw},P_w)_{w\in\mathcal D_v}\bigr).
\end{equation}

\item Finally, (P1) requires an unmet quota at exhaustion to
have weight greater than one. For (P2)--(P3), we also need
small Black increments and controlled growth of the remaining
point scores. We will check these conditions after choosing
the weights.
\end{enumerate}
\medskip
\paragraph{Designing the potential weight.}
We first determine the basic dependence of $p_Q$ on the
outstanding quota $r_Q$ and the number $n_Q$ of usable
points, temporarily ignoring descendant potentials.

Write $z_Q=2r_Q-n_Q$, and suppose first that
$p_Q=f_v(z_Q,n_Q)$. If an unused point of $Q$ is claimed
by Black, then $(z_Q,n_Q)$ changes to
$(z_Q+1,n_Q-1)$; if the same point is claimed by White,
it changes to $(z_Q-1,n_Q-1)$. Hence the basic
Black--White comparison asks for
\[
f_v(z+1,n-1)+f_v(z-1,n-1)\le2f_v(z,n).
\]

We seek a positive function satisfying this discrete
averaging inequality. Motivated by separation of variables
for the heat equation, we try an exponential function.
This leads to the basic weight
\begin{equation}\label{eq:cf-basic-weight}
f_v(z,n)=e^{\lambda_v(z+\mu_v n)},
\end{equation}
provided $\lambda_v\mu_v\ge\log\cosh(\lambda_v)$.
Thus, if descendant regions are ignored, the weight of
$Q$ would be
$\exp\{\lambda_v[2r_Q-(1-\mu_v)n_Q]\}$.

Here $0<\mu_v<1$ and $\lambda_v>0$ will be chosen only
at the end. The parameter $\mu_v$ gives White a slight
advantage over Black, whereas $\lambda_v$ controls the
size of the change caused by one move. Moreover, if
$n_Q=0$ and $r_Q>0$, this weight is greater than one,
as required by (P1).

\medskip
\noindent\emph{Why the basic weight is not sufficient.}
The basic weight does not satisfy the delegation condition
(P4). Suppose a descendant $u$ becomes active and, together
with the triggering Black claim, removes $k$ usable points
from a requirement $Q$ at an ancestor. The outstanding
quota $r_Q$ is unchanged, while $n_Q$ decreases by $k$.
Consequently the basic weight is multiplied by
$e^{\lambda_v(1-\mu_v)k}>1$.

Thus delegation makes the ancestor's requirement more
dangerous: the ancestor has lost usable points without
receiving any additional White credit. We need the weight
to have already reserved some allowance for this possible
loss.

Introduce a nonnegative correction $A_Q$ and replace the
basic weight by $p_Q=f_v(z_Q+A_Q,n_Q)$. If this correction
changes from $A_Q$ to $A_Q'$ when $k$ usable points are
lost, then $p_Q$ does not increase provided
$A_Q-A_Q'\ge(1-\mu_v)k$.

For a descendant $u$, write $m_{Qu}=|Q\cap E_u|$.
Activation of $u$ can remove at most $m_{Qu}$ usable
points from $Q$. It is therefore natural to give $u$
an allowance proportional to $m_{Qu}$ and to its current
danger. We seek a correction of the form
\begin{equation}\label{eq:design-form}
A_Q=\sum_{u\in\mathcal D_v}m_{Qu}g_{l(u)}(P_u),
\end{equation}
where $l(u)$ denotes the level of $u$.

We have not yet chosen $g_j$. Its required properties
follow directly from the proof.

First, when a descendant is dangerous enough to activate,
namely when $P_u\ge1/12$, we need
$g_{l(u)}(P_u)\ge1$. Then removal of its allowance pays
for all of the usable points that the ancestor may lose.

Second, $g_j$ should be increasing and concave. Increasing
ensures that a safer descendant cannot make its ancestor
more dangerous, while concavity allows White's advantage
over Black in a descendant potential to pass to the
corresponding allowance.

Third, the allowance should react much more slowly than
the potential itself. At level $j$, a small relative
change in $P$ should change $g_j(P)$ by only about that
relative change divided by $h_j$. This is the quantitative
saving that will allow the cutoff to approach the
logarithmic scale.

Finally, $g_j(P)$ must be very small when $P$ is initially
very small, so that the allowances of all descendants do
not destroy the small initial potential.

The modified downward decomposition is important precisely
here. By \eqref{eq:allscale-light}, if $u$ lies at level
$j$, then $m_{Qu}<\gamma_j$. Thus the influence of a
descendant is bounded using the threshold of its own
level, where the corresponding $1/h_j$ saving is also
available.

\begin{claim}
\label{claim:cf-potential-existence}
Fix $d\ge1$, $0<\varepsilon\le1/4$, and $\tau>0$.
The retained scales $H_j=2^{h_j}$ and thresholds
$0<\alpha_j<\gamma_j$ can be chosen, depending only on
$d,\varepsilon,\tau$, so that the following hold, and thus all the assumptions in Proposition \ref{prop:cf-conditional} are met. 

\begin{enumerate}
\item The total discarded fraction is controlled: 
$\sum_{j\ge0}\frac{\alpha_j}{\gamma_j}
\le\frac{\varepsilon}{4}$.

\item For some constant $c_{d,\varepsilon,\tau}$, 
$\gamma_0\le c_{d,\varepsilon,\tau}$, and  for all $j \geq 1$, it holds that 
$\gamma_j\le
c_{d,\varepsilon,\tau}h_{j-1}^{1+\tau}$.

\item For the local requirements constructed above, there
are potentials $P_v$ satisfying properties
\textup{(P1)--(P4)}.
\end{enumerate}

The choices are uniform in the finite set $E$, the rectangle
packet, and the number of retained levels.
\end{claim}
Assuming the claim, all three conditions of
Proposition~\ref{prop:cf-conditional} are satisfied.
Therefore Lemma~\ref{lem:local-rectangle-strategy}
follows.

\begin{proof}[Proof of Claim~\ref{claim:cf-potential-existence}]
We can now make all choices at once.

\paragraph{Parameter roadmap.}
The parameters are chosen in the following order:
\[
(d,\varepsilon,\tau)
\longrightarrow
\mu
\longrightarrow
(\sigma,\rho_j,\delta_j)
\longrightarrow
(C,a)
\longrightarrow
B
\longrightarrow
\Lambda_j
\longrightarrow
(\lambda_j,h_j,H_j,\alpha_j,\gamma_j).
\]
Here \(\delta_j\) controls discarded mass, \(\rho_j\) controls the
relative effect of one move, \(h_j\) is the retained logarithmic scale,
\(\lambda_j\) is the exponential sensitivity at that scale,
\(\alpha_j\) is the minimum retained local population, and
\(\gamma_j\) is the eligibility threshold. The constant \(B\) is
chosen last and sufficiently large to make all required inequalities
hold simultaneously.

Put $\mu=\varepsilon/2$ and take
\begin{equation}\label{eq:cf-correction-function}
g_j(P)=(12P)^{1/h_j}.
\end{equation}
This is increasing and concave, and it is at least one
whenever $P\ge1/12$. Moreover, when $P\le1/4$, changing
$P$ by a small relative amount $\rho$ changes $g_j(P)$
by at most a constant times $\rho/h_j$. This is exactly
the saving described above.

For the remaining parameters, set
$\sigma=\mu/64$ as in (P2)--(P3), and put
$\rho_j=\mu/[1024(j+1)^2]$ and
$\delta_j=\varepsilon/[8(j+1)^2]$.
Thus both sequences are summable and
$\rho_j/\delta_j$ is a fixed constant.

Choose
$C=\frac4\varepsilon
\left(2d\log2+1+\log\frac{16}{\varepsilon}\right)$ and 
$a=\left\lceil\frac2\tau\right\rceil$.
Next choose a sufficiently large integer
$B=B(d,\varepsilon,\tau)$ and define
$\Lambda_j=B^{j+1}((j+1)!)^3$. Finally set
\begin{equation}\label{eq:cf-parameters}
\lambda_j=\Lambda_j^{-1},\qquad
h_j=\Lambda_j^a,\qquad
H_j=2^{h_j},\qquad
\alpha_j=C(j+1)h_j\Lambda_j,\qquad
\gamma_j=\alpha_j/\delta_j.
\end{equation}
By increasing \(B\) if necessary, we may assume that
\(h_j\ge10\) and 
$\lambda_j\mu\ge\log\cosh(\lambda_j)$
at every level. We also require
\begin{equation}\label{eq:cf-move-bound}
D_j:=
\lambda_j\left(
1+\mu+
4\sum_{k<j}\frac{\gamma_k\rho_k}{h_k}
\right)
\le\frac{\rho_j}{8}
\qquad(j\ge0).
\end{equation}
These requirements are compatible with the preceding choices. Indeed,
$\frac{\gamma_k\rho_k}{h_k}
=
\frac{C}{256}(k+1)\Lambda_k.$
For \(B\ge2\), the rapid growth
\(\Lambda_{k+1}/\Lambda_k=B(k+2)^3\) gives
$\sum_{k<j}(k+1)\Lambda_k\le2j\Lambda_{j-1}.$
Hence, for \(j\ge1\),
$D_j
\le
\frac{1+\mu+C/32}{B(j+1)^2}$.
The same bound, without the \(C/32\) term, holds for \(j=0\).
Thus \eqref{eq:cf-move-bound} holds after increasing \(B\)
by a constant depending only on \(d,\varepsilon,\tau\).

At a level-$j$ vertex we take $\lambda_v=\lambda_j$ and
$\mu_v=\mu$. Our potential is therefore
\begin{equation}\label{eq:cf-potential}
P_v=\sum_{Q\in\mathcal A_v}
\exp\!\left\{\lambda_j\left[
2r_Q-(1-\mu)n_Q+
\sum_{u\in\mathcal D_v}
m_{Qu}(12P_u)^{1/h_{l(u)}}
\right]\right\}.
\end{equation}
The potentials are defined recursively from the lowest
level upward. These potentials satisfy the locality properties required above.
Indeed, a claim outside \(E_v^*\) affects neither the local
accounting at \(v\) nor any descendant potential monitored by \(v\).
Moreover, activating \(u\) changes the effective sets and monitored
descendants only at strict ancestors of \(u\), leaving all other
potentials unchanged.

\paragraph{Verification of \textup{(P2)--(P3)}.}
We prove slightly stronger estimates by induction on the level.
Let
$R_j=\sum_{k=0}^j\rho_k, 
\eta_j=\mu-16R_j, 
\beta_j=16R_j,$
and set \(\eta_{-1}=\mu\).
Since $\sum_{k\ge0}\rho_k
=
\frac{\mu}{1024}\sum_{m\ge1}\frac1{m^2}
<\frac{\mu}{600}$,
we have for every \(j\),
\begin{equation}\label{eq:cf-slack-bounds}
\eta_j>\frac{\mu}{2}>8\sigma,
\text{ and }
e^{\beta_j}<1+2\sigma.
\end{equation}

We prove by induction. We inductively prove the following four assertions at every level \(j\).
Whenever all relevant subtree potentials are at most \(1/4\), for every
unused point \(x\),
\begin{align}
0&\le P_v^{B_x}-P_v\le\rho_jP_v,
\label{eq:cf-ind-black}\\
0&\le P_v-P_v^{W_x}\le\rho_jP_v,
\label{eq:cf-ind-white}\\
P_v-P_v^{W_x}
&\ge(1+\eta_j)(P_v^{B_x}-P_v),
\label{eq:cf-ind-advantage}
\end{align}
and, after any White claim at \(y\), every remaining point \(x\) satisfies
\begin{equation}\label{eq:cf-ind-score}
s_v^{W_y}(x)\le e^{\beta_j}s_v(x).
\end{equation}

Assume these assertions have been proved at all lower levels.
For 
$g_k(P)=(12P)^{1/h_k}$,
the mean-value theorem, \(h_k\ge10\), and \(P\le1/4\) give
\begin{align}
0\le g_k(P^{B})-g_k(P)
\le\frac{2\rho_k}{h_k},\text{ and }
0\le g_k(P)-g_k(P^{W})
\le\frac{4\rho_k}{h_k}.
\label{eq:cf-g-white}
\end{align}
The case \(P=0\) is immediate.

Fix a level-\(j\) requirement \(Q\) and an unused point \(x\).
Write 
$p_Q^{B_x}=p_Qe^{b_Q}$ and 
$p_Q^{W_x}=p_Qe^{-w_Q}$,
where \(b_Q,w_Q\ge0\). A point belongs to at most one
descendant cube at each lower level, and
\eqref{eq:allscale-light} gives
\(m_{Qu}<\gamma_k\) for a level-\(k\) descendant.
Thus \eqref{eq:cf-g-white} implies
\[
0\le b_Q,w_Q\le
\lambda_j\left(
1+\mu+
4\sum_{k<j}\frac{\gamma_k\rho_k}{h_k}
\right)
=D_j
\le\frac{\rho_j}{8}.
\]
Consequently,
$P_v^{B_x}-P_v
=\sum_Qp_Q(e^{b_Q}-1)
\le 2D_jP_v
\le\rho_jP_v$,
and
$P_v-P_v^{W_x}
=\sum_Qp_Q(1-e^{-w_Q})
\le D_jP_v
\le\rho_jP_v$.
This proves \eqref{eq:cf-ind-black} and
\eqref{eq:cf-ind-white}. In particular, a White move never
increases a potential.

We next prove the Black--White advantage.
For a descendant \(u\) at level \(k<j\), the induction hypothesis gives
$P_u-P_u^{W_x}
\ge
(1+\eta_k)(P_u^{B_x}-P_u)$.
Since \(g_k\) is increasing and concave, we have 
$g_k(P_u)-g_k(P_u^{W_x})
\ge
(1+\eta_k)
\bigl(g_k(P_u^{B_x})-g_k(P_u)\bigr)$.
Also,
$\frac{1+\mu}{1-\mu}\ge1+\mu$.
Since \(\eta_k\ge\eta_{j-1}\) for \(k<j\), it follows term by term
in the exponent that
$w_Q\ge(1+\eta_{j-1})b_Q$.
Using \(b_Q,w_Q\le D_j\),
\[
1-e^{-w_Q}
\ge
(1+\eta_{j-1})
\frac{1-D_j}{1+2D_j}(e^{b_Q}-1)
\ge
(1+\eta_j)(e^{b_Q}-1).
\]
where the last inequality follows from
\(D_j\le\rho_j/8\) and
\(\eta_{j-1}-\eta_j=16\rho_j\).
Multiplying by \(p_Q\) and summing over \(Q\) gives
$P_v-P_v^{W_x}
\ge
(1+\eta_j)(P_v^{B_x}-P_v)$,
which proves \eqref{eq:cf-ind-advantage}.

It remains to control the scores of the remaining points.
Suppose White has claimed \(y\), and let \(x\ne y\) remain unused.
Consider a level-\(k\) descendant \(u\). By
\eqref{eq:cf-ind-white},
$P_u^{W_y}\ge(1-\rho_k)P_u$,
while by the induction hypothesis
$s_u^{W_y}(x)\le e^{\beta_k}s_u(x),
s_u(x)\le\rho_kP_u$.
Since
$g_k'(P)=
\frac{12^{1/h_k}}{h_k}P^{1/h_k-1}$,
concavity and the mean-value theorem give
\[\frac{
g_k(P_u^{W_y}+s_u^{W_y}(x))-g_k(P_u^{W_y})
}{
g_k(P_u+s_u(x))-g_k(P_u)
}
\le
e^{\beta_k}
\left(\frac{1+\rho_k}{1-\rho_k}\right)^{1-1/h_k}
\le e^{\beta_k+3\rho_k}.\]
If the denominator is zero, the numerator is zero as well.

For \(j\ge1\), let 
$G_j=e^{\beta_{j-1}+3\rho_{j-1}}$.
Thus, if \(b_Q\) is the logarithmic Black increment of \(p_Q\)
at \(x\) before White's move and \(\widetilde b_Q\) is the
corresponding increment afterwards, then
$\widetilde b_Q\le G_jb_Q$.
The weight \(p_Q\) itself does not increase under White's move.
Hence
$p_Q^{W_y}(e^{\widetilde b_Q}-1)
\le
p_QG_j e^{(G_j-1)b_Q}(e^{b_Q}-1)$.
Now \(\rho_{j-1}\le4\rho_j\), \(G_j<2\), and
\(b_Q\le D_j\le\rho_j/8\), so
$\log\!\left(
G_j e^{(G_j-1)b_Q}
\right)
\le
\beta_{j-1}+3\rho_{j-1}+\frac{\rho_j}{8}
<
\beta_{j-1}+16\rho_j
=
\beta_j$.
Thus every summand of the Black score grows by at most
\(e^{\beta_j}\), and therefore
$s_v^{W_y}(x)\le e^{\beta_j}s_v(x)$.
For \(j=0\), the Black exponent increment is unchanged by a
White move and the weight itself only decreases, so the same
conclusion is immediate. This completes the induction.

Finally,
$\rho_j\le\rho_0<\sigma,$ 
$\eta_j>8\sigma,$ and 
$e^{\beta_j}<1+2\sigma$
by \eqref{eq:cf-slack-bounds}. Hence
\eqref{eq:cf-ind-black}--\eqref{eq:cf-ind-score} give exactly
(P2) and (P3).

\paragraph{Verification of \textup{(P4)}.}

Suppose Black's claim triggers the activation of a
descendant $u$ whose pre-move potential is at least
$1/12$. For a requirement $Q$ at any relevant ancestor \(v\) of \(u\), the claim
and activation remove at most $m_{Qu}$ usable points.
The resulting increase of the basic deficit is therefore
at most $m_{Qu}$.

On the other hand,
$g_{l(u)}(P_u)\ge1$, so activation removes an allowance
of at least $m_{Qu}$. Thus every weight at the parent
does not increase.

Proceeding upward, all surviving intermediate descendant
potentials have already been shown not to increase.
Since every $g_j$ is increasing, the same comparison
holds at every ancestor. Hence ancestor potentials do
not increase under delegation.

It remains to compare their reserves. Let $y$ be any
point still counted by an ancestor after the activation.
Then $y\notin E_u$. Consider additionally making a
hypothetical Black claim at $y$ before the triggering
move. This does not alter the potential of $u$, so the
same delegation comparison still applies. The two Black claims commute, and therefore
$(P_v')^{B_y}\le P_v^{B_y}$
for every unused point \(y\) still counted by \(v\) after the
activation. Here \(P_v^{B_y}\) denotes the potential obtained from the
pre-activation state by a hypothetical Black claim at \(y\), and
\((P_v')^{B_y}\) denotes the corresponding potential starting from the
post-activation state.

Recall that
$M_v
=
\max\left\{
0,\,
\max_{{y\in E_v^*, y\text{ unused}}}
\bigl(P_v^{B_y}-P_v\bigr)
\right\}.$
Consequently,
\[I_v=P_v+M_v
=
\max\left\{
P_v,\,
\max_{{y\in E_v^*, y\text{ unused}}}
P_v^{B_y}
\right\}.\]
The effective board after activation is contained in the effective
board before activation. Since \(P_v'\le P_v\) and
\((P_v')^{B_y}\le P_v^{B_y}\) for every point \(y\) that remains
available to \(v\), we obtain
$I_v'=
\max\left\{
P_v',\,
\max_{{y\text{ still counted by }v,
                 y\text{ unused}}}
(P_v')^{B_y}
\right\} \le
\max\left\{
P_v,\,
\max_{{y\in E_v^*, y\text{ unused}}}
P_v^{B_y}
\right\}
=
I_v$.
This proves \textup{(P4)}.

At exhaustion, if a requirement has $r_Q>0$ and $n_Q=0$,
its weight in \eqref{eq:cf-potential} is greater than one.
Thus an unmet quota forces $P_v>1$, giving the
failure-detection part of (P1).

\paragraph{Verification of \textup{(P1)}: initial potentials.}
It remains to verify the initial part of (P1). We prove by induction
on the level \(j\) that every level-\(j\) vertex satisfies
\begin{equation}\label{eq:cf-initial-induction}
(12P_v(0))^{1/h_j}
\le
\frac{\varepsilon}{8(j+1)^2}.
\end{equation}

Assume this has been proved at all lower levels. For a requirement
\(Q\in\mathcal A_v\), cubes at any fixed lower level are disjoint, so
$\sum_{\substack{u\in\mathcal D_v\\l(u)=k}}m_{Qu}\le |Q|$.
Hence, by the induction hypothesis,
\[
\sum_{u\in\mathcal D_v}
m_{Qu}(12P_u(0))^{1/h_{l(u)}}
\le
|Q|\sum_{k<j}\frac{\varepsilon}{8(k+1)^2}
<\frac{\varepsilon}{4}|Q|.
\]
Initially 
$r_Q=\left(\frac12-\frac{\varepsilon}{2}\right)|Q|$, and 
$n_Q=|Q|$,
so the direct part of the exponent in
\eqref{eq:cf-potential} equals
$-\frac{\varepsilon}{2}|Q|$.
After the descendant correction, the bracket in the exponent is
therefore at most \(-\varepsilon|Q|/4\), and
$p_Q(0)\le
e^{-\varepsilon\lambda_j|Q|/4}$.

There are at most \(H_j^{2d}\) rectangle labels in an
\(H_j\)-cube, and every retained requirement has
\(|Q|\ge\alpha_j\). Consequently
\[
P_v(0)
\le
H_j^{2d}
e^{-\varepsilon\lambda_j\alpha_j/4}
=
\exp\left\{
\left(
2d\log2-\frac{\varepsilon C}{4}(j+1)
\right)h_j
\right\}.
\]
Since
$\frac{\varepsilon C}{4}
=
2d\log2+1+\log\frac{16}{\varepsilon}$,
and \(h_j\ge10\), the last bound implies
$12P_v(0)
\le
\left(\frac{\varepsilon}{8(j+1)^2}\right)^{h_j}$.
This is \eqref{eq:cf-initial-induction} and closes the induction.
In particular \(P_v(0)<1/16\) at every level, proving the initial
part of (P1).

\paragraph{Verification of \textup{(G1)--(G2)}.}
We have
$\gamma_j=(8C/\varepsilon)(j+1)^3h_j\Lambda_j$.
Since $h_j=\Lambda_j^a$ and
$\Lambda_j=B(j+1)^3\Lambda_{j-1}$, division by
$h_{j-1}^{1+\tau}$ leaves, apart from a constant and
a polynomial factor in $j$, the factor
$\Lambda_{j-1}^{\,1-a\tau}$.
Our choice $a\tau>1$ makes this tend rapidly to zero.
Therefore
\[
\gamma_j\le
c_{d,\varepsilon,\tau}h_{j-1}^{1+\tau}
\]
uniformly in $j$. This is (G2). By construction,
$\alpha_j/\gamma_j=\delta_j$ and
$\sum_j\delta_j\le\varepsilon/4$, giving (G1).
This proves Claim~\ref{claim:cf-potential-existence}.
\end{proof}

\section{Structural values and profile realization}
\label{app:zero-value}

We use the notation of \cite[Theorem~2.4]{KWBanach}; the computation is
included only to show explicitly how the structural invariant recovers
the earlier zero-density example.
\begin{prop}
\label{prop:prior-zero-value}
Let \(\X_0\) be the language family constructed in
\cite[Theorem~2.4]{KWBanach}, and let
\(K_0=\mathbb N\in\X_0\). Then
$b_n(K_0)=0$ for every $n\geq 1$.
Consequently, \(b(K_0)=0\).
\end{prop}

\begin{proof}
Throughout this construction, we interpret
$[\ell]=\{0,1,\ldots,\ell-1\}$.
Fix \(n\ge1\). We construct an admissible obstruction tree \(T_n\)
rooted at \(K_0=\mathbb N\).

The children of the root are the languages
$L_{\ell,n,0,\emptyset}$ for $\ell=0,1,2,\ldots.$
More generally, if a vertex is labelled
$L_{\ell,a,d,(h_1,\ldots,h_\tau)}$ with $a > 0$,
its children are
$L_{\ell,a-1,d+1,(k,h_1,\ldots,h_\tau)}$
for all sufficiently large \(k\). Vertices whose label has \(a=0\)
are leaves.

These are precisely the perfect-tower relations in the construction
of \(\X_0\). Thus every child is a proper sublanguage of its parent,
and every finite subset of the parent is contained in some child.
At the root, every finite subset of \(\mathbb N\) is contained in
\(L_{\ell,n,0,\emptyset}\) once \(\ell\) is sufficiently large.
The parameter \(a\) decreases at every subsequent edge, so \(T_n\)
has no infinite path. Hence \(T_n\) is admissible.

Now fix a root-to-leaf path \(H\). Apart from its root, its successive
labels have the form
$L_{\ell,n-j,j,\mathbf h_j}$ for $0\le j\le n$,
for suitable finite tuples \(\mathbf h_j\).
Recall that \(P[r]\) denotes the first \(r\) elements of the set \(P\)
from the construction in \cite{KWBanach}.  By the increasing-gap
property of \(P\), choose \(p\in P\) arbitrarily far out so that
$\bigl((P\setminus\{p\})+[0,n]\bigr)\cap[p,p+n]=\emptyset$,
and so that the interval
$J=[p,p+n]\cap\mathbb Z$
is disjoint from \([\ell]\) and from every finite exceptional set
\(P[h]+[0,n]\) appearing in the labels along \(H\).

The first condition ensures that, inside \(J\), the base term
\(P+[0,n-j]\) contributes only the translate
\(p+[0,n-j]\); the second condition removes the initial segment and all
finite exceptional terms.  Therefore for $0 \leq j \leq n$,
\[
L_{\ell,n-j,j,\mathbf h_j}\cap J
=
\{p,p+1,\ldots,p+n-j\}.
\]

Thus the first layer
\(\mathbb N\setminus L_{\ell,n,0,\emptyset}\)
does not meet \(J\); each subsequent difference layer meets \(J\)
in exactly one point; and the final language also meets \(J\) in
exactly one point. Every layer of \(H\) therefore contains at most
one point of \(J\), and hence
\[
p(H,J)
=
\frac1{|J|}
\sum_{D\text{ a layer of }H}
\left\lfloor\frac{|D\cap J|}{2}\right\rfloor
=0.
\]

Such intervals \(J\) can be chosen arbitrarily far from the origin,
and \(|J|=n+1\ge n\). Therefore \(c_n(H)=0\).
This holds for every root-to-leaf path of \(T_n\), so
\[
b_n(K_0)
\le
\sup_{H\text{ a root-to-leaf path of }T_n}c_n(H)
=0.
\]
Since pairing scores are nonnegative, \(b_n(K_0)=0\).
Letting \(n\to\infty\) gives \(b(K_0)=0\).
\end{proof}

In fact, every value between these two extremes occurs as an exact
optimum. We prove this fact by using the definition of $b_n(K)$. 


\begin{lem}[Every optimal value occurs in one dimension]
\label{cor:full-range}
For every \(\alpha\in[0,1/2]\), there is a nonempty countable
family \(\collec\) of infinite subsets of \(\mathbb Z\) and
\(K\in\collec\) whose optimal lower Banach-density guarantee is
\(b_{\collec}(K)=\alpha\).
\end{lem}

\begin{proof}
For \(\alpha=1/2\), take \(\collec=\{\mathbb N\}\).
Otherwise let \(\rho=2\alpha\) and define the {\it common core}
\(C=\{j\ge0:\lfloor(j+1)\rho\rfloor-\lfloor j\rho\rfloor=1\}\).
Telescoping gives
\(\bigl||C\cap J|-\rho|J|\bigr|<1\)
for every integer interval \(J\subseteq\mathbb N\).

Let \(\mathcal X_0\) be the family from
Proposition~\ref{prop:prior-zero-value}.
Enumerate numbers in \(\mathbb N\setminus C\) increasingly as
\(b_0<b_1<\cdots\), and define \(\phi(k)=b_k\).
For each language \(L\in\mathcal X_0\), define a new language 
\(\widetilde L=C\cup\phi(L)\), where
\(\phi(L)=\{\phi(k):k\in L\}\).
Thus we place a copy of each original language outside \(C\)
and add the same common core \(C\) to every language.
Our new family is
\(\mathcal X_\alpha=\{\widetilde L:L\in\mathcal X_0\}\),
\(C\cup\phi(\mathbb N)=\mathbb N\).
The distinguished target remains \(K = C\cup\phi(\mathbb N)=\mathbb N\).

Every language in the new family contains \(C\). Hence, for
any descending history \(H\), its terminal language contains
\(C\), so all points of \(C\cap J\) lie in the terminal layer for any fixed interval $J$.
Consequently,
\(p(H,J)\ge\lfloor|C\cap J|/2\rfloor/|J|
\ge\alpha-1/|J|\).
Since this holds for every history and every relative interval
\(J\) with \(|J|\ge n\), we have
\(b_n(K)\ge\alpha-1/n\).

We now prove the converse direction. Fix \(n\ge1\) and
\(\eta>0\), and choose
$m\ge \max\{n,\lceil 1/\eta\rceil\}$. Let \(T_m\) be the admissible obstruction tree constructed
in the proof of Proposition~\ref{prop:prior-zero-value}.
Keep the same tree, but replace each vertex label \(L\)
by \(\widetilde L=C\cup\phi(L)\).
The resulting tree remains admissible. Indeed, strict
inclusions are preserved. Moreover, for any finite
\(F\subseteq\widetilde L\), the original child condition
provides a child \(M\subsetneq L\) containing
\(\phi^{-1}(F\setminus C)\).
Its new label \(C\cup\phi(M)\) therefore contains \(F\).
Finally, relabelling does not change the well-foundedness
of the tree.

For each terminal history, the original construction supplies
arbitrarily distant intervals \(I=[p,p+m]\cap\mathbb N\) meeting
each layer in at most one point.
Set \(J=[\phi(p),\phi(p+m)]\cap\mathbb N\).
Every nonterminal layer meets \(J\) in at most one point, while
the terminal layer contains \(C\cap J\) and at most one other point.
Thus \(p(H,J)\le
\lfloor(|C\cap J|+1)/2\rfloor/|J|
\le\alpha+1/|J|<\alpha+\eta\).
Since \(|J|\ge m+1\ge n\) and these intervals occur arbitrarily far
out, every terminal history has \(c_n(H)\le\alpha+\eta\).
Therefore \(b_n(K)\le\alpha+\eta\).
Letting \(\eta\downarrow0\) and then \(n\to\infty\) proves the claim
by Theorem~\ref{thm:main-oneD}.
\end{proof}

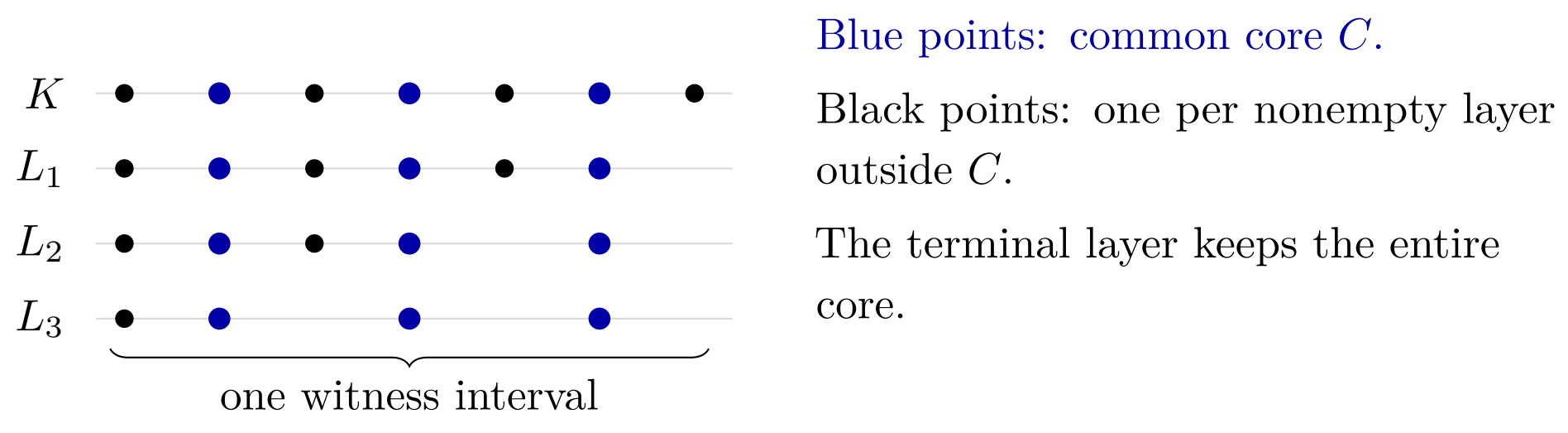
\begin{figure}[htbp]
\centering
\begin{tikzpicture}[x=.76cm,y=.60cm,font=\small]
  \foreach \y/\lab in {3/{K},2/{L_1},1/{L_2},0/{L_3}} {
    \draw[gray!30] (-.3,\y)--(6.4,\y);
    \node[left] at (-.45,\y) {$\lab$};
    \foreach \x in {1,3,5}
      \fill[blue!65!black] (\x,\y) circle (2.5pt);
  }
  \foreach \x in {0,2,4,6}
    \fill (\x,3) circle (2.1pt);
  \foreach \x in {0,2,4}
    \fill (\x,2) circle (2.1pt);
  \foreach \x in {0,2}
    \fill (\x,1) circle (2.1pt);
  \fill (0,0) circle (2.1pt);
  \draw[decorate,decoration={brace,mirror,amplitude=4pt}]
    (-.15,-.4)--(6.15,-.4);
  \node[below] at (3,-.58) {one witness interval};
  \node[anchor=west,align=left,text width=6.0cm] at (7.1,2.0)
    {\textcolor{blue!65!black}{Blue points: common core \(C\).}\\[3pt]
     Black points: one per nonempty layer outside \(C\).\\[3pt]
     The terminal layer keeps the entire core.};
\end{tikzpicture}
\caption{The scalar construction adds the same balanced core to
every language. The other points can still be separated into
singleton layers. A core of local density \(2\alpha\) therefore
contributes asymptotic pairing score \(\alpha\).
The picture shows one terminal history, with an empty layer omitted.}
\label{fig:scalar-padding}
\end{figure}

\begin{lem}[Realization of prescribed optimal profiles]
\label{prop:profile-realization}
Fix \(d\ge2\) and a positive nondecreasing threshold
\(h:\mathbb N_{\ge1}\to(0,\infty)\) satisfying
\(\sup_N N/h(N)=\infty\).
For every sequence \((\alpha_i)_{i\ge1}\in[0,1/2]^{\mathbb N}\),
there is a countable family of distinct infinite languages
\(\collec=\{K_i:i\ge1\}\) in \(\mathbb Z^d\) whose region of
simultaneously achievable \(h\)-filtered density profiles is
exactly \(\prod_{i\ge1}[0,\alpha_i]\).
In particular, one universally eventually valid generator attains
the greatest profile \((\alpha_i)_i\).
The family may depend on \(h\).
\end{lem}

\begin{proof}
We construct the languages in \(\mathbb N\); the higher-dimensional
statement will follow by embedding them on a coordinate ray.

In one dimension, the guaranteed density is governed by the pairing
score of a refinement chain. We use singleton and pair atoms, defined below, to realize
prescribed pairing scores.

\paragraph{Building blocks: witness words.}
An \emph{atom} is a block of one or two consecutive integers.
An atom of size one is a \emph{singleton atom}, denoted
\(\mathsf S\), and an atom of size two is a \emph{pair atom},
denoted \(\mathsf P\). Every language \(L_q\) constructed below
will be a union of whole atoms.

An \emph{atom word} is a finite sequence of the symbols
\(\mathsf S\) and \(\mathsf P\). If \(W\) has \(m\) atoms and
\(p(W)\) of them are pair atoms, then \(W\) occupies
$v(W)=m+p(W)$
integers. When the atoms of \(W\) lie in distinct layers of a
refinement history, every singleton atom contributes zero to the
pairing numerator and every pair atom contributes one. Thus the
pairing score of the word is \(p(W)/v(W)\).

We construct witness words whose pairing scores approach the
prescribed value \(\alpha_q\). Since \(v(W)=m+p(W)\), the relation
$\frac{p(W)}{v(W)}\approx\alpha_q$
is equivalent to
$\frac{p(W)}{m}\approx
\theta_q:=\frac{\alpha_q}{1-\alpha_q}.$
Thus a word of \(m\) atoms should contain approximately
\(m\theta_q\) pair atoms.

For every positive integer \(m\), let \(W_{q,m}\) be a word of
\(m\) atoms with 
\(\lfloor m\theta_q\rfloor\) pair atoms. To make the pair atoms distribute as evenly as possible, let its \(r\)-th atom be a pair precisely when
$\lfloor(r+1)\theta_q\rfloor-\lfloor r\theta_q\rfloor=1$ for $0 \leq r < m$,
and is a singleton otherwise.

The floor differences telescope, so 
\[
\frac{p(W_{q,m})}{v(W_{q,m})}\le\alpha_q,
\]
and this ratio tends to \(\alpha_q\) as \(m\to\infty\).

Moreover, every consecutive subword \(U\) of \(W_{q,m}\)
satisfies $p(U)-\alpha_q v(U)>-1$.
This local balance will be used for the lower bound, since lower
Banach density must be controlled on every interval, not only on
complete witness words.

\paragraph{Partition $\mathbb N$ into atoms and construction of languages.}
We now construct a partition of \(\mathbb N\) into singleton and pair atoms in stages, assigning to each atom a rational label in \((0,1)\). The candidate languages are then defined by thresholding these labels.

Fix an enumeration 
$\mathbb Q\cap(0,1)=\{q_1,q_2,\ldots\}$,
and transfer the prescribed density sequence to the rational cutoffs by setting
\(\alpha_{q_i}:=\alpha_i\).
After the construction is complete, for each
\(q\in\mathbb Q\cap(0,1)\), define
\[
L_q
=
\bigcup\{A:A\text{ is an atom with rational label }r<q\}.
\]
Thus each rational cutoff \(q\) determines a candidate language \(L_q\), and the resulting family is
\[
\collec=\{L_q:q\in\mathbb Q\cap(0,1)\}.
\]

We use a positive integer \(s\) to control the construction:
roughly speaking, at parameter \(s\), we consider the first \(s\) candidate languages $L_{q_1}, \dots, L_{q_s}$. We 
insert \(s\) pair atoms, called local padding pairs, after each base atom, and use
a witness word of length \(m_s\), chosen sufficiently large below.

For each positive integer \(s\), choose an integer \(m_s\),
strictly increasing with \(s\), such that
$m_s\ge s$ and 
$\frac{m_s}{h((2+2s)m_s)}\ge s$.
Such a choice is possible because \(h\) is nondecreasing and
\(\sup_N N/h(N)=\infty\).
The parameter \(s\) determines both the witness length \(m_s\)
and the number of local padding pairs used below.

Fix \(q\in\mathbb Q\cap(0,1)\) and \(s\ge1\).
Among \(q_1,\ldots,q_s\), let
\[
a_{q,s}
=
\max\bigl(\{0\}\cup\{q_j:j\le s,\ q_j<q\}\bigr).
\]
Thus \(a_{q,s}\) is the largest cutoff among the first \(s\)
languages that are proper subsets of \(L_q\), or zero if there
is no such language. The labels determine exactly which languages contain each atom:
an atom labelled \(r\) belongs to \(L_q\) precisely when \(r<q\).

For each pair \((q,s)\), consider every strictly decreasing sequence
$C=(u_0,\ldots,u_{m_s-1})$
of rational numbers satisfying
$q=u_0>u_1>\cdots>u_{m_s-1}>a_{q,s}$.
It determines the descending language chain
$H_C=
\bigl(
L_q=L_{u_0}\supsetneq L_{u_1}\supsetneq\cdots
\supsetneq L_{u_{m_s-1}}
\bigr).$
The witness block associated with \(C\) will force this chain to have
pairing score at most \(\alpha_q\) on a suitable relative interval.

There are countably many triples \((q,s,C)\) satisfying these
conditions. Enumerate them as \(R_1,R_2,\ldots\), and schedule them
in the sequence
\[
R_1;\quad
R_1,R_2;\quad
R_1,R_2,R_3;\quad
R_1,R_2,R_3,R_4;\quad\ldots.
\]
Thus every triple \((q,s,C)\) is scheduled infinitely often.

\paragraph{Inductive construction.}
At stage \(k\), let
$(q^\ast,s,C)$
be the \(k\)-th scheduled triple, where
$C=(u_0,\ldots,u_{m_s-1})$ and 
$q^\ast=u_0>\cdots>u_{m_s-1}>a_{q^\ast,s}$.
Let
\[
H=
(L_{q^\ast} = L_{u_0}\supsetneq L_{u_1}\supsetneq\cdots
\supsetneq L_{u_{m_s-1}})
\]
be the corresponding descending language chain.

Suppose the first \(k-1\) stages have partitioned
\(\{0,\ldots,N-1\}\) into labelled singleton and pair atoms.
For \(k=1\), take \(N=0\).
Stage \(k\) begins at the next unused integer \(N\).

The witness word \(W_{q^\ast,m_s}\) prescribes \(m_s\) atoms.
We call them the \emph{base atoms} of the stage.
Any additional pair atoms introduced below are called
\emph{padding pairs}; they are not among the \(m_s\) base atoms.

\paragraph{Base atoms and local padding.}
Read \(W_{q^\ast,m_s}\) from left to right.
For each symbol, use the next unused integer for a singleton,
or the next two unused integers for a pair.

For \(0\le r<m_s-1\), give the \(r\)-th base atom a rational label $\ell_r\in(u_{r+1},u_r)$.
Give the final base atom a rational label
$\ell_{m_s-1}\in(a_{q^\ast,s},u_{m_s-1})$.

Next, let
\[
b_{q^\ast,s}
=
\min\bigl(\{1\}\cup
\{q_j:j\le s,\ q_j>q^\ast\}\bigr).
\]
Immediately after each base atom, insert exactly \(s\) pair
atoms, all having rational labels in
\((q^\ast,b_{q^\ast,s})\).
These are the \emph{local padding pairs}.

Let \(B\) be the integer interval occupied by the base atoms and
their local padding pairs. We call \(B\) the
\emph{witness block}. See Figure \ref{fig:one-stage-spp} for an illustration.

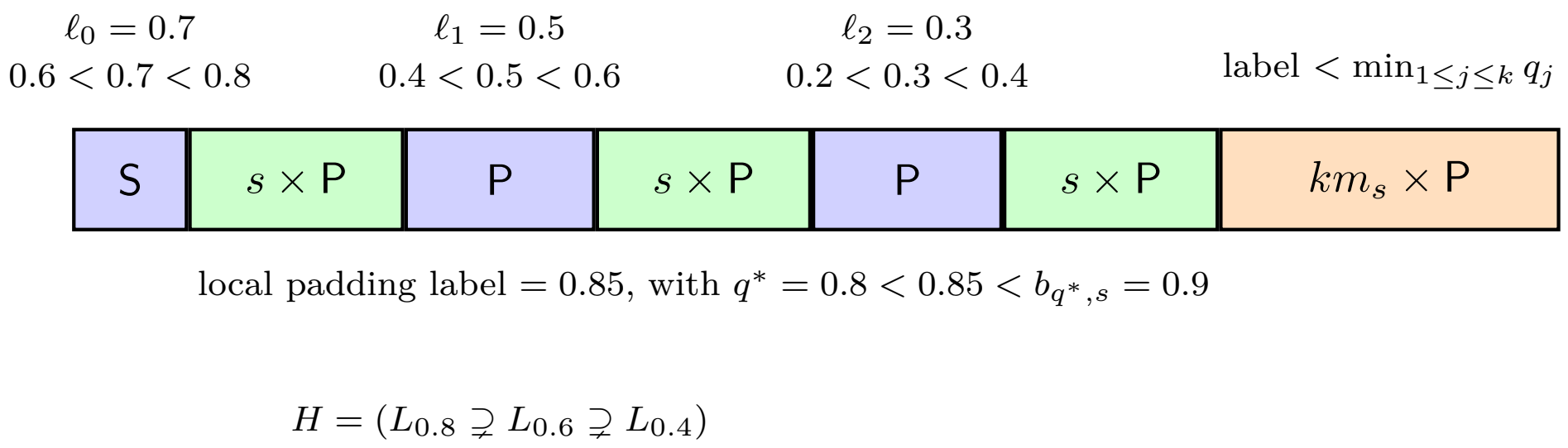
\begin{figure}[t]
\centering
\begin{tikzpicture}[
    atom/.style={
        draw,
        thick,
        minimum height=0.8cm,
        font=\small,
        inner sep=0pt
    },
    witnessS/.style={
        atom,
        minimum width=0.9cm,
        fill=blue!18
    },
    witnessP/.style={
        atom,
        minimum width=1.5cm,
        fill=blue!18
    },
    localpad/.style={
        atom,
        minimum width=1.7cm,
        fill=green!20
    },
    endpad/.style={
        atom,
        minimum width=2.7cm,
        fill=orange!25
    },
    note/.style={
        font=\scriptsize,
        align=center
    }
]


\node[witnessS,anchor=west] (A1) at (0,0)
    {\(\mathsf S\)};

\node[localpad,anchor=west] (G1) at (A1.east)
    {\(s\times\mathsf P\)};

\node[witnessP,anchor=west] (A2) at (G1.east)
    {\(\mathsf P\)};

\node[localpad,anchor=west] (G2) at (A2.east)
    {\(s\times\mathsf P\)};

\node[witnessP,anchor=west] (A3) at (G2.east)
    {\(\mathsf P\)};

\node[localpad,anchor=west] (G3) at (A3.east)
    {\(s\times\mathsf P\)};

\node[endpad,anchor=west] (E) at (G3.east)
    {\(k m_s\times\mathsf P\)};

\node[note,above=5pt of A1]
    {\(\ell_0=0.7\)\\
     \(0.6<0.7<0.8\)};

\node[note,above=5pt of A2]
    {\(\ell_1=0.5\)\\
     \(0.4<0.5<0.6\)};

\node[note,above=5pt of A3]
    {\(\ell_2=0.3\)\\
     \(0.2<0.3<0.4\)};

\node[note,below=5pt of G2]
    {local padding label \(=0.85\), with
     \(q^\ast=0.8<0.85<b_{q^\ast,s}=0.9\)};

\node[note,above=5pt of E]
    {label \(<\min_{1\le j\le k}q_j\)};

\node[note,below=1.25cm of A2]
    {\(H=(L_{0.8}\supsetneq L_{0.6}\supsetneq L_{0.4})\)};

\end{tikzpicture}

\caption{
Illustration of one stage for
\(W_{q^\ast,m_s}=\mathsf S\mathsf P\mathsf P\).
Here
\(q^\ast=u_0=0.8\), \(u_1=0.6\), \(u_2=0.4\),
and \(a_{q^\ast,s}=0.2\).
The three witness atoms receive labels
\(0.7,0.5,0.3\), placing them respectively in
\(L_{0.8}\setminus L_{0.6}\),
\(L_{0.6}\setminus L_{0.4}\), and
\(L_{0.4}\).
The local padding pairs have label \(0.85\), so they do not
belong to \(L_{0.8}\), while the stage-ending padding is assigned
a sufficiently small label.
}
\label{fig:one-stage-spp}
\end{figure}

The rational labels and the construction of languages $L_q$ based on $q$ have the following  immediate consequences.
\begin{enumerate}[label=(\roman*),leftmargin=*]
\item
Every base atom belongs to \(L_{q^\ast}\).
More precisely, for \(0\le r<m_s-1\), the \(r\)-th base atom
belongs to 
$L_{u_r}\setminus L_{u_{r+1}}$,
while the final base atom belongs to the terminal language
\(L_{u_{m_s-1}}\).
Thus the \(m_s\) base atoms lie in distinct layers of the
language chain \(H\).

\item
If \(j\le s\) and \(q_j<q^\ast\) (thus $L_{q_j} \subset L_{q^\ast}$), then no atom in the witness
block belongs to \(L_{q_j}\).
Indeed, every base label is larger than
\(a_{q^\ast,s}\ge q_j\), and every local-padding label is
larger than \(q^\ast>q_j\).

\item
If \(j\le s\) and \(q_j>q^\ast\) (thus $L_{q_j} \supset L_{q^\ast}$), then every base atom and
every local padding pair belongs to \(L_{q_j}\).
The base labels are smaller than \(q^\ast<q_j\), while the
local-padding labels are smaller than
\(b_{q^\ast,s}\le q_j\).

\item
No local padding pair belongs to \(L_{q^\ast}\), since its
label is greater than \(q^\ast\).

Write the  interval (window)
$J=L_{q^\ast}\cap B$.
Then \(J\) consists exactly of the base atoms of
\(W_{q^\ast,m_s}\).
Because \(B\) is an integer interval, \(J\) is a relative
interval of \(L_{q^\ast}\).

Each base atom lies in a distinct layer of \(H\), so every
singleton contributes zero and every pair contributes one.
Consequently,
\[
p(H,J)
=
\frac{p(W_{q^\ast,m_s})}{v(W_{q^\ast,m_s})}
\le\alpha_{q^\ast}.
\]
\end{enumerate}

\paragraph{Stage-ending padding.} Recall that we are at stage \(k\), processing the \(k\)-th
scheduled triple \((q^\ast,s,C)\).
After the witness block, append exactly \(k m_s\) further pair
atoms, all having rational labels smaller than
$\min_{1\le j\le k}q_j$.

These pairs belong to all of
\(L_{q_1},\ldots,L_{q_k}\).
For every \(j\le k\), among the atoms constructed at stage \(k\)
that belong to \(L_{q_j}\), at most \(m_s\) are singleton atoms
and at least \(k m_s\) are pair atoms.
Hence the ratio of pair atoms to the total number of integers
in these atoms is at least
$\frac{k m_s}{m_s+2k m_s} =\frac{k}{2k+1}$,
which tends to \(1/2\).

The stage-ending padding lies outside \(B\), so it does not
change \(J=L_{q^\ast}\cap B\) or its pairing score.
This completes stage \(k\), and stage \(k+1\) begins at the
next unused integer.

Every stage is finite and contains at least one atom.
Since successive stages begin at the next unused integer, their
union partitions all of \(\mathbb N\).

Every language \(L_q\) is infinite.
Indeed, triples with first coordinate \(q\) occur infinitely
often, and every corresponding stage contributes all its base
atoms to \(L_q\).

The languages are distinct and ordered by their rational cutoffs.
If \(p<q\), choose \(s\) large enough that \(p\) occurs among
\(q_1,\ldots,q_s\).
Then \(a_{q,s}\ge p\).
Every base atom in a stage constructed for \(L_q\) with this
parameter \(s\) has label strictly between \(a_{q,s}\) and \(q\).
It therefore belongs to \(L_q\setminus L_p\).
Such stages occur infinitely often, and hence
$p<q$ implies $L_p\subsetneq L_q$.

We now establish the corresponding lower bounds on the density.
\begin{claim}
    For each $q \in (0,1) \cap \mathbb Q$, the construction gives $b(L_q) \geq \alpha_q$. 
\end{claim}
\begin{proof}
Fix \(q\in\mathbb Q\cap(0,1)\).
If \(\alpha_q=0\), the conclusion is immediate.
Fix \(0\le c<\alpha_q\).
For a finite word \(U\) of atoms, define its \(c\)-surplus as
\(p(U)-c\,v(U)\).

We show that every sufficiently long relative
interval of \(L_q\) has pairing score close to \(\alpha_q\) from
below. A witness block either contains no atom of \(L_q\), contains
the balanced word \(W_{q,m_s}\), or contains enough local padding
pairs to have uniformly nonnegative \(c\)-surplus up to a bounded
endpoint error. The stage-ending padding prevents these bounded
losses from accumulating across successive stages.

Choose \(S\) larger than the position of \(q\) in the fixed
enumeration and sufficiently large that
$s(1-2c)\ge c$ for every $s\ge S$.
Consider a witness block constructed with parameter \(s\ge S\),
and let \(q^\ast\) be the target cutoff of that stage.

If \(q^\ast>q\), then the witness block contains no atom of
\(L_q\), by Item~(ii) above.

If \(q^\ast=q\), then the atoms in the witness block that belong
to \(L_q\) are exactly \(W_{q,m_s}\).
Every consecutive subword \(U\) therefore satisfies
$p(U)-c\,v(U)>-1$.

If \(q^\ast<q\), then every base atom and every local padding
pair belongs to \(L_q\).
Group each base atom with the \(s\) local padding pairs
following it.
Even if the base atom is a singleton, a complete group has
\(c\)-surplus at least
$-c+s(1-2c)\ge0$.
A consecutive subword meets at most two such groups partially,
so its negative \(c\)-surplus is bounded uniformly.

For the finitely many parameters \(s<S\), a witness block has
at most \(m_s\) singleton atoms, and all padding atoms are pairs.
Their negative \(c\)-surpluses are therefore uniformly bounded.

Stage-ending padding contains only pairs.
Moreover, every sufficiently late complete stage has pair-to-point
ratio at least \(c\) among its atoms belonging to \(L_q\).
It follows that there is a constant \(C_{q,c}\) such that every
consecutive block \(U\) of whole atoms belonging to \(L_q\)
satisfies
$p(U)-c\,v(U)\ge-C_{q,c}.$

Let \(G\) be any descending history beginning at \(L_q\), and let
\(J\) be any relative interval of \(L_q\).
Removing at most one endpoint point at each end leaves a
consecutive block of whole atoms.
No language can split a pair atom, so every complete pair lies
in one layer of \(G\) and contributes one to the pairing
numerator.
Consequently,
$p(G,J)
\ge
c-\frac{C'_{q,c}}{|J|}$
for a constant \(C'_{q,c}\).

Thus \(b_n(L_q)\ge c-C'_{q,c}/n\), and hence
\(b(L_q)\ge c\).
Letting \(c\to\alpha_q\) gives
$b(L_q)\ge\alpha_q$.
\end{proof}

We now prove the matching upper bounds. 
\begin{claim}
    For each $q \in (0,1) \cap \mathbb Q$, the construction gives $b(L_q) \leq \alpha_q$. 
\end{claim}
\begin{proof}
Fix \(q\in\mathbb Q\cap(0,1)\) and \(s\ge1\).
Define an obstruction tree \(T_{q,s}\) as follows.
A vertex at depth \(r<m_s\) is a strictly decreasing rational
sequence
\[
q=u_0>u_1>\cdots>u_r>a_{q,s},
\]
and is labelled by \(L_{u_r}\).
Every nonterminal vertex has one child for every rational
\(u_{r+1}\) satisfying
$a_{q,s}<u_{r+1}<u_r$.
The leaves have depth \(m_s-1\).

This is an admissible obstruction tree.
All inclusions are proper.
If \(F\subseteq L_{u_r}\) is finite, the finitely many atoms
meeting \(F\) have labels strictly smaller than \(u_r\).
Choose a rational \(u_{r+1}\) strictly between \(u_r\) and the
larger of \(a_{q,s}\) and those finitely many labels.
Then
$F\subseteq L_{u_{r+1}}\subsetneq L_{u_r}$.
The tree has finite height, so it has no infinite path.

Every terminal history of \(T_{q,s}\) corresponds to one of the
triples used in the construction.
That triple occurs infinitely often, producing arbitrarily distant
witness blocks \(B\).
For $J=L_q\cap B$, the interval \(J\) consists exactly of the atoms of \(W_{q,m_s}\),
one atom in each layer of the terminal history.
Therefore we have $|J|\ge m_s$ and 
$p(H,J) =
\frac{p(W_{q,m_s})}{v(W_{q,m_s})}
\le\alpha_q$. 

For every fixed \(n\), choose \(s\) with \(m_s\ge n\).
Every terminal history then has \(c_n(H)\le\alpha_q\), and hence
$b_n(L_q)\le\alpha_q$. Since this holds for every \(n\) and
\(b(L_q)=\lim_{n\to\infty}b_n(L_q)\), it follows that
\(b(L_q)\le\alpha_q\).
\end{proof}

We thus have $b(L_q)=\alpha_q$ for every $q\in\mathbb Q\cap(0,1)$.

Theorem~\ref{thm:main-oneD} supplies one universally eventually valid generator attaining all these values simultaneously in
ordinary lower Banach density.
The same generator attains them in \(h\)-filtered density, since
every sequence eligible for the \(h\)-filter has target population
tending to infinity.

For the matching filtered upper bound, fix \(q\) and any
universally eventually valid generator.
Run the mega-step construction from the proof of
Theorem~\ref{thm:upper} successively with
\(T_{q,1},T_{q,2},\ldots\). At the leaf of the \(s\)-th mega-step,
choose an unused repeated witness block for its terminal history.
The root refill steps across the successive mega-steps ensure that
the resulting input sequence is a full enumeration of \(L_q\).

Such a block has target population at least \(m_s\) and ambient length at most $(2+2s)m_s$.
Its filter level is therefore at least
$\frac{m_s}{h((2+2s)m_s)} \ge s$.
The layer races finish this interval with output fraction at most
\(\alpha_q\), and freshness prevents later outputs from increasing
that fraction. Thus no generator can guarantee filtered density
greater than \(\alpha_q\) at \(L_q\).

The achievable-profile region is therefore
$\prod_{q\in\mathbb Q\cap(0,1)}[0,\alpha_q]$.
Under the identification \(\alpha_{q_i}=\alpha_i\), this is exactly
the prescribed region
$\prod_{i\ge1}[0,\alpha_i]$.

For \(d\ge2\), replace every \(L_q\subseteq\mathbb N\) by
$\widetilde L_q = \{(n,0,\ldots,0):n\in L_q\}
\subseteq\mathbb Z^d$.
Every integer interval gives a thin axis-parallel rectangle with
the same target population and volume.
Conversely, every rectangle meeting the coordinate ray has the
same target and output points on that ray as a thin rectangle
of no greater volume.
Since \(h\) is nondecreasing, filtered densities are preserved.
Transporting generators through a bijection extending the
coordinate-ray embedding preserves freshness, full enumerations,
and eventual validity.
Hence the achievable-profile region remains
$\prod_{q\in\mathbb Q\cap(0,1)}[0,\alpha_q]$.
\end{proof}

\newpage

\section{Lean formalization and verification}
\label{app:formalization}

\subsection{Verification summary}

Two separately organized Lean~4 formalization efforts
report \emph{Level~K} verification of all
65 numbered mathematical assertions in the certified
manuscript version, representing 64 distinct results
since Theorem~5.8 restates Theorem~2.6.
The verification includes Appendices~A and~B.

Level~K is our project-specific standard requiring
(i) clause-by-clause correspondence between the
manuscript and elaborated Lean statements;
(ii) complete kernel-checked proofs without proof
holes or nonstandard axioms;
(iii) formal representation bridges where required;
and (iv) no weakened conclusions or additional
unproved assumptions.

Statement correspondence was assessed through
AI-assisted audits, supplemented by author review
of selected statements. The certification concerns
mathematical statements, not a line-by-line
formalization of the written proofs.

Of the 14 numbered definitions, 13 have Lean
representations. Definition~1.2 introduces lower
asymptotic density only as background and is not
needed for the correctness of the results.
The five remarks and unnumbered explanatory material
are outside the certification scope.

\subsection{Verified statements}

Table~\ref{tab:lean-complete-coverage} lists the
numbered statements and definitions.
Here \textbf{K} denotes Level~K verification,
\textbf{D} a definition represented in Lean,
and \textbf{B} a background definition not used in the proof.

\begingroup
\small
\setlength{\LTleft}{0pt}
\setlength{\LTright}{\fill}
\renewcommand{\arraystretch}{1.07}
\begin{longtable}{@{}p{0.18\textwidth}p{0.075\textwidth}p{0.678\textwidth}@{}}
\caption{Complete coverage of the numbered mathematical assertions and definitions.}
\label{tab:lean-complete-coverage}\\
\toprule
Paper item & Level & Mathematical content \\
\midrule
\endfirsthead
\multicolumn{3}{@{}l}{\footnotesize Table~\thetable\ (continued)}\\
\toprule
Paper item & Level & Mathematical content \\
\midrule
\endhead
\midrule
\multicolumn{3}{r@{}}{\footnotesize Continued on next page}\\
\endfoot
\bottomrule
\endlastfoot

Question~1.1 & --- & Guiding research question; not an asserted mathematical result.\\
Def.~1.2 & B &
Background definition of lower asymptotic density;
not needed for the paper's mathematical results.\\
Def.~1.3 & D & Relative lower Banach density over lattice boxes.\\
Def.~1.4 & D & \(h\)-filtered lower Banach density.\\
Thm.~1.5 & K & One-dimensional structural and topological values; exact simultaneous optimality.\\
Cor.~1.6 & K & Simultaneous optimal density \(1/2\) for all finite-point-rank targets.\\
Thm.~1.7 & K & Higher-dimensional failure of simultaneous optimality and its density trade-off.\\
Thm.~1.8 & K & One-dimensional achievability and adversarial upper bound for \(b(K)\).\\
\midrule
Def.~2.1 & D & Histories, their disjoint layers, and the pairing score.\\
Claim~2.2 & K & Eventual relative-interval score bound for a fixed history.\\
Def.~2.3 & D & Admissible obstruction trees and their branching structure.\\
Claim~2.4 & K & Root-to-leaf witness in every admissible obstruction tree.\\
Thm.~2.5 & K & Exists a generator attaining the one-dimensional structural lower bound.\\
Thm.~2.6 & K & Equality of the combinatorial and topological \(b_n\) formulas.\\
\midrule
Def.~3.1 & D & Cantor--Bendixson point rank in the finite-evidence topology.\\
Lemma~3.2 & K & Rank-decreasing refinement witness.\\
Cor.~3.3 & K & Optimal value \(b(K)=1/2\) at finite-point-rank targets.\\
\midrule
Thm.~4.1 & K & Adversarial upper bound against every eventually valid generator.\\
Lemma~4.2 & K & Finite adversarial race and counting bound.\\
\midrule
Thm.~5.1 & K & One \(\varepsilon\)-dependent generator attaining approximate lower bounds for all targets.\\
Def.~5.2 & D & Safe and unsafe histories.\\
Def.~5.3 & D & The history topology.\\
Lemma~5.4 & K & Neighborhood basis and the topology of histories.\\
Lemma~5.5 & K & Finest topology compatible with refinement continuation.\\
Lemma~5.6 & K & Closure described by well-founded obstruction trees.\\
Cor.~5.7 & K & Safety as an interior/closure property.\\
Thm.~5.8 & K & Topological \(b_n\) formula; restatement of Theorem~2.6.\\
Thm.~5.9 & K & Exact optimal generation expressed through the topological value.\\
Cor.~5.10 & K & Finite protection of safe histories.\\
Def.~5.11 & D & Protected sets and their local certificates.\\
Lemma~5.12 & K & Finite board traces and quantitative board bounds.\\
Lemma~5.13 & K & Forest ancestry and protection invariant.\\
Lemma~5.14 & K & Uniform board quota and its freezing rule.\\
Lemma~5.15 & K & Record-language memory and containment.\\
Lemma~5.16 & K & Eventual capture by the forest's finite-evidence scan.\\
Lemma~5.17 & K & Eventual validity of the constructed generator.\\
Lemma~5.18 & K & Filtered pairing game and the rounded-half guarantee.\\
Cor.~5.19 & K & Density guarantee on regular boards.\\
Lemma~5.20 & K & Ancestor relation between successive accuracy copies.\\
Lemma~5.21 & K & Eventual capture of each specified copy.\\
Lemma~5.22 & K & Lower-density guarantee from a fixed accuracy copy.\\
Thm.~5.23 & K & A single one-dimensional generator attaining \(b(K)\) for every target.\\
\midrule
Thm.~6.1 & K & Explicit \(1/16\) gap from simultaneous optimality.\\
Thm.~6.2 & K & Exact \(1/4\) and \(1/2\) optima and  unavoidable trade-off.\\
Lemma~6.3 & K & Finite commitment following eventual validity.\\
Lemma~6.4 & K & Forced blocks from distinct adversarial inputs.\\
Lemma~6.5 & K & Horizontal--vertical test windows density trade-off.\\
Claim~6.6 & K & Horizontal tests produced by a forced block.\\
\midrule
Thm.~7.1 & K & Necessary and sufficient protected-tree conditions for fixed quotas.\\
Thm.~7.2 & K & Exact achievable profiles via dyadic accuracy-copy trees.\\
Thm.~7.3 & K & Near-half guarantees and impossibility at the exact half endpoint.\\
Def.~7.4 & D & Finite geometric stopping game and quota feasibility.\\
Def.~7.5 & D & Geometric \(q\)-protected trees.\\
Def.~7.6 & D & Good geometric decorations and neighborhoods.\\
Thm.~7.7 & K & Equivalence of protected trees and decorated topological neighborhoods.\\
Cor.~7.8 & K & Global geometric assignments and all admissible obstruction trees.\\
Cor.~7.9 & K & Finite protection in a good neighborhood.\\
Lemma~7.10 & K & Capture in the higher-dimensional forest.\\
Lemma~7.11 & K & Logarithmic and weighted geometric sewing estimates.\\
Lemma~7.12 & K & Simulation of local games by one actual generator.\\
Lemma~7.13 & K & Commitment, including the output at the last committing input.\\
Def.~7.14 & D & Protected trees on indexed accuracy copies.\\
Lemma~7.15 & K & Ancestry of successive copies in the nonstrict forest.\\
Lemma~7.16 & K & Eventual capture of a specified higher-dimensional copy.\\
Lemma~7.17 & K & Copy-specific filtered density for the actual dyadic generator.\\
\midrule
Lemma~8.1 & K & One causal strategy meeting a finite packet of rectangle quotas, \(d\ge1\).\\
Thm.~8.2 & K & Simultaneous \(1/2-\varepsilon\) for all finite-point-rank targets, with no common rank bound.\\
\midrule
Thm.~9.1 & K & Impossibility of exact half-density under the endpoint threshold condition.\\
Lemma~9.2 & K & Adversarial strip strategy, including arbitrary White responses.\\
\midrule
Def.~A.1 & D & Potential-triggered activation rule.\\
Lemma~A.2 & K & One potential-based causal strategy meeting all local quotas.\\
Claim~A.3 & K & Bounding discarded small dyadic pieces.\\
Claim~A.4 & K & Local quotas imply the global rectangle bound.\\
Prop.~A.5 & K & Conditional rectangle strategy from hypotheses (G1)--(G3).\\
Claim~A.6 & K & Uniform construction of scales and potentials satisfying (P1)--(P4).\\
\midrule
Prop.~B.1 & K & Zero structural density for the specific family from earlier work, reconstructed in Lean.\\
Lemma~B.2 & K & Every optimal value in \([0,1/2]\) occurs in one dimension.\\
Lemma~B.3 & K & Realization of every prescribed product as the exact achievable-profile frontier.\\
Claim~B.4 & K & Lower bound on \(b(L_q)\) for the constructed rational-cut languages.\\
Claim~B.5 & K &
Matching upper bound $b(L_q)\le\alpha_q$
for the constructed languages.\\
\end{longtable}
\endgroup

The complete certificate identifies the corresponding
Lean declarations for every numbered assertion.
In particular, Corollary~1.6 includes the empty-family
case, and Proposition~B.1 formalizes the required
properties of the specific family from earlier work.
No unformalized theorem from that work is used
as an additional axiom or unproved assumption.

\subsection{Lean formalization and reproducibility}

Two separately coordinated Lean formalization efforts
were carried out: an author-coordinated Codex development
and a development coordinated by Daniel Zhou. The latter
incorporated, with attribution, two auxiliary components
from the former.

The verification reports, certificates, and
statement-correspondence records are available at the second author's homepage
\begin{center}
\url{https://sites.google.com/view/fan-wei/language-generation-nov-3-formalization}
\end{center}
The author-coordinated Lean sources and reproducibility
scripts will also be made available there. Daniel Zhou's
Lean sources remain in his separate repository.

The author-coordinated verification passed clean macOS
and Linux builds, standard-axiom checks, and kernel
replay. Further details and qualifications are provided
in the accompanying reports.

\newpage
\paragraph{Acknowledgments and support.}

Jon Kleinberg was supported in part by NSF MFAI 2601664, a Vannevar
Bush Faculty Fellowship, AFOSR grant FA9550-23-1-0410, a Simons
Collaboration grant, and a grant from the MacArthur Foundation.
Fan Wei was supported by NSF grant DMS-2401414 and NSF MFAI 2601663.

\paragraph{AI methodology.}
Generative AI tools assisted in the development of the multiscale sensitivity discrepancy lemma in Appendix \ref{app:geometry}. Our initial approach used a term linear in $P_v$ on the right-hand side of \eqref{eq:cf-potential}, yielding a degeneracy bound with $h$ given by a higher-degree polynomial in $\log V$. To obtain the sharper bound $h(V)=\log^{1+\tau}(2+V)$, ChatGPT suggested replacing this linear term with the nonlinear expression $(12P_u)^{1/h_{l(u)}}$ and adjusting the accompanying parameter choices. ChatGPT then provided a proof of Claim \ref{claim:cf-potential-existence}, establishing the existence of a potential function satisfying the conditions of Proposition \ref{prop:cf-conditional}. This was a key observation enabling the improvement from a higher-degree polylogarithmic bound to $\log^{1+\tau}$. Generative AI tools were also used to polish the introduction
and mathematical proofs and to produce the figures.
The authors carefully verified the mathematical arguments and
take full responsibility for the correctness and originality
of the work.

\paragraph{Formalization and verification.}
We thank Daniel Zhou for sharing his formalization and auditing protocol, which we adapted for this work. With the assistance of Claude and ChatGPT, Lean 4 formalizations in Mathlib and systematic correspondence audits were carried out in two separately organized efforts. The final verification reports document complete, kernel-checked Lean proofs covering all 64 distinct numbered mathematical statements in the certified manuscript version, including the results in Appendices A and B, under the recorded model conventions. The formal proofs contain no proof holes or additional nonstandard axioms. Separate AI-based correspondence audits compared the formal statements with the manuscript's definitions, hypotheses, quantifiers, and conclusions. Appendix~\ref{app:formalization} describes the respective coverage of the two efforts, their shared auxiliary components, the certified source versions, and the manuscript-to-formalization correspondence qualifications. Detailed reports and certificates are collected at \url{https://github.com/fan-wei-math/characterizing-local-breadth-lean}.

\bibliographystyle{alpha}
\bibliography{AIbib}

\end{document}